\documentclass[prd,superscriptaddress,unsortedaddress,twocolumn,showpacs,floatfix]{revtex4}

\usepackage{epsfig}
\usepackage{dcolumn}
\usepackage{amsmath}

\def\ket#1{\left| #1\right\rangle}

\def\nscat#1{N_{\rm scat}^{#1}}
\def\ncoh#1{N_{\rm coh}^{#1}}

\newcommand{\Kz}{\mbox{$K^{0}$}}
\newcommand{\Kzbar}{\mbox{$\overline{K^{0}}$}}
\newcommand{\KL}{\mbox{$K_{L}$}}
\newcommand{\KS}{\mbox{$K_{S}$}}
\newcommand{\KSKL}{\mbox{$K_{S}$-$K_{L}$}}
\newcommand{\ptsq}{\mbox{$p_T^2$}}
\newcommand{\ZK}{\mbox{$z$}}
\newcommand{\PK}{\mbox{$p$}}
\newcommand{\vw}{vacuum-window}
\newcommand{\brems}{Bremsstrahlung}
\newcommand{\tgtKS}{target-$K_S$}

\newcommand{\tauS}{\mbox{$\tau_{S}$}}
\newcommand{\tauL}{\mbox{$\tau_{L}$}}
\newcommand{\delm}{\mbox{$\Delta m$}}
\newcommand{\delg}{\mbox{$\Delta \Gamma$}}
\newcommand{\epe}{\mbox{$\epsilon'\!/\epsilon$}}
\newcommand{\treg}{\mbox{$T_{reg}$}}
\newcommand{\epspr}{\mbox{$\epsilon^\prime$}}
\newcommand{\epsrat}{\mbox{$\epsilon^{\prime}\!/\epsilon$}}
\newcommand{\reepoe}{\mbox{$Re(\epsrat)$}}
\newcommand{\imepoe}{\mbox{$Im(\epsrat)$}}
\newcommand{\ktev}{\mbox{KTeV}}
\newcommand{\etapm}{\mbox{$\eta_{+-}$}}
\newcommand{\etazz}{\mbox{$\eta_{00}$}}
\newcommand{\phipm}{\mbox{$\phi_{+-}$}}
\newcommand{\phizz}{\mbox{$\phi_{00}$}}
\newcommand{\phisw}{\mbox{$\phi_{SW}$}}
\newcommand{\delphi}{\mbox{$\Delta \phi$}}

\newcommand{\Kpp}{\mbox{$K\rightarrow\pi\pi$}}
\newcommand{\Kpm}{\mbox{$K\rightarrow\pi^{+}\pi^{-}$}}
\newcommand{\Kzz}{\mbox{$K\rightarrow\pi^{0}\pi^{0}$}}
\newcommand{\KzzD}{\mbox{$K\rightarrow\pi^{0}\pi^{0}_D$}}
\newcommand{\Kpmg}{\mbox{$K\rightarrow\pi^{+}\pi^{-}\gamma$}}

\newcommand{\Kethree}{\mbox{$K_{e3}$}}
\newcommand{\Kmuthree}{\mbox{$K_{\mu 3}$}}
\newcommand{\KLzgg}{\mbox{$K_L\rightarrow\pi^{0}\gamma\gamma$}}
\newcommand{\KLpienu}{\mbox{$K_{L}\to\pi^{\pm}e^{\mp}\nu$}}
\newcommand{\KLpimunu}{\mbox{$K_{L}\to\pi^{\pm}{\mu}^{\mp}\nu$}}
\newcommand{\KLzzz}{\mbox{$K_{L}\rightarrow \pi^{0}\pi^{0}\pi^{0}$}}
\newcommand{\KLpmz}{\mbox{$K_{L}\rightarrow \pi^{+}\pi^{-}\pi^{0}$}}

\newcommand{\KLpm}{\mbox{$K_{L}\rightarrow\pi^{+}\pi^{-}$}}
\newcommand{\KSpm}{\mbox{$K_{S}\rightarrow\pi^{+}\pi^{-}$}}
\newcommand{\KLzz}{\mbox{$K_{L}\rightarrow\pi^{0}\pi^{0}$}}
\newcommand{\KSzz}{\mbox{$K_{S}\rightarrow\pi^{0}\pi^{0}$}}

\newcommand{\Lppi}{\mbox{$\Lambda \rightarrow p \pi^-$}}
\newcommand{\LBARppi}{\mbox{$\overline{\Lambda} \rightarrow \bar{p} \pi^+$}}
\newcommand{\piz}{\mbox{$\pi^{0}$}}
\newcommand{\eeg}{\mbox{$e^+e^-\gamma$}}
\newcommand{\gamgam}{\mbox{$\gamma\gamma$}}
\newcommand{\ppc}{\mbox{$\pi^{+}\pi^{-}$}}

\newcommand{\ppn}{\mbox{$\pi^{0}\pi^{0}$}}
\newcommand{\zzz}{\mbox{$\pi^{0}\pi^{0}\pi^{0}$}}

\newcommand{\ring}{{\tt RING}}
\newcommand{\chisqzz}{\chi^2_{\pi^0}}
\newcommand{\chisqvtx}{\chi^2_{vtx}}
\newcommand{\dzvtx}{\Delta z_{vtx}}
\newcommand{\zcsi}{Z_{\rm CsI}}
\newcommand{\npredict}{N_{p,z}^{\pi\pi}}

\newcommand{\eu}{ \times 10^{-4}}
\newcommand{\uptsq}{ {\rm MeV}^2/c^2 }
\newcommand{\upk}{ {\rm GeV}/c }
\newcommand{\magrho}{\vert\rho\vert}
\newcommand{\degs}{^{\circ}}
\newcommand{\delmunits}{\mbox{$\times 10^{6}~\hbar {\rm s}^{-1}$}}
\newcommand{\tausunits}{\mbox{$\times 10^{-12}~{\rm s}$}}
\newcommand{\fminus}{\mbox{$|f_{-}(70~{\rm GeV}/c)|$}}
\newcommand{\sigdz}{\mbox{$\sigma_{\Delta z}$}}
\newcommand{\zregeff}{\mbox{$z_{\rm eff}$}}
\newcommand{\Rreg}{\mbox{${\sf R}_{\sf reg}$}}

\def\EopCut{0.85}
\def\MinpCut{\mbox{$8~{\rm GeV}/c$}}

\def\XClusEcut{\mbox{$1.0~{\rm GeV}$}}
\def\XClusTrkcut{\mbox{$20~{\rm cm}$}}
\def\MassResln{\mbox{$1.6~{\rm MeV/c}^2$}}
\def\VacKelBkg{0.036}
\def\VacKmuBkg{0.054}
\def\VacColBkg{0.010}
\def\VacSctBkg{---}
\def\VacTotBkg{0.100}
\def\RegKelBkg{0.001}
\def\RegKmuBkg{0.002}
\def\RegColBkg{0.010}
\def\RegSctBkg{0.074}
\def\RegTotBkg{0.087}

\def\PtsqSyst{0.25}
\def\BkgdChrgSyst{0.20}
\def\TrCellSyst{0.22} 
\def\TrCellVar{$(-0.16 \pm 0.12) \eu$}
\def\RegEdgeSyst{0.20}
\def\ChrgZSlope{$(-0.70 \pm 0.30) \eu~{\rm m}^{-1}$}
\def\ChrgZSlopeSyst{0.79}
\def\ReepoeDelmSyst{0.05}
\def\ReepoeTausSyst{0.06}
\def\ReepoeDelmTausSyst{0.11}
\def\AttSyst{0.19}
\def\FitSyst{0.30}
\def\ZSlopeKzz{$(+0.60 \pm 0.53) \eu~{\rm m}^{-1}$}
\def\ZSlopeKzzz{$(+0.23 \pm 0.19) \eu~{\rm m}^{-1}$}
\def\NeutZSlopeSyst{0.39}
\def\BkgdNeutRegSyst{1.06}
\def\BkgdNeutSyst{1.07}
\def\TotSystMC{2.39}  
\def\L3CHRGSYST{0.54}
\def\KtevXReepoe{20.71}
\def\KtevReepoe{20.7}
\def\KtevReepoeA{23.2}
\def\KtevStat{1.48}
\def\KtevMCStatChrg{0.41}
\def\KtevMCStatNeut{0.40}
\def\KtevTErr{2.8}
\def\KtevTErrA{4.4}
\def\KtevAmp70andErr{\mbox{$1.2070\pm0.0003~{\rm mbarns}$}}
\def\KtevPwrSlpandErr{\mbox{$-0.5426\pm0.0008$}}

\def\KtevDelm{5261}
\def\KtevDelmTerr{15}
\def\KtevDelmFITerr{13.3} 
\def\KtevTaus{89.65}
\def\KtevTausTerr{0.07}
\def\KtevTausFITerr{0.045}
\def\PhiSW{43.38}
\def\PhiSWErr{0.10} 
\def\Phipm{44.12}
\def\PhipmErr{0.72}
\def\PhipmSyst{1.20}
\def\PhipmTerr{1.40}
\def\PhipmAll{\Phipm}
\def\PhipmAllErr{\PhipmErr}
\def\PhipmAllSyst{\PhipmSyst}
\def\PhipmAllTerr{\PhipmTerr}
\def\PhiAllDM{5288}
\def\PhiAllDMErr{23}
\def\PhiAllTs{89.58}
\def\PhiAllTsErr{0.08}
\def\PhiAllChi{223.6}
\def\PhiAllDOF{197}
\def\dPhiSW{+0.61}
\def\dPhiSWSTATerr{0.62}
\def\dPhiSWSYSTerr{1.01}
\def\dPhiSWTOTerr{1.19}
\def\DelPhi{+0.39}
\def\DelPhiSTATerr{0.22}
\def\DelPhiSYSTerr{0.45}
\def\DelPhiTOTerr{0.50}
\def\IMEPOE{\mbox{$-22.9$}}
\def\IMEPOEStatErr{\mbox{$12.8$}}
\def\IMEPOESystErr{\mbox{$26.2$}}
\def\IMEPOETotErr{\mbox{$29.1$}}
\def\IMEPOEpmErr{\mbox{$(\IMEPOE \pm \IMEPOETotErr) \eu$}}

\begin{document}     


\title{ 
       Measurements of Direct CP Violation, CPT Symmetry, \\
       and Other Parameters in the Neutral Kaon System
           }


\newcommand{\UAz}{University of Arizona, Tucson, Arizona 85721}
\newcommand{\UCLA}{University of California at Los Angeles, Los Angeles,
                    California 90095} 
\newcommand{\UCSD}{University of California at San Diego, La Jolla,
                   California 92093} 
\newcommand{\EFI}{The Enrico Fermi Institute, The University of Chicago, 
                  Chicago, Illinois 60637}
\newcommand{\UB}{University of Colorado, Boulder, Colorado 80309}
\newcommand{\ELM}{Elmhurst College, Elmhurst, Illinois 60126}
\newcommand{\FNAL}{Fermi National Accelerator Laboratory, 
                   Batavia, Illinois 60510}
\newcommand{\Osaka}{Osaka University, Toyonaka, Osaka 560-0043 Japan} 
\newcommand{\Rice}{Rice University, Houston, Texas 77005}
\newcommand{\Rutgers}{Rutgers University, Piscataway, New Jersey 08854}
\newcommand{\UVa}{The Department of Physics and Institute of Nuclear and 
                  Particle Physics, University of Virginia, 
                  Charlottesville, Virginia 22901}
\newcommand{\UW}{University of Wisconsin, Madison, Wisconsin 53706}

\affiliation{\UAz}
\affiliation{\UCLA}
\affiliation{\UCSD}
\affiliation{\EFI}
\affiliation{\UB}
\affiliation{\ELM}
\affiliation{\FNAL}
\affiliation{\Osaka}
\affiliation{\Rice}
\affiliation{\Rutgers}
\affiliation{\UVa}
\affiliation{\UW}

\author{A.~Alavi-Harati}  \affiliation{\UW}
\author{T.~Alexopoulos}   \affiliation{\UW}
\author{M.~Arenton}       \affiliation{\UVa}
\author{K.~Arisaka}       \affiliation{\UCLA}
\author{S.~Averitte}      \affiliation{\Rutgers}
\author{R.F.~Barbosa}     \altaffiliation[Permanent address]
             {University of S\~{a}o Paulo, S\~{a}o Paulo, Brazil} 
                          \affiliation{\FNAL}
\author{A.R.~Barker}      \affiliation{\UB}
\author{M.~Barrio}        \affiliation{\EFI}
\author{L.~Bellantoni}    \affiliation{\FNAL}
\author{A.~Bellavance}    \affiliation{\Rice}
\author{J.~Belz}          \affiliation{\Rutgers}
\author{D.R.~Bergman}     \affiliation{\Rutgers}
\author{E.~Blucher}       \affiliation{\EFI}
\author{G.J.~Bock}        \affiliation{\FNAL}
\author{C.~Bown}          \affiliation{\EFI}
\author{S.~Bright}        \affiliation{\EFI}
\author{E.~Cheu}          \affiliation{\UAz}
\author{S.~Childress}     \affiliation{\FNAL}
\author{R.~Coleman}       \affiliation{\FNAL}
\author{M.D.~Corcoran}    \affiliation{\Rice}
\author{G.~Corti}         \affiliation{\UVa}
\author{B.~Cox}           \affiliation{\UVa}
\author{A.R.~Erwin}       \affiliation{\UW}
\author{R.~Ford}          \affiliation{\FNAL}
\author{A.~Glazov}        \affiliation{\EFI}
\author{A.~Golossanov}    \affiliation{\UVa}
\author{G.~Graham}        \affiliation{\EFI}
\author{J.~Graham}        \affiliation{\EFI}
\author{E.~Halkiadakis}   \affiliation{\Rutgers}
\author{J.~Hamm}          \affiliation{\UW}
\author{K.~Hanagaki}      \affiliation{\Osaka}
\author{S.~Hidaka}        \affiliation{\Osaka}
\author{Y.B.~Hsiung}      \affiliation{\FNAL}
\author{V.~Jejer}         \affiliation{\UVa}
\author{D.A.~Jensen}      \affiliation{\FNAL}
\author{R.~Kessler}       \affiliation{\EFI}
\author{H.G.E.~Kobrak}    \affiliation{\UCSD}
\author{J.~LaDue}         \affiliation{\UB}
\author{A.~Lath}          \affiliation{\Rutgers}
\author{A.~Ledovskoy}     \affiliation{\UVa}
\author{P.L.~McBride}     \affiliation{\FNAL}
\author{P.~Mikelsons}     \affiliation{\UB}

\author{E.~Monnier}
   \altaffiliation[Permanent address ]{C.P.P. Marseille/C.N.R.S., France}
   \affiliation{\EFI}

\author{T.~Nakaya}       \affiliation{\FNAL}
\author{K.S.~Nelson}     \affiliation{\UVa}
\author{H.~Nguyen}       \affiliation{\FNAL}
\author{V.~O'Dell}       \affiliation{\FNAL}
\author{R.~Pordes}       \affiliation{\FNAL}
\author{V.~Prasad}       \affiliation{\EFI}
\author{X.R.~Qi}         \affiliation{\FNAL}
\author{B.~Quinn}        \affiliation{\EFI}
\author{E.J.~Ramberg}    \affiliation{\FNAL}
\author{R.E.~Ray}        \affiliation{\FNAL}
\author{A.~Roodman}      \affiliation{\EFI}
\author{S.~Schnetzer}    \affiliation{\Rutgers}
\author{K.~Senyo}        \affiliation{\Osaka}
\author{P.~Shanahan}     \affiliation{\FNAL}
\author{P.S.~Shawhan}    \affiliation{\EFI}
\author{J.~Shields}      \affiliation{\UVa}
\author{W.~Slater}       \affiliation{\UCLA}
\author{N.~Solomey}      \affiliation{\EFI}
\author{S.V.~Somalwar}   \affiliation{\Rutgers}
\author{R.L.~Stone}      \affiliation{\Rutgers}
\author{E.C.~Swallow}    \affiliation{\EFI}\affiliation{\ELM}
\author{S.A.~Taegar}     \affiliation{\UAz}
\author{R.J.~Tesarek}    \affiliation{\Rutgers}
\author{G.B.~Thomson}    \affiliation{\Rutgers}
\author{P.A.~Toale}      \affiliation{\UB}
\author{A.~Tripathi}     \affiliation{\UCLA}
\author{R.~Tschirhart}   \affiliation{\FNAL}
\author{S.E.~Turner}     \affiliation{\UCLA}\affiliation{\EFI}
\author{Y.W.~Wah}        \affiliation{\EFI}
\author{J.~Wang}         \affiliation{\UAz}
\author{H.B.~White}      \affiliation{\FNAL}
\author{J.~Whitmore}     \affiliation{\FNAL}
\author{B.~Winstein}     \affiliation{\EFI}
\author{R.~Winston}      \affiliation{\EFI}
\author{T.~Yamanaka}     \affiliation{\Osaka}
\author{E.D.~Zimmerman}  \affiliation{\EFI}

\collaboration{The KTeV Collaboration}


\date{August 6, 2002}

\begin{abstract}
  We present a series of measurements based on $K_{L,S}\to \ppc$
  and $K_{L,S}\to \ppn$ decays collected in 1996-1997 by the 
  {\ktev} experiment (E832) at Fermilab.  
  We compare these four \Kpp\ decay rates
  to measure the direct CP violation
  parameter $\reepoe = ( \KtevReepoe \pm \KtevTErr)\eu$. 
  We also test CPT symmetry by measuring the relative phase
  between the CP violating and CP conserving decay amplitudes
  for \Kpm\ ($\phipm$) and for \Kzz\ ($\phizz$).
  We find the difference between the relative phases 
  to be 
  $\delphi \equiv \phizz-\phipm = 
   \left( \DelPhi \pm \DelPhiTOTerr \right)\degs$, 
  and the deviation of $\phipm$ from the superweak phase to be 
  $\phipm - \phisw = (\dPhiSW \pm \dPhiSWTOTerr)\degs$; 
  both results are consistent with CPT symmetry.
   In addition, we present new measurements of
   the $K_L$-$K_S$ mass difference and $K_S$ lifetime:
   $\delm  = ( \KtevDelm  \pm \KtevDelmTerr ) \delmunits$ and
   $\tauS  = ( \KtevTaus  \pm \KtevTausTerr ) \tausunits$~. \\
\end{abstract}


\maketitle


%
%

 \section{Introduction}
 \label{sec:intro}

The discovery of the $\KLpm$ decay in 1964 \cite{ccft64}
demonstrated
that CP symmetry is violated in weak interactions.
Subsequent experiments showed that the
effect is mostly due to a small asymmetry between
the $\Kz \to \Kzbar$ and $\Kzbar \longrightarrow \Kz$
transition rates,
which is referred to as indirect CP violation.
Over the last three decades, significant effort has been
devoted to searching for direct CP violation in a
decay amplitude.

Direct CP violation can be detected by comparing the level of CP
violation for different decay modes.  The parameters $\epsilon$ and
$\epsilon'$ are related to the ratio of CP violating to CP conserving
decay amplitudes for $\Kpm$ and $\Kzz$:
\begin{equation}
 \begin{array}{lcccc}
  \etapm & \equiv & \frac{\textstyle A \left( \KLpm\right)}
                    {\textstyle A \!\left(\KSpm\right)} &
  = & \epsilon + \epspr, \\
  \etazz & \equiv &\frac{\textstyle A \left( \KLzz\right)}
                   {\textstyle A \!\left(\KSzz\right)} & =
  & \epsilon -  2\epspr.
 \end{array}
\end{equation}
$\epsilon$ is a measure of indirect CP violation, which is common
to all decay modes. 
If CPT symmetry holds, the phase of $\epsilon$
is equal to the ``superweak'' phase:
\begin{equation}
  \phisw \equiv \tan^{-1} \left( 2 \delm / \Delta \Gamma \right),
\end{equation}
where $\delm\equiv m_L - m_S$ is the $K_L$-$K_S$ mass difference and 
$\Delta\Gamma =\Gamma_S - \Gamma_L$ is the difference in the 
decay widths.

The quantity $\epsilon^{\prime}$ is a measure of direct CP violation,
which contributes differently to the $\pi^+\pi^-$ and $\pi^0\pi^0$ decay
modes,
and is proportional to the difference between the decay amplitudes
for $\Kz\to\pi^+\pi^-(\pi^0\pi^0)$ and  
$\Kzbar\to\pi^+\pi^-(\pi^0\pi^0)$.
Measurements of $\pi\pi$ phase shifts~\cite{ochs} show that,
in the absence of CPT violation, the phase of $\epsilon'$ is
approximately equal to that of $\epsilon$. Therefore, $\reepoe $ is a
measure of direct CP violation and $\imepoe $ is a measure of CPT
violation.

Experimentally,  $\reepoe$ is determined from the double ratio of the
two pion decay rates of $K_L$ and $K_S$:
\begin{eqnarray}
  \frac{\Gamma\!\left(\KLpm\right)/\,\Gamma\!\left(\KSpm\right)}{
        \Gamma\!\left(\KLzz\right)/\,\Gamma\!\left(\KSzz\right)} 
                  \nonumber \\
        = \left| \frac{\etapm}{\etazz} \right|^2 
        \approx  1 + 6 \reepoe.  
              &  & 
  \label{eq:reepoe}
\end{eqnarray}
For small $|\epsilon'/\epsilon|$, $\imepoe$ is related
to the phases of $\etapm$ and $\etazz$ by
\begin{equation}
  \Delta\phi \equiv \phizz-\phipm   \approx -3  \imepoe ~.
  \label{eq:delphimpe}
\end{equation}

The Standard Model accommodates both direct and indirect
CP violation \cite{ckm,ellis,gilman_wise}.  
Unfortunately, there are large
hadronic uncertainties in the calculation of
$\reepoe$.
Most recent Standard Model predictions 
for $\reepoe$ are less than $30 \times 10^{-4}$
\cite{nbp:hambye,Cheng:1999dj,jhep:bijnens,Pallante:2001he,prd:wu,Buras:2000qz,Bertolini:2000dy,npb:narison,npps:ciuchini01,Aoki:2001dw,Blum:2001yx}.
The superweak model \cite{superweak}, 
proposed shortly after the discovery of $\KLpm$,
also accommodates indirect CP violation,
but not direct CP violation.

Previous measurements have established that
\reepoe\ is non-zero \cite{prl:731,pl:na31,prl:pss,na48:reepoe}.
This paper reports an improved measurement of $\reepoe$ by the
KTeV Experiment (E832) at Fermilab.
This measurement is based on 40 million reconstructed \Kpp\ decays 
collected in 1996 and 1997, 
and represents half of the total KTeV data sample. 
The 1996+1997 sample is
four times larger than, and contains, our previously
published sample \cite{prl:pss}.
We also present measurements of 
the kaon parameters $\delm$ and $\tauS$,
and tests of CPT symmetry based on measurements of 
$\delphi$ and $\phipm -\phisw$.

The outline of the paper is as follows.
Section~\ref{sec:exp} describes the \ktev\ measurement technique,
including details about the neutral hadron beams (Sec.~\ref{sec:beam})
and the detector used to identify $\Kpp$ decays
(Sec.~\ref{sec:det}).
The detector description also includes the calibration procedures
and the performance characteristics relevant to the
$\reepoe$ measurement. 
The Monte Carlo simulation of the kaon beams and detector is 
described in Section~\ref{sec:mc}.
Section~\ref{sec:ana} explains the reconstruction and event selection
for the $\Kpm$ and $\Kzz$ decay modes,
and also the background subtraction for each mode.
Section~\ref{sec:extract} describes the acceptance correction,
and the fit used to extract $\reepoe$ and other 
physics parameters.
Each of these sections is followed by a discussion of
systematic uncertainties related to that part
of the $\reepoe$ measurement.
Section~\ref{sec:reepoe_measure} presents the $\reepoe$ result
along with several crosschecks.
Finally, Section~\ref{sec:kaonpar} presents our measurements
of the kaon parameters $\delm$, $\tauS$, $\delphi$, and $\phipm$.
Additional details of the work presented 
here are given in~\cite{reepoe_theses}.

 \section{Measurement Technique and Apparatus}
 \label{sec:exp}

  \subsection{Overview}
  \label{sec:overview}

The measurement of \reepoe\ requires a source of $K_L$ and $K_S$
decays, and a detector to reconstruct the charged ($\ppc$)
and neutral ($\ppn$) final states.  
The strategy of the \ktev\ experiment is to produce two identical 
$K_L$ beams, and then to pass one of the beams through a 
thick ``regenerator.''
The beam that passes through the regenerator is called the 
regenerator beam,
and the other beam is called the vacuum beam.
The regenerator creates a coherent 
$\ket{K_L}+\rho\ket{K_S}$ state,
where $\rho$ is the regeneration amplitude chosen
such that most of the \Kpp\ decay rate
downstream of the regenerator is from the $K_S$ component.
The measured quantities are the numbers of \Kpm\ and \Kzz\
decays in the vacuum and regenerator beams.
The vacuum-to-regenerator ``single ratios'' for \Kpm\ and \Kzz\ 
decays are proportional to
$|\etapm/\rho|^2$ and $|\etazz/\rho|^2$, 
and the ratio of these two quantities gives $\reepoe$ 
via Eq.~\ref{eq:reepoe} (also see Appendix~\ref{app:exp details}).
The effect of \KSKL\ interference in the regenerator beam
is accounted for in a fitting program used to extract \reepoe.

To reduce systematic uncertainties related to left-right
asymmetries in the detector and beamline, the regenerator 
position alternates between the two beams once per minute.
Decays in both beams are collected
simultaneously to reduce sensitivity to
time-dependent variations in the beam intensity and in 
detector efficiencies.
The fixed geometry of the beamline elements,
combined with alternating the regenerator position,
ensures a constant vacuum-to-regenerator kaon flux ratio.

\Kpm\ decays are detected in a spectrometer
consisting of four drift chambers and a dipole magnet;
the well-known kaon mass is used to determine the momentum scale.
The four photons from \Kzz\ decays are detected in a pure 
Cesium Iodide (CsI) calorimeter;
electrons from $\KLpienu$ decays ($\Kethree$)  
are used to calibrate the CsI energy scale.
An extensive veto system is used to reject events coming from 
interactions in the regenerator, 
and to reduce backgrounds from kaon decays into
non-$\pi\pi$ final states
such as $\KLpimunu$ and $\KLzzz$.

A Monte Carlo simulation is used to correct for the acceptance
difference between \Kpp\ decays in the two beams,
which results from the very different $K_L$ and $K_S$ lifetimes.  
The simulation includes details of detector geometry and efficiency,
as well as the effects of ``accidental'' activity from the high flux
of particles hitting the detector.
The decay-vertex distributions provide a
critical check of the simulation.
To study the detector performance,
and to verify the accuracy of the Monte Carlo simulation, 
we also collect samples of $\KLpienu$ and \KLzzz\ decays 
with much higher statistics than the $\Kpp$ signal samples.

 \subsection{The Kaon Beams}
 \label{sec:beam}

\begin{figure}
  \centering
  \epsfig{file=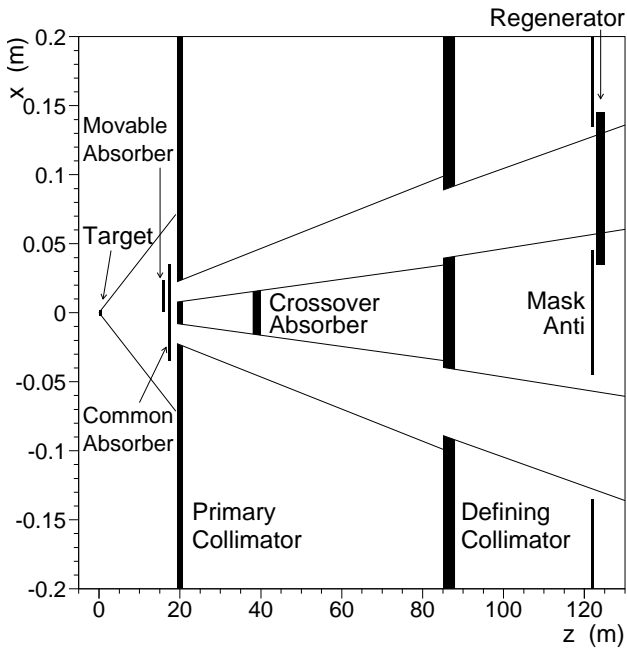,width=\linewidth}
  \caption{
      Top view of the KTeV beamline, showing the BeO target, 
      absorbers, collimating system, Mask Anti, and regenerator. 
      The common absorber is a 7.62~cm thick Pb slab followed by
      a 52.3~cm  slab of beryllium.
      The movable absorber is a 45.7~cm slab of beryllium
      that alternates beam 
      positions once every minute along with the regenerator.
      The crossover absorber is a 2~m thick iron slab.
      Note the different horizontal and vertical scales.
          }
  \label{fig:beamline}
\end{figure}

Neutral kaons are produced by a proton beam hitting a fixed
target (Fig.~\ref{fig:beamline}).  The Fermilab Tevatron provides
$3 \times 10^{12}$ 800~GeV/$c$ protons in a 20~s extraction cycle (``spill'')
once every minute. The proton beam has a 53~MHz RF micro-structure such
that protons arrive in $\sim 1$~ns wide ``buckets'' and at 19~ns intervals.
The bucket-to-bucket variations in beam intensity are typically 10\%.

The target is a narrow beryllium oxide (BeO) rod, $3 \times 3$~mm$^2$
in the dimensions transverse to the beam,
and 30~cm long corresponding to about
one proton interaction length. 
The proton beam is incident on the target at a downward
angle of  $4.8$~mrad  with respect to the line joining the target 
and the center of the detector. 
This targeting angle is chosen as a compromise between
higher kaon flux at small angles, and a smaller 
neutron-to-kaon ratio ($n/K$)
at large angles.

The center of the BeO target defines the origin
of the KTeV right-handed coordinate system.
The positive $z$-axis is directed from the target
to the center of the detector. 
The positive $y$-axis is directed up.

The particles produced in the BeO target include very few 
kaons compared to other hadrons and photons.
A $\sim100$~m long beamline is used to remove unwanted particles
and to collimate two well-defined kaon beams. 
In this beamline,
charged particles are removed with sweeping magnets,
photons are absorbed by a 7.62~cm Pb slab located at $z = 19$~m,
and most of the hyperons decay near the target.
To reduce the neutron-to-kaon ratio,
neutrons (kaons) are attenuated by a factor of 4.6 (2.6) in a
beryllium absorber that is common to both beams.
An extra ``movable'' absorber, synchronized with the regenerator 
position, provides an additional attenuation factor of 
3.8 (2.3) for neutrons (kaons) in the regenerator beam.

Each neutral kaon beam is defined by two collimators:
a 1.5~m long primary collimator ($z = 20$~m),
and a 3~m long ``defining'' collimator ($z = 85$-$88$~m)
that defines the size and solid angle of each beam.
Each collimator has two square holes, 
which are tapered to reduce scattering. 
A ``crossover'' absorber at $z=40$~m prevents
kaons that scatter in the upstream absorbers from 
crossing over into the other beam.
At the defining collimator, the two beams have the same size 
($4.4 \times 4.4$~cm$^2$) and solid angle ($0.24~\mu$str). 
The beam centers are separated by 14.2~cm,
and the horizontal angle between the two beams is 1.6~mrad.

The two beams pass through an evacuated volume, held at $10^{-6}$ Torr,
which extends from 28~m to 159~m from the target.
This evacuated region includes the 110~m to 158~m range used in the analysis.
The downstream end of the evacuated volume is sealed with a 
0.14\% radiation length ($X_0$) \vw\ made of kevlar and mylar.

Most of the $K_S$ component decays near the BeO target.
Downstream of the defining collimator,
the small ``{\tgtKS}'' component that remains in the vacuum beam 
increases the \Kpp\ decay rate by 0.4\%
compared with a pure $K_L$ beam of equal intensity.
This contribution from \tgtKS\ is
essentially zero for kaon momenta below $100~\upk$
and rises to 15\% at $160~\upk$.

In the regenerator beam, most of the \Kpp\ decays
are from the regenerated $K_S$ component. 
Decays from the $K_L$ component and from 
\KSKL\ interference 
account for about 20\% of the \Kpp\ decay rate,
ranging from 5\% near the regenerator to 90\% at the \vw.

The composition of the neutral hadron beams is as follows.
At 90~m from the BeO target, 
the vacuum beam has a 2.0~MHz flux of kaons  with $n/K = 1.3$,
and an average kaon momentum of about $70~\upk$.
Upstream of the regenerator, 
the kaon flux is 0.9~MHz with $n/K =0.8$; 
downstream of the regenerator the flux of unscattered kaons is 0.15~Mhz, 
or $13$ times smaller than in the vacuum beam. 
The flux of hyperons at 90~m from the target 
is about $1$~kHz in the vacuum beam.
In addition to hadrons, there are muons that come from the 
BeO target, the proton beam dump, and kaon decays;
the total muon flux hitting the \ktev\ detector is about 200~kHz.

\begin{figure*}
  \centering
  \epsfig{file=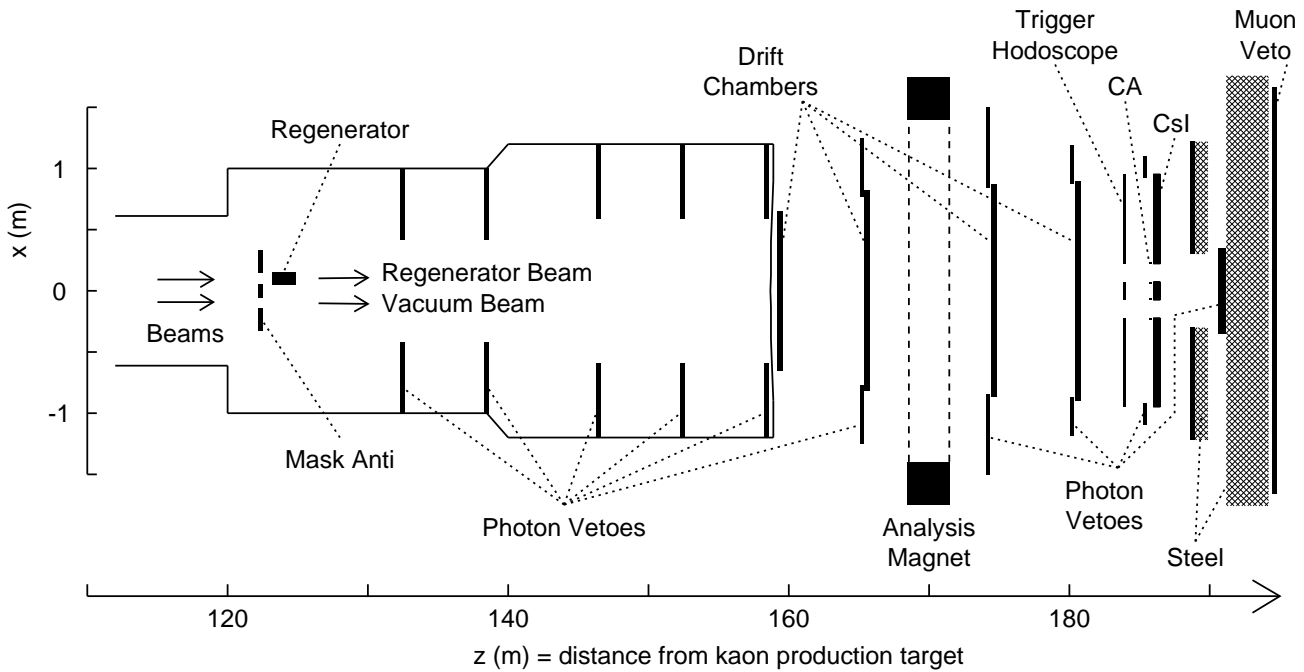,width=\linewidth}
  \caption{ 
     Top view of the KTeV (E832) detector.
     The evacuated decay volume extends to $z = 159$~m. 
         }
  \label{fig:detector}
\end{figure*}

  \subsection{The Detector}
  \label{sec:det}

Kaon decays downstream of the defining collimator 
($z=85$~m, Fig.~\ref{fig:beamline}) 
are reconstructed by the \ktev\ detector (Fig.~\ref{fig:detector}),
which includes a magnetic spectrometer, CsI calorimeter,
and veto system.
Downstream of the \vw,
the space between detector components is filled with helium 
to reduce interactions from the neutral beam,
and to reduce multiple scattering and photon conversions 
of decay products.
The total amount of material upstream of the CsI calorimeter
corresponds to $4\%$ of a radiation length;
about 60\% of the material is in the trigger hodoscope just upstream
of the CsI calorimeter, 
and 10\% of the material is upstream of the first drift chamber.
Each of the two neutral beams passes through holes in the
Mask Anti veto, trigger hodoscope, and CsI calorimeter.
The beams finally strike a beam-hole photon veto
5 meters downstream of the CsI.
The following sections describe the detector components in more detail.

  \subsubsection{Spectrometer}
  \label{sec:det_spec}

The \ktev\ spectrometer includes four drift chambers (DCs)
that measure charged particle positions.
The two downstream chambers are separated from the upstream chambers 
by a $3\times 2~{\rm m}^2$ aperture dipole magnet.  
The magnet produces a field that is uniform to better than 1\% 
and imparts a $0.41~\upk$
kick in the horizontal plane.  
During data taking, the magnet polarity was reversed every 1-2 days.

\begin{figure}
  \centering
  \epsfig{file=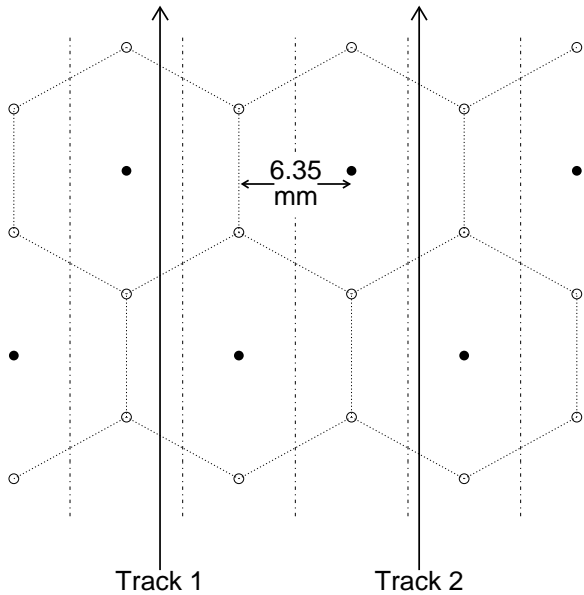,width=\linewidth}
  \caption{
      Drift chamber  hexagonal cell geometry showing six field wires
      (open circles) around each sense wire (solid dots).
      The vertical dashed lines 
      denote the boundaries of the ``offset-cells'' used in the 
      track-separation cut (see Section~\ref{sec:chrg_evtsel}). 
      For example, two offset-cells are between 
      tracks 1 and 2.
        }
  \label{fig:dccell}
\end{figure}

The DC planes have a hexagonal cell geometry formed by six 
field-shaping wires surrounding each sense wire 
(Fig.~\ref{fig:dccell}).   
The cells are $6.35$~mm wide,
and the drift velocity is about $50 \mu$m/ns
in the 50-50 argon-ethane gas mixture.
The maximum drift time across a cell is 150 ns,
and defines the width of the ``in-time'' window.  
A chamber consists of two planes of horizontal wires
to measure $y$ hit positions, 
and two planes of vertical wires to measure $x$ hit positions;
the two $x$-planes, as well as the two $y$-planes,
are offset by one half-cell to resolve the 
left-right ambiguity.
The $x$ and $y$ hits cannot be associated to each other using only
drift chamber information, but can be matched using
CsI clusters as explained in Section~\ref{sec:chrg_evtsel}.

There are a total 
of 16 planes and $1972$ sense wires in the four DCs.
The transverse chamber size increases with distance from the target.
The smallest chamber (DC1) is $1.26\times 1.26~{\rm m}^2$
and has 101 sense wires per plane;
the largest chamber (DC4) is $1.77\times 1.77~{\rm m}^2$
and has 140 sense wires per plane.

Lecroy 3373 multi-hit Time-to-Digital Converters (TDCs) are used to
measure the drift times relative to the RF-synchronized Level 1
trigger.  
The TDC resolution is 0.25 ns, 
which contributes 13~$\mu$m to the position resolution.
The  total TDC time window is $2.5$ times longer than the 
in-time window, and is centered on the in-time window.
The track reconstruction software uses only
the earliest in-time hit on each wire.
Hits prior to the in-time window are recorded to study their
influence on in-time hits.

Each measured drift time, $t$, is converted 
into a drift distance, $x$, with a non-linear $x(t)$ map. 
The maps are measured separately for each of the sixteen planes 
using the uniform hit-illumination across each cell.  
The $x(t)$ calibrations are performed
in roughly 1-day time periods.

\begin{figure}
    \centering
    \epsfig{file=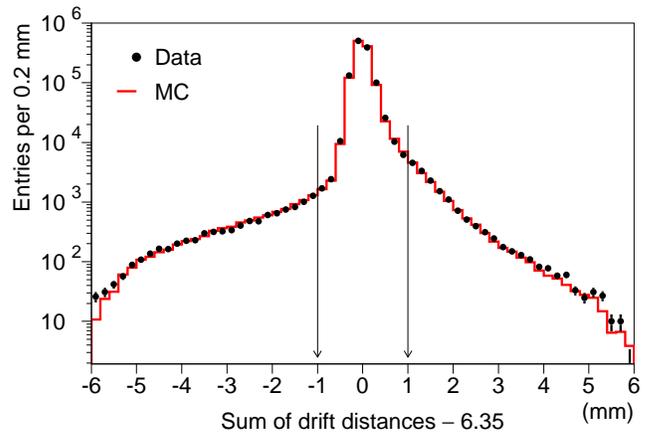,width=\linewidth}
    \caption{
        Deviation of the sum of drift distances (SOD) 
        from the nominal cell size of 6.35~mm,
        for the $y$-views using a sample of \Kpm\ decays.
        The arrows indicate the
        $\pm 1$~mm requirement for a good hit-pair.
        The data are shown as dots.
        The histogram is the Monte Carlo (MC) prediction 
        (Sec.~\ref{sec:mc_dc}).
           }
    \label{fig:sod}
\end{figure}

A charged-particle track produces a hit in each of the 
two $x$ and $y$ planes.
The two $x$ hits (or two $y$ hits) in each plane are referred
to as a ``hit-pair.''
For a track that is perpendicular to a drift chamber
with perfect resolution, 
the sum of drift distances (SOD) from each hit-pair
would equal the cell width of 6.35 mm.
The measured SOD distribution is shown in Fig.~\ref{fig:sod}.
For inclined tracks, an angular correction is applied to the SOD.
The track-finding software requires the SOD to be within 
$1$~mm of the cell width.
The mean SOD is stable to within $10~\mu$m
during the run.
The single-hit position resolution is typically $110~\mu$m,
corresponding to a SOD resolution of $150~\mu$m
and a hit-pair resolution of $80~\mu$m.
Using the tracking algorithm described in Section~\ref{sec:chrg_evtsel},
the momentum resolution is 
$\sigma_p/p \simeq [1.7 \oplus (p/14) ] \times 10^{-3}$,
where $p$ is the track momentum in $\upk$.

The average inefficiency for reconstructing a hit-pair
is 3.7\%. 
Delta rays and accidental hits contribute to the low-side
tail of the SOD distribution (Fig.~\ref{fig:sod}),
and result in a hit-pair loss of 0.5\% and 0.7\%, respectively.
The remaining 2.5\% loss is from missing hits and 
from SOD values more than 1~mm larger than the cell size
(high-side tail in Fig.~\ref{fig:sod}).
More details on sources of hit-pair inefficiency 
are given in the description of the Monte Carlo simulation
(Sec.~\ref{sec:mc_dc}).

To reconstruct the trajectories of charged particles accurately,
the alignment of the drift chambers relative to each other,
to the target, and to the CsI calorimeter must be known.
This alignment is determined in roughly 1-day periods
using data samples described below.

The transverse alignment of the drift chambers is based on 
muons from dedicated runs with the analysis magnet turned off. 
The muon intensity is raised to $\sim~1$~MHz by reducing the field 
in the upstream sweeping magnets, and the neutral hadron beam is 
absorbed by a 2~m long steel block placed in the beam at $z=90$~m. 
These muon runs were performed every 1-2 days
when the magnet polarity was reversed.
Software calibration results in a transverse alignment of 
each $x$ and $y$ plane to within $10~\mu$m, 
and relative rotations known to within $20~\mu$rad. 

The transverse target position is determined with a precision of 
$35~\mu$m using $\Kpm$ decays in the vacuum beam and 
projecting the reconstructed kaon trajectory back to the target.
The CsI offset relative to the drift chambers is measured using
electrons from $\KLpienu$ decays.

  \subsubsection{The CsI Calorimeter}\label{sec:det_csi}

\begin{figure}
    \centering
    \epsfig{file=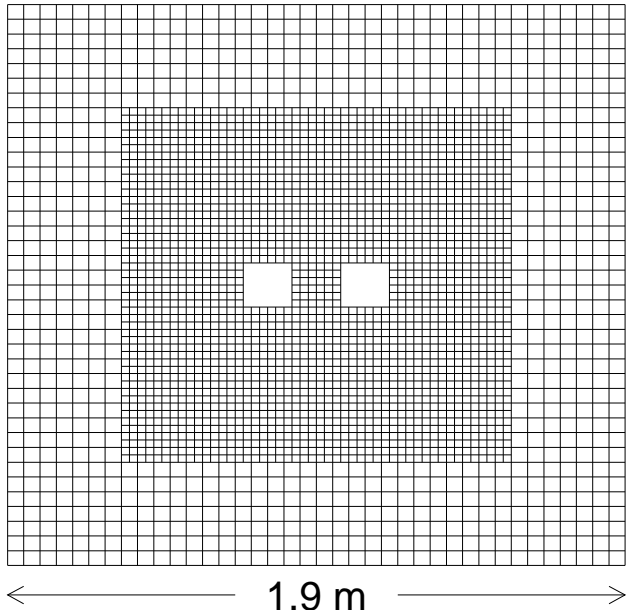,width=\linewidth}
    \caption{
        Beamline view of the KTeV CsI calorimeter, 
        showing the 868 larger outer
        crystals and the 2232 smaller inner crystals. 
        Each beam hole size is  $15\times 15$~cm$^2$ and
        the two beam hole centers are separated by 0.3~m.
           }
    \label{fig:csilayout}
  \end{figure}

The KTeV electromagnetic calorimeter consists of 3100 pure CsI
crystals as shown in Fig.~\ref{fig:csilayout}.
There are 2232 $2.5 \times 2.5$ cm$^2$ crystals in the central region, 
each viewed by a 1.9~cm Hamamatsu R5364 photomultiplier tube (PMT).
There are also 868 $5 \times 5$ cm$^2$ crystals, 
each viewed by a 3.8~cm Hamamatsu  R5330 PMT.
The transverse size of the calorimeter is $1.9 \times 1.9$~m$^2$,
and the length of each crystal is 50~cm (27 $X_0$).  
Two $15\times 15$~cm$^2$ carbon fiber beam pipes allow the 
few MHz of beam particles to
pass through the calorimeter without striking any material.

The crystals were individually wrapped and tested with a 
$^{137}$Cs source to ensure that the response over the length 
of each crystal is uniform to within $\sim 5$\% \cite{csical}.  
This longitudinal uniformity requirement is necessary to 
obtain sub-percent energy resolution for electrons and photons.

The CsI scintillation light has two components:
(i) a fast component with decay times of 10~ns and 36~ns,
and maximum light output at a wavelength of 315~nm; 
(ii) a slow component with a decay time of $\sim 1\mu$s and
maximum light output at $480$~nm.
To reduce accidental pile-up effects from the slow component, 
a Schott UV filter is placed between the crystal and the PMT.
The filter reduces the total light output by $\sim 20$\%,
but increases the fast component fraction from about 80\%  to 90\%.
The average light yield with the filter gives
20 photo-electrons per MeV of energy deposit.  
The CsI signal components discussed above include
the effect of the PMT spectral response.

Digitizing electronics are placed directly behind the PMTs to minimize
electronic noise ($<1$~MeV).  
Each PMT is equipped with a custom 8-range
digitizer to integrate the charge delivered by the PMT.
This device, which is referred to as a ``digital PMT'' 
(DPMT)~\cite{dpmt},
has 16 bits of dynamic range with 8-bit resolution, 
and allows the measurement of energies from a few MeV to 100~GeV. 
In 1997, the digitization and readout operated at the Tevatron 
RF frequency of 53~MHz,
and the PMT signal integration time was 114 ns 
(6 RF ``buckets''), which permitted collection of 
approximately $96\%$ of the fast scintillation component.
In 1996, the readout frequency was 18~MHz (RF/3)
and the integration time was a factor of two longer than
in 1997.

To convert measurements of integrated charge to energy,
a laser system is first used to calibrate the response from 
each DPMT.
Then momentum-analyzed electrons from $\KLpienu$ decays 
are used to calibrate the energy scale of each channel.

The laser system consists of a pulsed Nd:YAG laser,
a fluorescent dye, and a variable filter 
to span the full dynamic range of the readout system.
The laser light ($360$~nm) 
is distributed via 3100 quartz fibers
to each individual crystal;
the light level for each quadrant of the calorimeter
is monitored with a PIN diode read out by a 20-bit ADC. 
Throughout the data taking,
special hour-long laser scans to calibrate the readout system 
were performed roughly once per week during periods
when beam was not available.
Using these laser scan calibrations, 
deviations from a linear fit of the combined DPMT plus PMT 
response versus light level are
less than 0.1\% (rms) for each channel.
During nominal data-taking, the laser operated at a fixed intensity 
with $1$~Hz repetition rate to correct for short-term 
gain drifts that were typically $< 0.2$\% per day.

To determine the energy scale in the calorimeter,
we collected 600 million  $\KLpienu$ decays during the experiment;
this number of events allows us to determine
the energy scale for each channel with $\sim 0.03$\%  
precision every 1-2 days.
The electron energy is determined by  summing energies 
from a $3 \times 3$ ``cluster'' of large crystals or a $7 \times 7$ 
cluster of small crystals centered on the crystal with maximum energy.
The cluster energy is corrected for shower leakage outside the 
cluster region,
leakage at the beam-holes and calorimeter edges, 
and for channels with energies below the 
$\sim 4$~MeV readout threshold.

\begin{figure}
 \epsfig{file=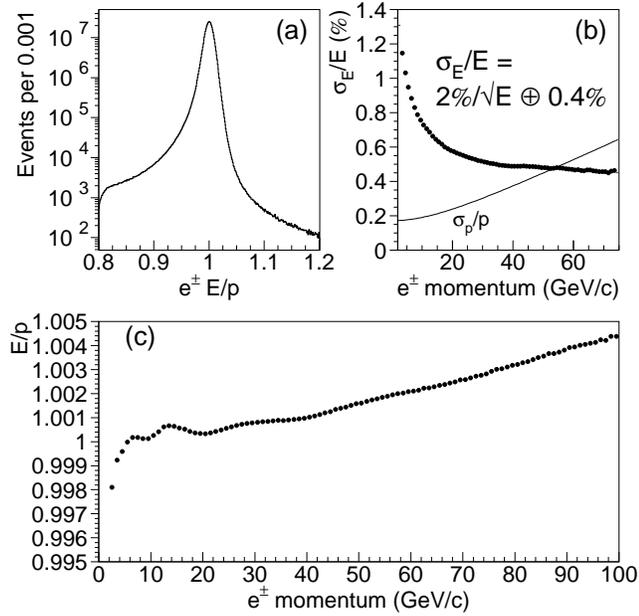,width=\linewidth}
 \caption {
       (a) $E/p$ distribution based on electrons from $\KLpienu$ decays,
       where $E$ is the energy measured in the CsI calorimeter and
       $p$ is the momentum measured in the spectrometer;
       (b) electron energy resolution vs. 
       momentum (dots) obtained by subtracting the tracking resolution,
       shown by the smooth curve, from the $E/p$-resolution;
       (c) $E/p$-mean vs. momentum before
       the linearity correction is applied.
     }
 \label{fig:linres}
\end{figure}

Figure~\ref{fig:linres}a shows the $E/p$ distribution
for electrons from \Kethree\ decays, where  
$E/p$ is the ratio of cluster 
energy measured in the CsI calorimeter
to momentum measured in the spectrometer.
To avoid pion shower leakage into the electron shower,
the $\pi^{\pm}$ is required to be at least 50~cm away
from the $e^{\mp}$ at the CsI.
The $E/p$ resolution has comparable contributions from 
both $E$ and $p$.
The CsI energy resolution is obtained by
subtracting the momentum resolution from the
$E/p$ resolution, 
and is shown as a function of momentum in
Fig.~\ref{fig:linres}b.
The energy resolution can be parameterized as
$\sigma_E/E \simeq 2\%/\sqrt{E} \oplus 0.4\%$,
where $E$ is in GeV;
the resolution is 1.3\% at 3~GeV, 
the minimum cluster energy used in the \Kzz\ analysis, 
and is better than 0.6\% for energies above 20~GeV.
The momentum dependence of the mean $E/p$ 
(Fig.~\ref{fig:linres}c) shows that the average energy
nonlinearity is 0.5\% between 3 and 75~GeV.
This energy nonlinearity is measured for each CsI channel
and used as a correction.

Electromagnetic cluster positions are determined from the
fraction of energy in neighboring columns and rows.
The conversion from energy fraction to position is
done using a map 
based on the uniform photon illumination across each crystal
for \Kzz\ data.
The average position resolution for electrons is 1.2~mm  
for clusters in small crystals,
and 2.4~mm for large crystals.

  \subsubsection{The Regenerator} 
  \label{sec:regenerator}

\begin{figure}
  \epsfig{file=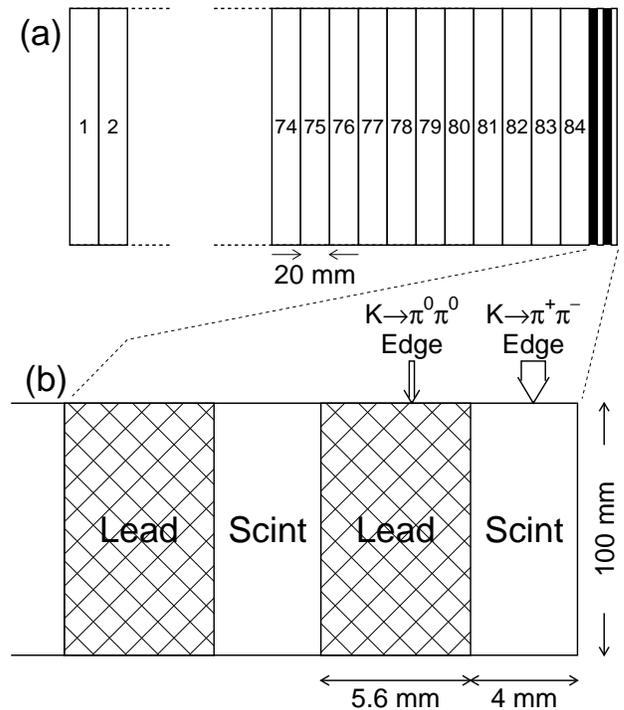,width=\linewidth}
  \caption{
      (a) Diagram of the regenerator and (b) magnified view
      of the downstream end of the regenerator.
      PMTs are not shown.
      The dimension of each element is given in the figure.
      The kaon beam enters from the left.
      Vertical arrows above the lead and scintillator 
      show the  effective edge locations for \Kzz\ and \Kpm\ decays;
      the width of each arrow indicates the uncertainty.
           } 
  \label{fig:regdiagram}
\end{figure}

KTeV uses an active regenerator as a source of $K_S$ decays.
It consists mainly of 84 ~~$10\times10\times 2~{\rm cm}^3$ scintillator 
modules (Fig.~\ref{fig:regdiagram}a).
Each module is viewed by two photomultiplier tubes (PMTs), 
one from above and one from below.
The downstream end of the regenerator (Fig.~\ref{fig:regdiagram}b)
is a lead-scintillator sandwich, 
which is also viewed by two PMTs.
This last module of the regenerator is used to define a sharp
upstream edge for the kaon decay region in the regenerator beam. 

At the average kaon momentum of $70~\upk$, 
the magnitude of the regeneration amplitude is $\magrho \sim 0.03$.
The isoscalar carbon in the regenerator accounts for
about 95\% of the regeneration amplitude, which simplifies the 
model used to describe $\rho$ when extracting physics parameters
(Sec.~\ref{sec:fitting}).

One can distinguish three main processes that contribute 
to $K_S$ regeneration. 
These are 
(i) ``coherent'' regeneration, which occurs in the forward direction,
(ii) ``diffractive'' regeneration, in which target nuclei do not 
disintegrate but kaons scatter at finite angle, and 
(iii) ``inelastic'' regeneration, characterized by nuclear break-up 
and often by production of secondary particles.
Only the decays of coherently regenerated kaons
are used in the $\reepoe$ analysis. 
The other processes are treated as  background.

The 170~cm length of the regenerator corresponds to about two 
hadronic interaction lengths. 
This length maximizes coherent regeneration and 
suppresses diffractive regeneration~\cite{pr:good}. 
The diffractive-to-coherent ratio is 0.09 for \Kpp\ decays
downstream of the regenerator, 
and is reduced by kinematic cuts in the analysis. 
The corresponding inelastic-to-coherent ratio is about $100$.
Since inelastic regenerator interactions typically leave 
energy deposits  of a few~MeV to 100~MeV 
from the recoil nuclear fragments, 
this source of background
is reduced using the regenerator PMT signals;
events with more than 8~MeV in any scintillator element
are rejected.
Inelastic interactions with 
production of secondary particles are suppressed further
by other elements of the veto system.
After all veto requirements, 
the level of inelastic scattering is reduced
by a factor of a  few thousand,
making its contribution smaller than that of diffractive scattering.

The downstream edge of the regenerator defines the beginning
of the regenerator-beam decay region for both the \Kpm\ and \Kzz\
modes. 
A small fraction of decays inside the regenerator enters the 
signal sample because photons can pass through scintillator and lead,
and because charged pions can exit the last regenerator module 
without depositing enough energy to be vetoed.
In the fit to extract $\reepoe$ (Sec.~\ref{sec:fitting}),
the median of the distribution of \Kpp\ decays inside the regenerator 
is used to define a perfectly sharp ``effective edge,''  $\zregeff$.
For the neutral decay mode, 
$\zregeff$ is calculated using the known
geometry and regeneration properties, 
and is $(-6.2\pm 0.1)$~mm from the
downstream end of the regenerator; 
it is shown by the arrow above the lead in Fig.~\ref{fig:regdiagram}b.
For the charged decay mode, $\zregeff$  is determined by
the veto threshold in the last  regenerator module.
The threshold is measured using muons collected
with a separate trigger, and results in a \Kpm\ edge that is 
$(-1.65 \pm 0.45)$~mm from the downstream  edge of the regenerator;
it is shown by the arrow above the scintillator in 
Fig.~\ref{fig:regdiagram}b.
The uncertainty comes from the geometry of the PMT-scintillator
assembly and the threshold measurement from muons.

  \subsubsection{The Veto System} 
  \label{sec:det_veto}

The \ktev\ detector uses veto elements to reduce trigger rates, 
to reduce backgrounds, and to define sharp apertures and edges 
that limit the detector acceptance. 
The regenerator veto was discussed in the previous section.

\begin{figure}
  \centering
  \epsfig{file=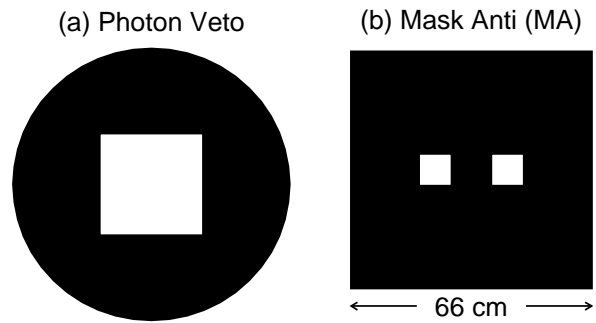,width=\linewidth}
  \caption{
      (a) Transverse layout for one of five photon 
      veto detectors between the regenerator and the \vw\
      (Fig.~\ref{fig:detector}).
      The square aperture size varies, but is roughly 1~meter.
      (b) MA photon veto detector ($z=123$~m).
      The two MA beam hole sizes are $9\times 9~{\rm cm}^2$
      and their centers are separated by 20~cm.
      For both detectors
      the shaded region indicates the active veto area.
      The PMTs are at the outer edges of each detector.
          }
  \label{fig:RCMA}
\end{figure}

Nine lead-scintillator (16~$X_0$) photon veto counters 
are positioned along the beam-line, with five located upstream of
the {\vw} (Fig.~\ref{fig:RCMA}a) 
and four located downstream of the \vw.
These nine veto counters detect escaping particles 
that would miss any of the drift chambers or the CsI calorimeter.
Another $10~X_0$ photon veto is placed behind the CsI to detect 
photons that go through the beam-holes; 
this ``beam-hole veto'' mainly suppresses background from $\KLzzz$ decays.
A scintillator bank behind 4~m of steel ($z=192$~m in 
Fig.~\ref{fig:detector}) is used to veto muons,
primarily from $\KLpimunu$ decays.

\begin{figure}
  \centering
  \epsfig{file=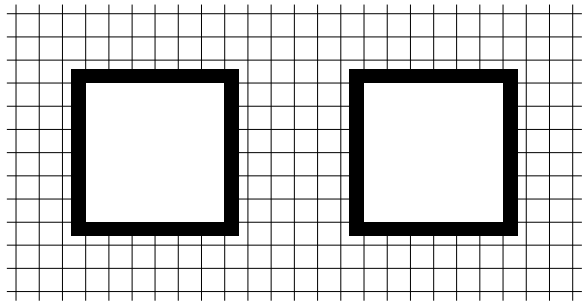,width=\linewidth}
  \caption{
      The Collar Anti (CA) is shown as the two shaded 
      ``picture-frame'' regions 
      that cover the inner 60\% (1.50~cm) of the CsI crystals
      surrounding the beam holes. 
      The neutral beams go into the page, 
      and the two beam-hole centers are separated by 30~cm.
      Wavelength shifting fibers transmit the scintillation light
      to PMTs at the edges of the calorimeter.
         }
  \label{fig:CA}
\end{figure}

The upstream distribution of reconstructed kaon decays
is determined mainly by the 
``Mask Anti''  (MA,  Fig~\ref{fig:RCMA}b), 
which is a 16 $X_0$ lead-scintillator sandwich 
located at $z = 123$~m. 
The MA has two $9 \times 9$~{\rm cm}$^2$ holes through 
which the neutral beams pass.
At the downstream end of the detector,
the CsI crystals around the beam-holes are partially 
covered by an 8.7~$X_0$ tungsten-scintillator 
``Collar Anti'' (CA, Fig~\ref{fig:CA}).
In addition to defining a sharp edge, 
the CA veto
rejects events in which more than 10\% of a photon's energy
is lost in a beam hole.


  \subsubsection{Trigger and Data Acquisition}
  \label{sec:trigger}

\ktev\ uses a three-level trigger system to reduce the total
rate of stored events to approximately 2~kHz, while efficiently
collecting \Kpp\ decays. 
The Level~1 trigger (L1) has no deadtime and makes a decision 
every 19~ns (corresponding to the beam RF-structure) 
using fast signals from the detector.  
The Level~2 trigger (L2) is based on more sophisticated processors and
introduces a deadtime of 2-$3\mu$s.
When an event passes the Level~2 trigger, the entire detector 
is read out with an average deadtime of $15~\mu s$.
Each event is then sent to one of
twenty-four 200-MHz SGI processors running 
a Level~3 (L3) software filter.
An event passing Level~3 selection is written to a 
Digital Linear Tape for permanent storage.
An independent set of ten 150-MHz processors is used for
online monitoring and calibration.

For rate reduction, the most important trigger element 
is the regenerator veto, which uses the signal from 
the downstream lead-scintillator sandwich plus 
signals from 3 of the 84 scintillator modules.
This  veto is applied in Level~1 triggers
to reject events from the 2~MHz of hadrons
that interact in the regenerator.
After applying the regenerator veto,
there is still a 100~kHz rate of kaon decays
and another 100~kHz rate of hadron interactions 
in the \vw\ and drift chambers.
Additional trigger requirements are used to 
reduce this 200~kHz rate by about a factor of 100
to match the bandwidth of the data acquisition system.

Separate triggers are defined for the signal $\ppc$ and 
$\ppn$ modes.
Each trigger is identical for the two beams 
and for both regenerator positions.
The charged, neutral, and special purpose triggers are described below. 
Each section includes a brief summary of the trigger 
inefficiency, which is defined as the fraction of events
that pass all analysis cuts, but fail the trigger.
The inefficiencies are measured using decays
collected in separate minimum-bias triggers.

\bigskip{\it Charged Decay Trigger:}
The Level~1 trigger for charged decays requires hits in
the two drift chambers upstream of the magnet, 
and requires hits in the ``trigger hodoscope'' 
located 2~m upstream of the CsI. 
Each drift chamber is required to have
at least one hit in both the $x$ and $y$ views.
The trigger hodoscope
consists of two 5~mm thick scintillation planes,
each with 31 individually wrapped counters. 
There are small gaps between the counters,
representing 1.1\% of the area of each scintillation plane.
The hodoscope counters are arranged to minimize the impact of
these gaps;
a particle traversing a gap in one plane cannot pass
through a gap in the other plane.
Each plane has two $14\times 14~{\rm cm}^2$
holes to allow the neutral beams to pass through without interacting.
The trigger requires 1 or more hits in both planes,
and 2 or more hits in at least one plane;
this requirement allows for one of the 
charged particles to pass through a gap between the counters.
The two hodoscope hits must also include both the upper and
lower ``regions,'' as well as the left and right regions.
The defined hodoscope regions have sufficient overlap to prevent 
losses for $\Kpm$ events, 
except for decays in which a pion passes through a 
scintillator gap in the central region defined by 
$|y| < 7~$cm.
To reduce the trigger rate from non-$\ppc$ kaon decays, 
signals from the veto system are used to reject events
at the first trigger level.
The muon veto is used to suppress $\KLpimunu$ decays,
and the photon vetos downstream of the \vw\ are used to
reject decays with a photon in the 
final state.

The Level 2 trigger requires that the drift chamber hits in the $y$-view 
be consistent with two tracks from a common vertex;
to reduce signal loss from inefficient wires,
a missing hit is allowed.
The L3 filter selects \Kpm\ candidates by reconstructing two charged
tracks in the spectrometer;
to reject $\KLpienu$ and $\KLpmz$ decays,
the $\ppc$ mass is required to be greater than $450~{\rm MeV}/c^2$
and $E/p$ is required to be less than $0.9$ for both tracks.
For CsI calibration and detector acceptance studies,
$1/7$ of the \Kethree\ candidates are kept by requiring 
$E/p > 0.9$ for one of the tracks.

The total charged trigger inefficiency is 0.5\%, 
and is mainly from the 0.3\% loss due to 
gaps between the scintillator counters.
The drift chamber requirements result in a 0.1\% inefficiency.  
Accidental effects result in a 0.06\% loss, 
and there is a 0.04\% loss from the trigger hardware.
The Level~3 inefficiency is 0.09\%.

\bigskip{\it Neutral Decay Trigger:}
The Level~1 trigger for neutral decays
is based on the total energy deposited in the CsI calorimeter.
Using a 3100 channel analog sum,
the threshold is 24~GeV and the 10-90\% turn-on width is 7~GeV.
The photon vetos downstream of the \vw\ are used 
in the Level~1 trigger to reduce the
rate from \KLzzz\ decays.
The Level~2 trigger
uses a Hardware Cluster Counter~\cite{nim:hcc} to count clusters 
of energy above 1~GeV in the CsI.  
Four clusters are required for \Kzz\ decays, 
and a separate six-cluster trigger is pre-scaled
by five to collect \KLzzz\ decays. 
The Level~3 filter requires that the 
invariant mass be greater than $450~{\rm MeV}/c^2$ for both 
the $\ppn$ and $\zzz$ final states.

The neutral energy-sum trigger inefficiency
has two components: 
a 0.6\% inefficiency from early accidental effects
and an inefficiency of $4\times 10^{-5}$ from the trigger hardware.
The Level~2 trigger inefficiency of 0.4\%
comes from the Hardware Cluster Counter.
The Level~3 inefficiency is 0.01\%.

\bigskip{\it Other Triggers:}
In addition to the triggers used to select specific decay modes,
several special-purpose triggers are used for monitoring. 
These include 
(i) a ``pedestal'' trigger, which reads out the
entire CsI calorimeter a few times per 20~s spill,
(ii) a laser trigger to monitor the CsI (Sec.~\ref{sec:det_csi}), 
(iii) a muon trigger that requires a muon hodoscope signal
instead of using it in veto,
(iv) a charged mode trigger that does not use the regenerator veto, 
(v) a trigger using only the Level~1 CsI analog sum
with no Level~2 requirement,
and
(vi) an ``accidental'' trigger to record random activity in the detector 
that is proportional to 
the instantaneous intensity of the proton beam;
triggers iii-vi are heavily pre-scaled.
For the accidental trigger, 
we use a telescope consisting of three scintillation counters, 
each viewed by a photomultiplier tube.
It is located 1.8~m from the BeO target and is oriented 
at an angle of $90\degs$ with respect to the beam axis.
The target is viewed by the counters through a 
$6.4 \times 6.4$~{\rm mm}$^2$ hole in the stainless steel shielding 
around the target.
A coincidence of signals in all three counters 
generates an accidental trigger.

\bigskip{\it Trigger Rates:}
Under nominal conditions, the total rates passing L1, L2, and L3 
are 40, 10, and 2~kHz, respectively; 
this L3 rate corresponds to  approximately $40,000$ events
written to tape each minute.
The deadtime, which is common to all triggers, is about $33\%$
with roughly equal contributions from Level~2 and readout.

  \subsection{Data Collection}
  \label{sec:datarun}

The data used in this analysis were collected
in two distinct ``E832'' periods: 
October-December in 1996 and April-July in 1997. 
In these periods, 
there were about 200 billion kaon decays 
between the  defining collimator and the CsI calorimeter,
of which 5 billion events were written to 3000 15-Gb tapes. 
About 5\% of the events are used to select the $\Kpp$ sample,
and the remaining 95\% of the events are used to understand
the detector.
The neutral mode data from 1996 and 1997 are used. 
For the charged mode, the 1996 sample is not used
because the Level~3 rejection of delayed hits in the 
drift chambers led to a 20\% signal loss that is 
difficult to simulate.
In 1997, the Level~3 tracking was modified to avoid this loss.
Excluding the 1996 charged mode data has a negligible 
effect on the overall statistical and systematic 
uncertainty in $\reepoe$.

To ensure that the final data sample is of high quality,
periods in which there are known detector 
problems are excluded.  
The CsI calorimeter readout suffered DPMT failures
at a rate of about one per day,
and was the most significant detector problem.
Each DPMT failure was identified immediately
by the online monitoring, and then repaired. 
Calibrations were frequent 
enough so that every repaired DPMT can be calibrated offline.
The final data sample does not include periods in which there
were dead CsI channels or dead cells in the drift chambers.  
Approximately 12\% of the data on tape are rejected because of
detector problems.

  \subsection{Monte Carlo Simulation}
  \label{sec:mc}

The Monte Carlo (MC) simulation consists of three main steps.
The first step is kaon generation at the BeO target and 
propagation along the beamline to the decay point. 
The second step is kaon decay into an appropriate final state, 
and tracing of the decay products through the detector. 
The last step is to simulate the detector response 
including digitization of the detector signals.   
The simulated event format and analysis
are the same as for the data.

The detector geometry used in the simulation comes from
survey measurements and data.
The survey measurements are used for the transverse dimensions and
$\ZK$ locations of beamline and detector elements.
The transverse offsets and rotations of the drift chambers,
relative to the CsI calorimeter,
are determined using various data samples as discussed in
Section~\ref{sec:det_spec}.
The Mask-Anti and Collar-Anti (Sec.~\ref{sec:det_veto}) 
aperture sizes and locations
are determined using electrons from $\KLpienu$ decays.

As a result of the high flux of kaons and neutrons in the 
\ktev\ apparatus, there can be
underlying accidental activity in the  detector
that is unrelated to the kaon decay.
After applying veto cuts, the average accidental energy 
under each CsI cluster is a few MeV, 
and there are roughly 20 extra in-time drift chamber hits.
To simulate these effects, we use data events from the 
accidental trigger (Sec.~\ref{sec:trigger}) 
to add the underlying accidental activity to each generated MC event. 
In the CsI calorimeter and veto system, 
activity from an accidental event
is added to the MC energy deposits in a
straightforward manner.
The procedure for including accidental activity in 
the drift chamber simulation, however, 
is more complicated because
an empirical model is needed to describe how 
an accidental hit can obscure a signal hit that arrives
later on the same wire.

 \subsubsection{Kaon Propagation and Decay}
 \label{sec:mc_kaonprop}

The kaon energy spectrum and the relative flux of $\Kz$ and $\Kzbar$ 
states produced at the target are based on a 
parameterization \cite{malensek} that
is tuned to match \ktev\ $\Kpm$ data. 
The $z$ position of each kaon decay is chosen based on the 
calculated $z$ distribution for the initial
\Kz\ or \Kzbar\ state, 
and accounts for interference between \KL\ and \KS.

The simulation propagates the \Kz\ and \Kzbar\ amplitudes 
along the beamline,
accounting for regeneration and scattering in the
absorbers and the regenerator.
Some small-angle kaon scatters in the absorbers 
($z\sim 19$~m, Fig.~\ref{fig:beamline})
can pass through the collimator system and 
satisfy the \Kpp\ analysis requirements.
The upstream collimators are modeled as perfectly absorbing, 
while scattering in the defining collimators 
and the regenerator are treated using models that are tuned to data
(Sections~\ref{sec:bkg_coscat}-\ref{sec:bkg_regscat}).

For $\Kpm$ decays, radiative corrections due to
inner \brems\ are included \cite{prl:belz}. 
We do not simulate the direct emission part of the 
radiative spectrum since the invariant mass cut used in the
analysis essentially eliminates this component.
For the \Kzz\ mode, only the four photon final state is considered.

GEANT \cite{geant} is used to parameterize scattering of 
final state charged
particles in the \vw, helium bags, drift chambers,
trigger hodoscope, and the steel.
Electrons can undergo \brems, 
photons can convert to $e^{+}e^{-}$ pairs,
and charged pions can decay into a muon and a neutrino;
these secondary particles are traced through the detector.

  \subsubsection{Simulation of the Drift Chambers} 
  \label{sec:mc_dc}

The Monte Carlo traces each charged particle through the 
drift chambers, 
and the hit position at each drift chamber plane is converted 
into a TDC value.
The position resolutions measured in data are used to smear the
hit positions, and the inverse of the $x(t)$ map is used to convert the
smeared hit position into a drift time.  

The simulation includes four effects that cause drift chamber
signals to be corrupted or lost.
\begin{enumerate}
  \item ``Wire inefficiency'' results in 
        no in-time TDC hit.
        The inefficiency is measured in 1~cm steps along each 
        wire of each chamber.
        The average single-hit inefficiency is
        less than 1\%. Since
        the inefficiency increases with distance from the
        wire, the measured inefficiency profile within the cell 
        is used in the simulation.
  \item A ``delayed hit'' results in a hit-pair with  
        a sum-of-distance (SOD)
        that is more than 1~mm greater than the nominal cell size,
        and therefore does not satisfy the hit-pair requirement
        (Sec.~\ref{sec:det_spec}).
        The delayed hit probability is a few percent in the
        regions where the neutral beams pass through the drift chambers, 
        and about 1\% over the rest of the chamber area.
        The effect is modeled by distributing primary drift 
        electrons along the track using a Poisson
        distribution with an average interval of  $340~\mu m$,
        and then generating a composite signal at the sense wire.
        A delayed hit occurs when the signal from the nearest ionization
        cluster is below threshold, but the composite pulse from
        all ionization clusters is above threshold.
        The MC threshold is determined empirically by matching the       
        MC delayed hit probability to data;        
        the delayed hit probability in data is measured in 1~cm 
        steps along each wire.
  \item When an ``in-time accidental'' hit arrives before a signal 
        hit on the same wire, the accidental hit
        is used instead of the signal hit because the tracking program 
        considers only the first in-time hit.
        For roughly 0.7\% of the hit-pairs,
        this effect causes a ``low-SOD'' that is more than 1~mm
        below the nominal cell size, and therefore does not pass
        the hit-pair requirement.
        An ``early accidental'' hit prior to the in-time window
        can also obscure a signal hit on the same wire for two reasons.
        First, the discriminator has a deadtime of $42$~ns during
        which the wire is 100\% inefficient.
        Second, large analog pulses can
        stay above the discriminator threshold longer than 42~ns; 
        the variation of the pulse-length is modeled and tuned to data. 
  \item Delta rays also cause a low-SOD for 0.5\% of the hit-pairs. 
        In the Monte Carlo, delta rays are generated 
        in the same cell as the track,
        and the rate is tuned to match the low-SOD
        distribution in data.
\end{enumerate}

The quality of the drift chamber simulation is illustrated
by the SOD distribution in Fig.~\ref{fig:sod}.
Both the low and high-side tails in the data
are well simulated.

  \subsubsection{Simulation of the CsI Calorimeter} 
  \label{sec:mc_csi}

The Monte Carlo simulation is used to predict the energy deposit 
in each crystal when kaon decay products hit the calorimeter.
In particular, the MC is needed to model energy leakage
in the beam-holes and at the outer edges of the calorimeter,
and to model nearby showers that share energy.
A library of  GEANT-based \cite{geant} 
electron and photon showers is used to simulate electromagnetic
showers in the calorimeter.   
In the shower generation, each electron or photon is incident on 
the central crystal of  a $13\times 13$ array of small crystals 
($32.5\times 32.5~{\rm cm}^2$).
The showers are generated in 6 energy bins from 2 to 64 GeV,
and in $x,y$ position bins.
The  position bin spacing varies from 7~mm at the crystal center
to 2~mm at the edge; this binning matches the variation in 
reconstructed position resolution, 
which is better for particles incident near the edge of a crystal.
Outside the $13\times 13$ array, 
a GEANT-based parameterization is used instead of a library of 
individual showers;
this parameterization models energy deposits in a $27\times 27$ array.
Energy leakage across the beam-holes is modeled based
on electron data from $\KLpienu$ decays.

To simulate the DPMT response, the energy deposit in each crystal
is distributed among
six consecutive RF buckets according to the measured 
time profile of the scintillation light output.
In each 19~ns wide RF bucket, the energy is smeared to account
for photo-statistics, and random activity from an accidental
trigger is added.
Each channel is digitized using the inverse
of the calibrations obtained from data.
A channel is processed if the digitized
signal exceeds the $4$~MeV readout threshold that
was applied during data-taking.

In addition to simulating electromagnetic showers,
we also simulate the calorimeter response to 
charged pions and muons.
The energy deposits from charged pions are based on a
library of GEANT-based showers using a $50\times 50$ 
array of small crystals.  A continuous energy distribution
is generated in $x,y$ position bins with 4~mm separation.
For muons, the average CsI energy deposit is 320~MeV, and is
simulated using the Bethe-Bloch energy 
loss formula.

 \subsubsection{Simulation of the Trigger} 
 \label{sec:mc_trg}

The \ktev\ Monte Carlo includes a simulation of the 
Level~1 and Level~2 triggers, and the Level~3 software filter.
For the $\Kpm$ trigger, the most important effect to simulate is the 
$0.3$\% inefficiency due to scintillator gaps 
in the hodoscope just upstream of the CsI calorimeter.
The gap sizes and positions are measured in data using the
$\Kethree$ sample.
The simulation also includes the drift chamber signals at both
Level~1 and Level~2.
For the  $\Kzz$ trigger, we use $\KLpienu$ decays to determine
the calorimeter energy-sum threshold and turn-on width,
and to measure the Hardware Cluster Counter
threshold for each CsI channel.

 \section{Data Analysis}
 \label{sec:ana}

\def\dels{\Delta_s}  

The analysis is designed to identify \Kpp\ decays 
while removing poorly reconstructed events that
are difficult to simulate,
and to reject background.
The following sections describe the analysis and the
associated systematic uncertainties in $\reepoe$.
When discussing systematic uncertainties, 
we typically estimate a potential shift $s\pm \sigma_s$, 
where $s$ is the shift in $\reepoe$ and $\sigma_s$ is the
accompanying statistical uncertainty.
We convert the shift to a symmetric systematic error, 
$\dels$, such that the range $[-\dels,+\dels]$ includes
68.3\% of the area of a Gaussian with 
mean $s$ and width $\sigma_s$:
\begin{equation}
  \frac{1}{\sigma_s\sqrt{2\pi}} 
       \int_{-\dels}^{+\dels} dx ~
       {\rm exp}{\left[-\frac{(x-s)^2}{2\sigma_s^2}\right]} 
                 = 0.683 ~.
  \label{eq:error}
\end{equation}
Note that $\dels = \sigma_s$ when $s=0$;
when $s > \sigma_s$, $\dels \approx s+\sigma_s/2$.

 \subsection{Common Features of \ppc\ and \ppn\ Analyses}
 \label{sec:ana common}

Although many details of the charged and neutral decay mode
analyses are different, several features are common
to reduce systematic uncertainties.
For each decay mode, the same cuts are applied to 
decays in the vacuum and regenerator beams,
so that most systematic uncertainties cancel in 
the single ratios used to measure
$|\etapm/\rho|^2$ and $|\etazz/\rho|^2$.  

Since the regeneration amplitude $\rho$ depends on the kaon momentum,
we select an identical $40$-$160~\upk$ kaon momentum range for 
both the charged and neutral decay modes. 
The $40~\upk$ cut is chosen because of the rapidly falling
detector acceptance at lower kaon momenta;
the higher momentum cut is a compromise between slightly higher 
statistics and \tgtKS\ contamination.
We also use the same $z$-vertex range of 110-158~m for each decay mode.
The 110~m cut is chosen to be well upstream of the Mask Anti
and therefore removes very few decays in the vacuum beam;
this cut removes no events in the regenerator beam.
The downstream $z$-vertex requirement avoids
background from beam interactions in the \vw.

To simplify the treatment of background from kaons that
scatter in the regenerator, the veto requirements for the
charged and neutral mode analyses
are as similar as possible.
In both analyses, the main reduction in background from 
regenerator scattering comes from the requirement that there be less
than 8~MeV in every regenerator module (Sec.~\ref{sec:regenerator}).
Interactions in the regenerator that are close in
time to a \Kpp\ decay can add significant activity
in the detector, resulting in events that are difficult
to reconstruct.
To avoid this problem,
the trigger signal from the regenerator is used
to reject events in a 57~ns wide window 
(3 RF buckets), which removes events 
immediately following or just prior to
an interaction in the regenerator.
In addition to the regenerator veto,
we require that there be less than 150 (300) MeV in the photon 
vetos upstream (downstream) of the \vw, 
and less than 300 MeV in the Mask Anti;
note that the photon veto thresholds refer to equivalent 
photon energy.

  \subsection{\Kpm\ Reconstruction and Selection} 
  \label{sec:chrg_evtsel}

The strategy to identify \Kpm\ decays is to
reconstruct two well-measured tracks in the spectrometer,
and to reduce backgrounds with particle identification
and kinematic requirements.

The spectrometer reconstruction begins by finding
$y$-tracks using all four drift chambers.
In the $x$-view, 
separate segments are found upstream of the magnet using DC1 and DC2,
and downstream of the magnet using DC3 and DC4.
The extrapolated upstream and downstream 
$x$-track segments typically match to within 0.5~mm at the
center of the magnet.
To reduce sensitivity to multiple-scattering
and magnetic fringe fields between the drift chambers, 
only a loose match of 6~mm is required at the magnet.
If two $x$-tracks and two $y$-tracks are found,
the reconstruction continues by extrapolating both sets
of tracks upstream to define an $x$-$z$ and $y$-$z$ vertex.
The difference between these two vertices, $\dzvtx$,
is used to define a vertex-$\chi^2$,
\begin{equation}
   \chisqvtx \equiv \left( {\dzvtx}/\sigdz \right)^2 ~,
   \label{eq:chisqvtx}
\end{equation}
where $\sigdz$ is the resolution of $\dzvtx$.
This resolution  depends on momentum and opening angle, 
and accounts for multiple scattering effects.
The two $x$-tracks and two $y$-tracks are assumed to originate
from a common vertex if $\chisqvtx < 100$;
this $\chisqvtx$ requirement is sufficiently loose to remain insensitive
to the tails in the multiple scattering distribution.
At this stage, the $x$ and $y$ tracks are independent. 
To determine the full particle trajectory,
the $x$ and $y$ tracks are matched to each other
based on their projections to CsI clusters;
the track projections to clusters must
match within 7~cm.

After combining the $x$ and $y$ tracks,
the reconstructed $z$-vertex resolution 
is about 30~cm near the regenerator and 
5~cm near the \vw.  
Each particle momentum is determined from the track bend-angle 
in the magnet and a precise B-field map.

An ideal event with two charged tracks results in 32 hits 
in the four chambers. 
About 40\% of the events have at least one missing hit
or a hit-pair that does not reconstruct a proper 
sum-of-distance (SOD) value.
The tracking program can reconstruct events with many
of these defects along the tracks.
The overall single-track reconstruction inefficiency is 
measured to be 1\% using $\KLpmz$ decays.

An event is assigned to the regenerator beam if the
regenerator $x$-position has the same sign as
the $x$-coordinate of the kaon trajectory 
at the downstream face of the regenerator;
the event is assigned to the vacuum beam 
if the signs are different.
For $\KLpienu$ decays, this beam assignment definition
cannot be used because of the missing neutrino;
instead we compare the $x$-coordinate of the decay vertex
to that of the regenerator.

The reconstruction described above is used
for kaon decays with two charged tracks in the final state.
The following selection criteria are specific to the \Kpm\ channel, 
and are designed mainly to reduce background from semileptonic decays
and kaon scattering.

Requiring $E/p < \EopCut$ for each track reduces the $\KLpienu$
background by a factor of 1000.
The $\KLpimunu$ background is rejected largely by the muon veto
in the Level~1 trigger.
Since low momentum muons may range out in the 
4~m of steel before depositing energy in the muon veto, 
we also require each track to have a momentum $p > \MinpCut$. 
To reject radiative \Kpmg\ decays, events  with
an isolated electromagnetic cluster above \XClusEcut\  are  removed if the
cluster is at least  \XClusTrkcut\ away from both extrapolated 
pion track positions at the CsI; the pion-photon separation requirement
avoids removing events that
have satellite clusters from hadronic interactions.
To remove background from \Lppi\ and \LBARppi decays, 
the higher momentum track is assumed to be from a proton 
(or antiproton)
and the event is rejected if the proton-pion invariant mass is within 
$3.5$~MeV of the known $\Lambda$ mass.

\begin{figure}[ht]
  \centering
  \epsfig{file=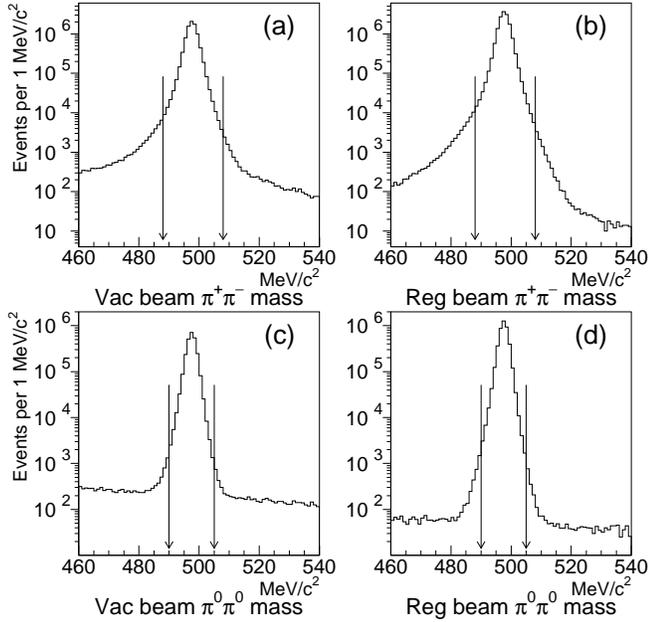,width=\linewidth}
  \caption{
      Invariant mass distributions for
      (a) vacuum-beam $\ppc$, (b) regenerator-beam $\ppc$,
      (c) vacuum-beam $\ppn$, (d) regenerator-beam $\ppn$.
      All other analysis cuts are applied.
      The vertical arrows indicate the accepted mass range.
          }
  \label{fig:mass}
\end{figure}

To provide additional background
rejection of semileptonic and \Kpmg\ decays,
the $\ppc$ invariant mass is required to be in the range
$488$-$508~{\rm MeV}/c^2$.
Figures~\ref{fig:mass}a,b show the invariant mass 
distributions for the vacuum and regenerator beams
after all other selection cuts.
The shapes of the vacuum and regenerator beam distributions 
are nearly the same, 
and have an RMS resolution of $\sim \MassResln$. 
The low-mass tail is mainly due to the presence of 
\Kpmg\ events in both beams. 
The tails in the vacuum beam distribution also include background
from semileptonic decays.

\begin{figure}[ht]
  \epsfig{file=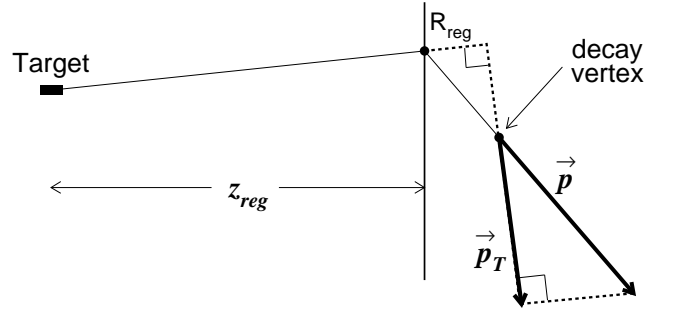,width=\linewidth}
  \caption{
      Illustration of the transverse momentum, $\vec{p}_T$, for a
      kaon that scatters in the regenerator and 
      then decays at the point labeled ``decay vertex.''
      The line connecting the target to $\Rreg$ represents
      the kaon trajectory upstream of the regenerator,       
      $\vec{p}$ is the measured momentum of the $\ppc$ system,
      and $z_{\rm reg}$ is the distance from the target to the
      downstream end of the regenerator.
         }
  \label{fig:pt2diag}
\end{figure}

\begin{figure}[ht]
  \epsfig{file=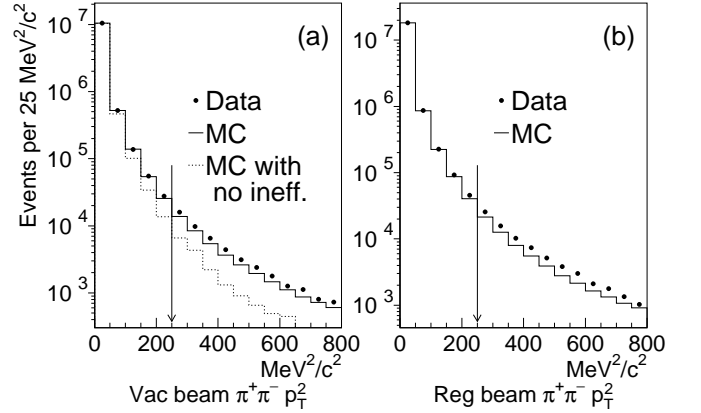,width=\linewidth}
  \caption{
      \ppc\ \ptsq\ distributions after all other \Kpm\ analysis cuts in
      (a) the vacuum beam and (b) the regenerator beam. 
      The background-subtracted data
      are shown as dots, and the histogram shows the MC prediction.
      The selection cut is shown by the arrows.
      The dotted histogram in the vacuum beam shows the 
      MC prediction without the effects of wire inefficiency,
      delayed hits, and accidentals (Sec.~\ref{sec:mc_dc}).
        }
  \label{fig:pt2datamc}
\end{figure}

Background from kaon scattering in the regenerator 
is suppressed mainly by the regenerator veto requirement.
To reduce this background further, 
a cut is made on the \ppc\ transverse momentum,
which also rejects events from kaons that scatter in the 
defining collimator.
As illustrated in Fig.~\ref{fig:pt2diag},
the total momentum of the \ppc\ system ($\vec{p}$)
is projected back to a point ($\Rreg$) in the plane
containing the downstream face of the regenerator,
and $\vec{p}_T$ is the momentum component that is transverse
to the line connecting $\Rreg$ to the target.
This definition of $\vec{p}_T$ is used for both beams,
and is optimized to distinguish between scattered and
unscattered kaons in the regenerator beam.
The data and MC \ptsq\ distributions in both beams are shown in
Fig.~\ref{fig:pt2datamc};
the selection cut, $\ptsq < 250~\uptsq$, is shown by the arrows.

Next, we describe selection requirements that define apertures.
For $\Kpm$, unlike \Kzz\ decays, the edges 
of the Mask Anti and Collar Anti vetos are not
useful apertures because of the difficulty in 
simulating the interactions of charged pions.
For the MA, a charged pion can pass through a
small amount of lead-scintillator near the edge
without leaving a veto signal,
and multiple scattering in the MA results in a
poorly reconstructed vertex.
For the CA, the difficulty is with pions that miss the CA,
but interact in the CsI calorimeter;
backscattering from these pion interactions can 
still deposit sufficient energy in the CA
to produce a veto signal.
To avoid these apertures,
we require that the track projections 
be away from the physical boundaries of the veto detectors.
In a similar fashion, the outer CsI aperture is defined by
track projections so that the acceptance does not depend
on the CsI energy profile from charged pions.

If two tracks pass within the same drift chamber 
cell or adjacent cells, 
a hit from one track can obscure a
hit from the other track.
Since this effect is difficult to simulate,
we use a track-separation requirement of 
two offset-cells  (Fig.~\ref{fig:dccell}) between the tracks
in both the $x$ and $y$ views for each chamber;
offset-cells are used so that tracks near a wire,
which have the poorest position resolution,
are not near the boundary which defines the
track-separation cut.
This track-separation requirement
results in an effective inner aperture.

  \subsection{Systematic Uncertainties from \ppc\ 
              Trigger, Reconstruction, and Selection} 
  \label{sec:chrg_syst}

\begin{table}
\caption{
    \label{tb:chrg_syst}
    Systematic uncertainties in $\reepoe$
    from the charged mode trigger and analysis.   
    The uncertainties in the 3rd column, which also appear in the
    systematics summary (Table~\ref{tb:syst_reepoe}),
    are the quadratic sum of contributions in the 2nd column.
       }
\begin{ruledtabular}
\begin{tabular}{ l | l | c }
Source of       &   \multicolumn{2}{c}{Uncertainty on $\reepoe$}   \\
uncertainty     &   \multicolumn{2}{c}{($\eu$)}  \\ 
\hline 
Trigger              &       & {\bf 0.58}  \\ 
~~~L1 \& L2          & 0.20  &      \\
~~~L3 filter         & $\L3CHRGSYST$  &      \\
 \hline  
Track reconstruction       &  & {\bf 0.32}  \\
~~~Alignment and calibration & 0.28 & \\
~~~Momentum scale          & 0.16 & \\
 \hline  
Selection efficiency      &      & {\bf 0.47} \\
~~~$p^2_T$ cut            & 0.25 & \\
~~~DC efficiency modeling & 0.37 & \\
~~~DC resolution modeling & 0.15 & \\
%
 \hline  
Apertures            &      & {\bf 0.30}  \\
~~~Wire spacing      & 0.22 &       \\
~~~Regenerator edge  & 0.20 &       \\
\end{tabular}
\end{ruledtabular}
\end{table}

In this section,
we discuss the $\reepoe$ systematic uncertainties from the
charged mode trigger, reconstruction, and selection, 
which are summarized in Table~\ref{tb:chrg_syst}.
Systematic uncertainties related to background and acceptance 
are discussed in 
Sections~\ref{sec:bkg_syst} and \ref{sec:acceptance}.

\bigskip{\it Trigger:}
The Level~1 and Level~2 trigger requirements are studied with  \Kethree\
events collected in a trigger based only on 
CsI calorimeter energy, 
and contribute a $0.20\eu$ uncertainty to $\reepoe$.
The L3 uncertainty is determined using a 1\% sample of charged 
triggers that pass L1 and L2, and are saved without requiring L3.
From this sample, $3\times 10^5$ 
events pass all \Kpm\ analysis cuts.
Applying the L3 requirement to this sample results in a 
$2.2\sigma$ shift in the vacuum-to-regenerator ratio;
the same analysis on MC events results in no shift from L3.
These shifts in the vacuum-to-regenerator ratio
correspond to a $(0.45\pm 0.20)\eu$ shift in $\reepoe$,
which leads to an uncertainty of $\L3CHRGSYST\eu$ using
Eq.~\ref{eq:error}.

\bigskip{\it Trajectory and Momentum Reconstruction:}
The effects of detector misalignment (Sec.~\ref{sec:det_spec}) 
are studied by evaluating the change in $\reepoe$ as the following
are changed within their measured uncertainties:
transverse chamber offsets and rotations,
non-orthogonality between the $x$ and $y$ wire planes,
and the $z$-locations of the drift chambers.
The time-to-distance calibration is varied to change
the average SOD value within its uncertainty.
A $\reepoe$ uncertainty of $0.28\eu$ is assigned
based on these tests.
The kaon mass is known to $0.031$~MeV~\cite{pdg00},
leading to a momentum-scale uncertainty of $1\eu$,
and a $\reepoe$ uncertainty of $0.16\eu$.

\bigskip{\it Selection Efficiency:}
In the charged mode analysis, \ptsq\ is the 
only variable for which $\reepoe$ is sensitive to the cut value.
Figure~\ref{fig:pt2datamc} shows the \ptsq\ data-MC comparison 
in both beams after background subtraction
(Sec.~\ref{sec:bkg}), 
and also illustrates the importance of simulating 
details of the drift chamber performance.
Increasing the \ptsq\ cut value from 250 to $500~\uptsq$ changes 
\reepoe\ by
$(-0.23 \pm 0.05) \eu$, leading to a systematic uncertainty
of $\PtsqSyst \eu$.  
There is no further statistically significant variation 
if the \ptsq\ cut value is increased beyond $500~\uptsq$.

The uncertainty in modeling the drift chamber efficiency
is related to the effects of delayed hits and early accidentals.
The delayed-hit probability (Sec.~\ref{sec:mc_dc})
predicted by the MC
is compared to data in various regions of the chambers 
and for different time periods. Residual data-MC differences
do not exceed $10\%$ of the effect, 
which corresponds to a \reepoe\ systematic
uncertainty of $0.21 \eu$.

There is a component of the early accidental inefficiency
that may not be modeled properly because the total TDC range
covers only about $2/3$ of the relevant early time window 
that is prior to the in-time window.
A systematic error of $0.30\eu$ is assigned based on the 
change in $\reepoe$ when accidental hits from
only half of the early time window are used
to obscure simulated in-time hits.
The total uncertainty from modeling the effects of
delayed hits and accidentals is $0.37\eu$ on $\reepoe$.

The modeling of the drift chamber resolutions is checked by comparing
the widths of the SOD distributions between data and MC;
they agree to within $5\%$ which corresponds to a systematic 
error of $0.15\eu$ on $\reepoe$.

\bigskip{\it Apertures:}
The drift chamber track-separation cut (Fig.~\ref{fig:dccell}) 
depends on the wire spacing, 
which is known to $20~\mu$m on average \cite{thesis:mbw}.
The bias in $\reepoe$ from variations in the wire spacing
is \TrCellVar\, leading to an uncertainty of 
$\TrCellSyst \eu$.
The $0.45$~mm uncertainty on the effective regenerator edge
(Fig.~\ref{fig:regdiagram}b) leads to a small uncertainty in the
expected number of \KS\ decays, and results in  a systematic error of 
$\RegEdgeSyst \eu$ on \reepoe.

  \subsection{\Kzz\ Reconstruction and Selection}
  \label{sec:neut_evtsel}

The strategy to identify \Kzz\ decays is to
reconstruct four photon clusters in the CsI calorimeter
that are
consistent with coming from two neutral pions,
and to reduce background with kinematic cuts.
Since the reconstruction of \Kzz\ and \KLzzz\ decays is
almost identical, most of the discussion will be valid
for both these decay modes.

The calorimeter reconstruction begins by  determining
the energy and position of each cluster found in 
the Level~2 trigger.
The neutral decay mode analysis requires
two additional corrections that are not relevant for
the $\KLpienu$ calibration.
First, a correction is applied to account for energy shared among 
nearby clusters. 
The second correction, described in detail at the end of this section, 
accounts for a small energy scale difference between 
data and Monte Carlo.

Each cluster is required to have a transverse
energy distribution consistent with that of a photon.
This requirement rejects decays that have significant accidental energy 
overlapping a photon cluster;
for \Kzz\ decays, it also rejects background from $\KLzzz$ decays 
in which two or more photons overlap at the CsI calorimeter.
Sensitivity to modeling the transverse energy distribution
is reduced by requiring that all clusters be separated from 
each other by at least 7.5~cm.
To reduce the dependence on the MC trigger simulation,
the energy of each cluster is required to be greater than 3~GeV.

To reconstruct each neutral pion, 
we assume that two photon clusters originate from a $\pi^0$ decay.
In the small angle approximation, 
the distance between the $\pi^0$ decay vertex
and the CsI calorimeter is given by
 \begin{equation}
   d_{\pi^0} = r_{12} \sqrt{E_{\gamma_1} E_{\gamma_2}} / {m_{\pi^0}} , 
   \label{eq:neut_z}
 \end{equation}
where $r_{12}$ is the distance between the two photon clusters,
$E_{\gamma_1}$ and $E_{\gamma_2}$ are the two photon energies,
and $m_{\pi^0}$ is the known mass of the neutral pion.
The number of possible photon pairings is
3 for $\Kzz$ and 15 for $\KLzzz$ decays.
For each possible photon pairing we compute the quantity
\begin{equation}
    \chisqzz \equiv 
        \sum_{j=1}^{N_{\pi^0}}
        \left[
               \frac{d_{\pi^0}^j - d_{avg}}{\sigma_d^j}
        \right]^2  ~, 
    \label{eq:chisqzz}
\end{equation}
where $N_{\pi^0}$ is the number of neutral pions (2 or 3),
$d_{\pi^0}^j$ is the distance between the CsI and the
vertex of the $j$'th $\pi^0$,  
$d_{avg}$ is the weighted average of all the $d_{\pi^0}^j$,
and $\sigma_d^j$ is the energy-dependent $\pi^0$-vertex resolution, 
which is  roughly 40~cm (30~cm) at the upstream (downstream) 
end of the decay region.  
The photon pairing with the minimum $\chisqzz$ value is used
because this pairing corresponds to the best agreement of the 
$\pi^0$ vertices.
To reduce the chance of choosing the wrong photon pairing,
we require that the minimum $\chisqzz$ value be less than 12
for \Kzz\ decays, and less than 24 for \KLzzz\ decays.
After all selection cuts,
the probability that the minimum $\chisqzz$ value gives the wrong 
pairing is 0.006\% for \Kzz\ and 0.02\% for $\KLzzz$.
The $\ZK$-vertex used in the analysis is given by 
$\zcsi - d_{avg}$, 
where $\zcsi$ is the $z$-position of the mean 
shower depth in the CsI.
For \Kzz\ decays, the $z$-vertex resolution is $\sim 30$~cm 
near the regenerator and $\sim 20$~cm near the \vw.

The beam assignment is made by comparing the $x$-component of the 
``center-of-energy'' to the regenerator $x$ position.
The center-of-energy is defined to be
 \begin{equation}
       x_{\rm coe} \equiv \frac{ \sum x_i E_i }{ \sum E_i }~,
         ~~~~~~~~
       y_{\rm coe} \equiv \frac{ \sum y_i E_i }{ \sum E_i } ~,
        \label{eq:coe}  
 \end{equation}
where $E_i$ are the cluster energies, $x_i$ and $y_i$ are 
the cluster coordinates at the CsI calorimeter, 
and the index $i$ runs over the photons.
The center-of-energy is the point at which the kaon 
would have intercepted the plane of the CsI 
if it had not decayed.
The value of $\{ x_{\rm coe},y_{\rm coe}\}$ typically
lies well inside the beam-holes,
except for kaons that are scattered by a large angle
in either the regenerator or the defining collimator
(Sec.~\ref{sec:bkg}).
The center-of-energy resolution is $\sim\!1$~mm,
which is much smaller than the beam separation.
An event is assigned to the regenerator beam
if  $x_{\rm coe}$ has the same sign as the
regenerator $x$-position,
or to the vacuum beam if $x_{\rm coe}$ has the opposite sign.

Significant background rejection comes from the invariant
mass cut, $490 < m_{\ppn} < 505~{\rm MeV}/c^2$, and from the
photon veto cuts. 
To calculate the invariant mass, the decay vertex is assumed
to be on the line joining the target and the center-of-energy,
at a distance $d_{\pi^0}$ from the calorimeter. 
Figures~\ref{fig:mass}c,d show the invariant mass distributions.
In the vacuum beam, events in the mass side-band regions are mostly from
$\KLzzz$ decays in which two of the six photons are not detected.
In the regenerator beam, the side-band regions include
comparable contributions from 
$\ppn$ pairs produced in the lead of the regenerator
(Section~\ref{sec:bkg_non2pi}),
\Kzz\ events with the wrong photon pairing,
and $\KLzzz$ decays.

The photon veto cuts are the same as in the charged mode analysis, 
with three additional cuts to reduce background:
(i) the energy of extra isolated EM clusters
    is required to be below 0.6~GeV instead of 1.0~GeV in the charged analysis,
(ii)  the photon-equivalent energy in the beam-hole veto behind the 
      CsI must be less than 5~GeV, and
(iii) the  photon-equivalent energy in the Collar Anti that surrounds the 
      CsI beam-holes (Fig.~\ref{fig:CA})
      must be less than  1~GeV.

Five apertures define the \Kzz\ acceptance. 
\begin{enumerate}
  \item The CsI inner aperture near the beam holes is defined 
        by the CA (Fig.~\ref{fig:CA}).
  \item The outer CsI aperture is defined by rejecting events 
        in which a photon hits the outer-most layer of crystals
        (Fig.~\ref{fig:csilayout}).
        This cut is applied based on the location of the ``seed''
        crystals that have the maximum energy in each cluster.
  \item The upstream aperture in the vacuum beam is defined by 
        the MA (Fig.~\ref{fig:RCMA}b).
  \item The upstream edge in the regenerator beam is defined by the 
        lead at the downstream end of the regenerator
        (Figure~\ref{fig:regdiagram}b).
  \item The requirement of at least 7.5~cm between photons 
        results in an effective inner aperture.
\end{enumerate}

\begin{figure}[ht] 
  \epsfig{file=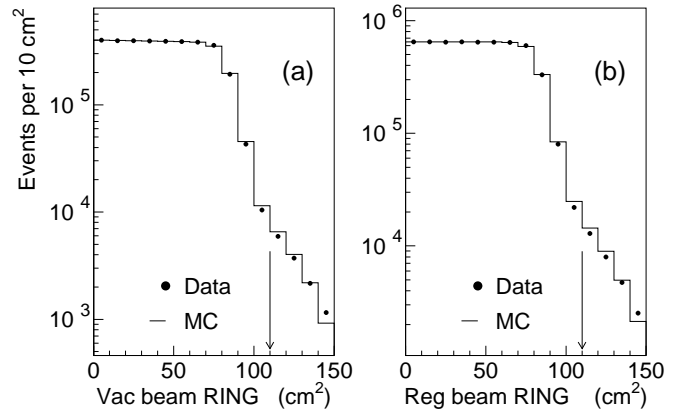,width=\linewidth}
  \caption{
      Vacuum and regenerator $\ring$ distributions for  \Kzz\ 
      events after all other analysis cuts 
      and after background subtraction.
      The data (MC) are shown by the dots (histogram).
      The vertical arrows show the signal selection of 
      $\ring < 110~{\rm cm}^2$.
            }
  \label{fig:ring}
\end{figure}

Since we do not measure photon angles, the transverse momentum
is unknown and therefore
cannot be used to reject events in which a kaon scatters
in the collimator or regenerator.
Instead, we use the
center-of-energy (Eq.~\ref{eq:coe}) to define a
variable called ``Ring Number:''
\begin{equation}
  \ring = 40000 \times 
        {\rm Max}(\Delta x^2_{\rm coe},\Delta y^2_{\rm coe})~,
  \label{eq:ring_def}
\end{equation}
where $\Delta x_{\rm coe}$ ($\Delta y_{\rm coe}$) 
is the $x$-distance ($y$-distance)
of the center-of-energy from the center of the closest
beam hole. 
The $\ring$ value is the area, in cm$^2$, of the smallest square
that is centered on the beam hole and contains the point 
$\{ x_{coe}, y_{coe}\}$.  
The vacuum and regenerator $\ring$ distributions are shown in
Fig.~\ref{fig:ring}.
The beam size at the CsI is approximately $9\times 9~{\rm cm}^2$,
corresponding to $\ring$ values less than $81~{\rm cm}^2$.
The $\ring$ distribution between 100 and $150~{\rm cm}^2$ is sensitive
to scattering in the upstream beryllium absorbers 
($z\simeq 19$~m, Fig~\ref{fig:beamline}).
We require $\ring  < 110~{\rm cm}^2$, which is a compromise
between background reduction and sensitivity to the 
beam halo simulation.

\bigskip

\begin{figure}[ht]
  \epsfig{file=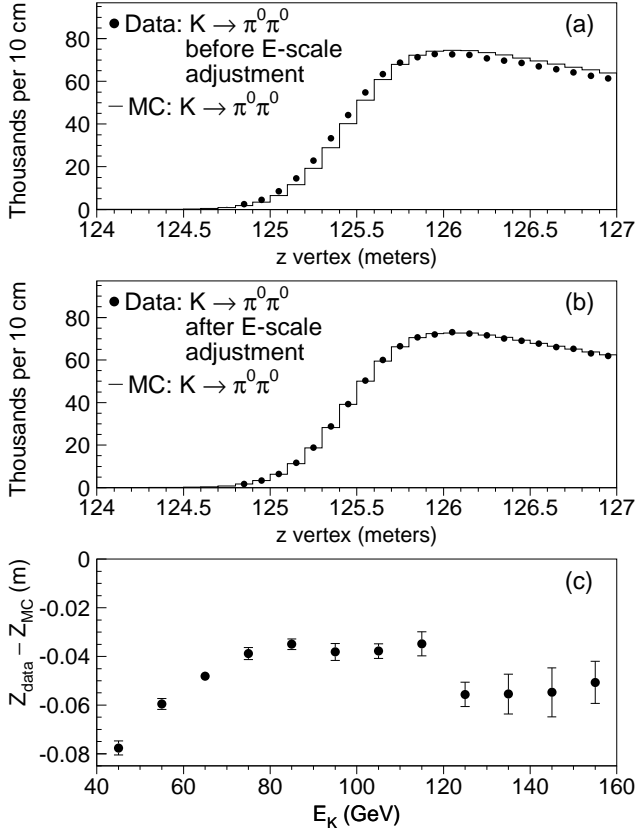,width=\linewidth}
  \caption{
       (a) Reconstructed \Kzz\ $z$-vertex near the 
       regenerator edge for data (dots)
       using the \Kethree\ electron calibration,
       and for MC (histogram).  
       (b) Same as (a) after the final energy scale adjustment 
       is applied.
       (c) Data-MC difference in the reconstructed $z$-vertex
           (meters) in 10 GeV kaon energy bins,
           before the final energy scale adjustment.
                  }
 \label{fig:regmatch}
\end{figure}

The final part of the \Kzz\ analysis is to match
the photon energy scale for data and Monte Carlo.
This matching is performed after the background subtraction
that will be described in Sec.~\ref{sec:bkg}.
Since the energy scale affects the determination of both the kaon
energy and the $z$-vertex position (through Eq.~\ref{eq:neut_z}),
events can migrate into and out of the selected event sample
depending on the energy scale.  The final energy scale is adjusted
to match the data and MC
reconstructed $z$-vertex distribution of $\Kzz$ decays at the
regenerator edge, as shown in Fig.~\ref{fig:regmatch}a,b.  
Using $\sim 10^6$ events near the regenerator edge,
the energy scale is determined
in 10 GeV wide kaon energy bins, which is the binning used 
to extract $\reepoe$.
On average, the regenerator edge in data lies $\sim 5$~cm upstream 
of the MC edge as seen in Fig.~\ref{fig:regmatch}c;
the data-MC regenerator edge difference varies
by a few centimeters depending on the kaon energy.

Since the vertex distance from the CsI is proportional to the 
cluster energy scale (Eq.~\ref{eq:neut_z}),
the multiplicative energy scale correction is $\sim 0.9992$,
which corresponds roughly to the $-5$~cm data-MC difference divided by the 
61~meter distance between the regenerator and the CsI.
The regenerator edge shift in each 10~GeV kaon energy bin
(Fig.~\ref{fig:regmatch}c) is converted to an energy scale
correction, and is applied to each cluster in data.

  \subsection{Systematic Uncertainties from \ppn\ 
              Trigger, Reconstruction, and Selection}
  \label{sec:neut_syst}

\begin{table}
\caption{ 
    \label{tb:neut_syst}
    Systematic uncertainties in $\reepoe$ 
    from the neutral mode trigger and analysis.
    The uncertainties in the 3rd column, which also appear in the
    systematics summary (Table~\ref{tb:syst_reepoe}),
    are the quadratic sum of contributions in the 2nd column.
         }
\begin{ruledtabular}
\begin{tabular}{ l | c | c} 
Source of       &     \multicolumn{2}{c}{Uncertainty on $\reepoe$}     \\
uncertainty     &     \multicolumn{2}{c}{ ($\eu$)}  \\ 
 \hline  
Trigger                   &        & {\bf 0.18}  \\ 
~~~ L1 trigger            & 0.10   &      \\
~~~ L2 trigger            & 0.13   &      \\
~~~ L3 filter             & 0.08   &      \\
 \hline  
Cluster reconstruction        &        & {\bf 1.47}  \\
~~~ Energy scale              &  1.27  &  \\
~~~ Non-linearity             &  0.66  &  \\
~~~ Position reconstruction   &  0.35  &  \\
 \hline  
Selection efficiency          &      & {\bf 0.37} \\
~~~ $\ring$ cut            & 0.24 &         \\
~~~ $\chisqzz$ cut         & 0.20 &         \\
~~~ Transverse shape       & 0.20 &         \\
 \hline  
Apertures                 &       & {\bf 0.48}  \\
~~~ Collar Anti           & 0.42  &       \\
~~~ Mask Anti             & 0.18  &       \\
~~~ Reg edge              & 0.04  &       \\
~~~ CsI size              & 0.15  &       \\
~~~ Photon separation     & $-$   &       \\
\end{tabular}
\end{ruledtabular}
\end{table}

In this section, 
we discuss the $\reepoe$ systematic uncertainties from the
neutral mode trigger, reconstruction, and selection,
which are summarized in Table~\ref{tb:neut_syst}.
Systematic uncertainties related to background and acceptance 
are discussed in 
Sections~\ref{sec:bkg_syst} and \ref{sec:acceptance}.

\bigskip{\it Trigger:}
The Level~1  CsI ``energy-sum'' trigger is studied
using the large $\KLpienu$ sample in the charged trigger
which does not have L1 CsI requirements (Sec.~\ref{sec:trigger}).
The Level~2 cluster-counter is studied using a half-million 
\KLzzz\ decays from a trigger that requires only Level~1.
The Level~3 systematic uncertainty is based on
$1.3\times 10^5$ \Kzz\ events that satisfy all analysis cuts
and that were accepted online without requiring L3.
The combined $\reepoe$ systematic uncertainty from the trigger
is $0.18\eu$.

\begin{figure}[htb]
  \epsfig{file=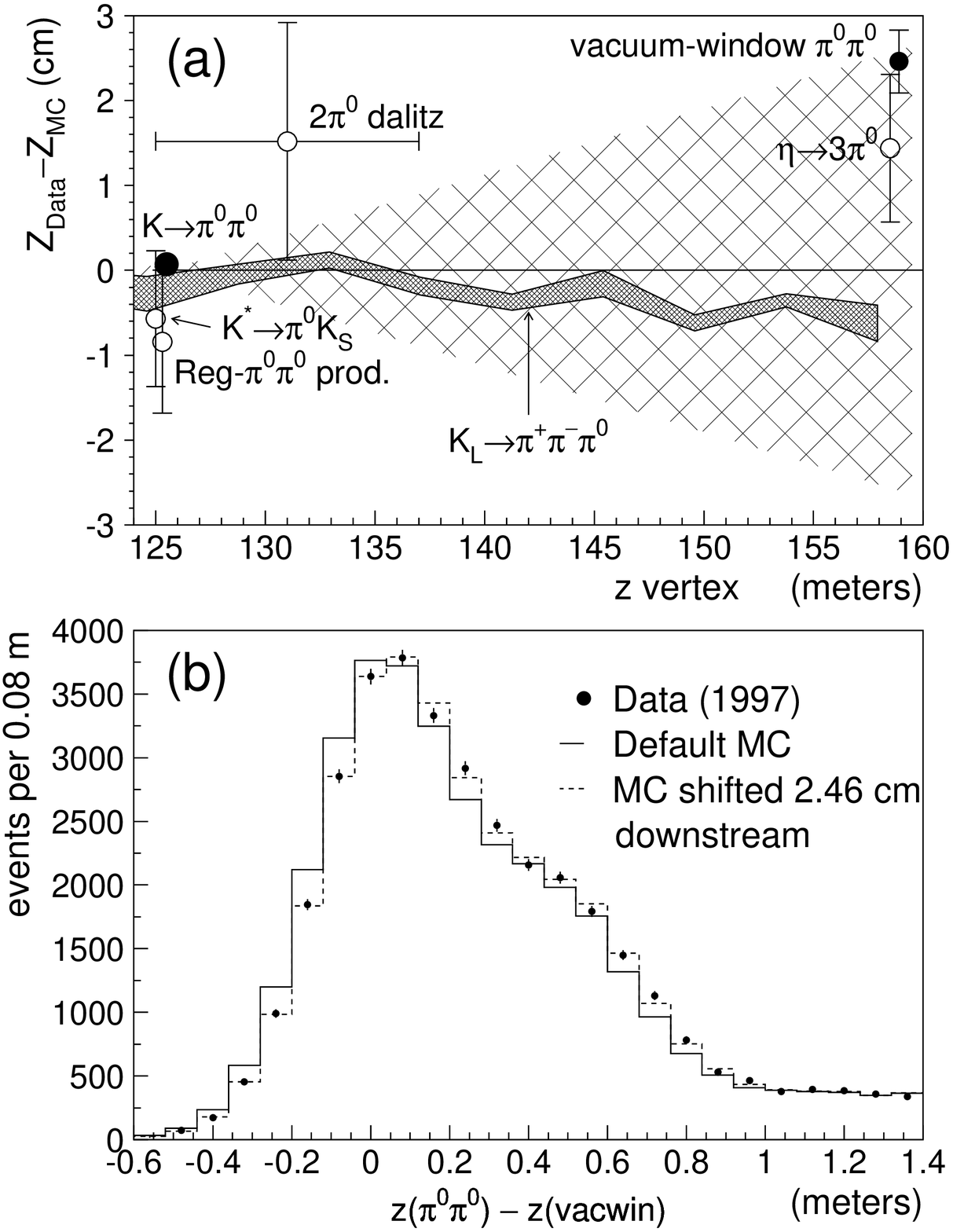,width=\linewidth}
  \caption{
       (a) The difference between the reconstructed data and MC
       neutral vertex at different locations within the 
       decay region (1997 sample).
       The \Kzz\ mode at $z=125$~m (solid dot) has a data-MC difference 
       of zero by definition since the regenerator edge has been matched.
       The \vw\ \ppn\ crosscheck at $z=159$~m (solid dot)
       shows the largest discrepancy; the open circles and 
       $\KLpmz$ band show other crosschecks. The hatched region shows the
       range of discrepancies covered by the assigned systematic 
       uncertainty.
       (b) The reconstructed \ppn\ vertex distribution relative to 
       the \vw\ for data after matching the regenerator edge (dots),
       for the default MC (histogram), and for the MC shifted 2.46~cm
       downstream (dashed histogram). 
            } 
 \label{fig:esclsyst}
\end{figure}

\bigskip{\it Cluster Reconstruction:}
The understanding of the cluster reconstruction
in the CsI calorimeter,
and in particular the energy scale,
results in the largest systematic uncertainty 
in the $\reepoe$ analysis.
After matching the data and MC at the regenerator edge,
we use a variety of other modes that
have one or more $\piz$s in the final state
to check how well the data and MC match at other $z$-locations.
All of these crosscheck samples are collected at the same time as
$\Kpp$ decays, and therefore the detector conditions and calibration
are precisely the same as for the \Kzz\ sample.
Data-MC comparisons of the reconstructed neutral vertex, 
relative to either a charged mode vertex or the known {\vw} location,
are made in the vacuum beam for the following:
\begin{enumerate}
   \item  $\KzzD$, where $\pi^0_D$ refers to $\pi^0\to\eeg$;
          the reconstructed $\pi^0\to\gamgam$ vertex is compared 
          to the $e^+e^-\gamma$ vertex;
   \item  $\KLpmz$ in which the  reconstructed $\pi^0\to\gamgam$ vertex
          is compared to the $\ppc$ vertex;          
   \item  $\eta\to\zzz$, where $\eta$ mesons are produced by 
          beam interactions in the \vw\ at $z=159$~m;
   \item  \ppn\ pairs produced by beam
          interactions in the \vw\ at $z=159$~m.
\end{enumerate}
Additional crosschecks at the regenerator edge include \ppn\ pairs
produced by neutron interactions in the regenerator,
and $K^{\star}\to\piz K_S$ with $K_S\to\ppc$.
A summary of the neutral vertex crosschecks is given
in Fig.~\ref{fig:esclsyst}a; 
they are all consistent with the nominal energy scale correction 
except for the \vw\ \ppn\ pairs.

Since the \vw\ \ppn\ pairs result in the largest discrepancy,
a discussion of the analysis of these events is given here.
The selection of \vw\ $\ppn$ pairs is similar to that for 
the \Kzz\ decay mode; 
the main difference is that \Kzz\ decays are
excluded by selecting events in which the \ppn\ invariant mass is at
least $15~{\rm MeV/c}^2$ away from the kaon mass. 
The \vw\ \ppn\ sample consists of 45000 events
in the vacuum beam.
The \vw\ $z$-location is known with $1$~mm 
precision using charged two-track events produced in the \vw.
The simulation of \ppn\ pairs is tuned to match
data and MC distributions of reconstructed energy, $\ring$,
and \ppn\ invariant mass.
Figure~\ref{fig:esclsyst}b shows the reconstructed \vw\ \ppn\ vertex
for data and MC after the regenerator edge $z$-distribution
has been matched.
The data-MC comparison is complicated by the helium and 
drift chamber immediately downstream of the \vw\ since this extra 
material is also a source of \ppn\ pairs.
The production of \ppn\ pairs is simulated separately in the 
\vw, helium, and drift chamber. 
To evaluate the data-MC discrepancy in the $z$-vertex distribution,
a fit is used to determine the relative 
contribution from each material along with the data-MC vertex shift.  
The result of this analysis is that the data $z$-vertex distribution 
at the \vw\ lies 
$[2.46\pm 0.16~({\rm stat}) \pm 0.33~({\rm syst})]$~cm downstream 
of the MC distribution for the 1997 sample; 
for the 1996 sample, the corresponding shift is 
$[1.89\pm 0.24~({\rm stat}) \pm 0.33~({\rm syst})]$~cm.
The data-MC shifts are evaluated separately in the 
1996 and 1997 samples
because of the different CsI PMT signal integration times.

The data-MC ``$z$-shift'' at the \vw\ 
is translated into a $\reepoe$ uncertainty
by introducing a CsI energy scale distortion 
to data such that the data and MC $z$-vertex distributions
match at both the regenerator and \vw\ edges;
this distortion changes \reepoe\ by 
$-1.08\eu$ and $-1.37 \eu$ for the 1996 and 1997 data samples,
respectively. 
The combined systematic uncertainty on $\reepoe$ is $1.27\eu$.
The energy scale distortion leading to this uncertainty
varies linearly with decay vertex in both beams, 
and corresponds to the hatched region in Fig.~\ref{fig:esclsyst};
non-linear energy scale variations as a function of decay vertex
are ruled out because they introduce data-MC discrepancies
in other distributions.

Some reconstructed quantities in the analysis do not depend on
the CsI energy scale, 
but are sensitive to  energy non-linearities.
For example, the $m_{\ppn}$ peak varies by 0.2~MeV/c$^2$ 
for kaon energies between 40 and 160 GeV,
and the data-MC $z$-difference at the regenerator edge 
(Fig.~\ref{fig:regmatch}c)
varies by 4~cm over the same kaon energy range.
Such data-MC discrepancies can be reproduced
with cluster-energy distortions based on 
energy, angle, position, and pedestal shift.
These distortions are not used in the final result,
but are used to determine a systematic uncertainty of 
$0.66\eu$ on $\reepoe$.

Cluster position reconstruction is studied using electrons from
\Kethree\ decays and comparing the position reconstructed from
the CsI to the position of the electron track projected to the 
mean shower depth in the CsI.
The position differences are parameterized and simulated; 
the maximum $\reepoe$ shift of $0.35\eu$ 
is assigned as a systematic uncertainty.

\bigskip{\it Selection Efficiency:}
There are three variables for which $\reepoe$ is 
sensitive to the cut value. 
(i) Varying the $\ring$ cut between 100 and $150~{\rm cm}^2$ changes 
$\reepoe$ by $0.24\eu$ (Fig.~\ref{fig:ring}). 
(ii) Relaxing the $\chisqzz$ requirement 
such that the inefficiency from this cut
is reduced by a factor of 3
changes \reepoe\ by $0.20\eu$.
(iii) Removing the transverse energy distribution requirement
changes $\reepoe$ by $0.20\eu$.
These three changes added in quadrature contribute an 
uncertainty of $0.37\eu$ to $\reepoe$.

\bigskip{\it Apertures:}
The aperture uncertainties are mainly from the
Collar-Anti (CA, Fig.~\ref{fig:CA}) and
Mask-Anti (MA, Fig.~\ref{fig:RCMA}b).
Their effective sizes and positions are measured
with $100~\mu m$ precision using $\KLpienu$ decays,
resulting in $\reepoe$ uncertainties of 
$0.42\eu$ and $0.18\eu$ from the CA and MA, respectively.
The CsI calorimeter size is known to better than 1~mm from surveys,
resulting in a $0.15\eu$ uncertainty on $\reepoe$.
The 0.1~mm uncertainty on the
effective regenerator edge (Fig.~\ref{fig:regdiagram}b) leads to a
$\reepoe$ uncertainty of $0.04\eu$.
Varying the minimum allowed photon separation between 5~cm and  20~cm 
results in no significant change in $\reepoe$.
The total $\reepoe$ uncertainty from the
neutral mode apertures is $0.48\eu$.

  \subsection{Background to \Kpm\ and \Kzz\ } 
  \label{sec:bkg}

Two types of background are relevant for this analysis. 
First, there is ``non-$\pi\pi$'' background from misidentification
of high branching-ratio decay modes such as 
semileptonic $\Kethree$ and $\Kmuthree$ in the charged mode,
and $\KLzzz$ in the neutral mode.
The second type of background is from 
kaons that scatter in the regenerator or the defining collimator
and then decay into two pions.
Kaon scattering is the same for both the charged and 
neutral decay modes;
it can be largely eliminated in the charged mode analysis
using the reconstructed 
total transverse momentum of the decay products,
but the lack of a photon trajectory measurement does not allow 
a similar reduction in the neutral mode analysis.

Regenerator and collimator scattering affect
the \Kzz\ reconstruction in two ways.
First, a kaon can scatter at a small angle and still reconstruct
within the same beam; 
this ``in-beam'' background
has different regeneration properties and acceptance compared
to unscattered kaons, 
which can cause a bias in $\reepoe$, 
as well as in kaon parameters such as
$\Delta \phi$ and $\Delta m$.
Second, a kaon that undergoes large-angle scattering can 
traverse from one beam to the other beam and then decay, 
leading to the wrong beam assignment in the neutral mode analysis;
this ``crossover background'' is mostly $K_S\to\ppn$.
Background from regenerator scattering is roughly ten times
larger than from scattering in the defining collimator.

All backgrounds are simulated, normalized to data, 
and then subtracted from the \Kpp\ signal samples.
The background-subtraction procedure is described in the 
following sections, and the background-to-signal ratios
are summarized in Table~\ref{tab:bkgd}.

  \subsubsection{Non-$\pi\pi$  Background} 
  \label{sec:bkg_non2pi}

Charged pion identification and kinematic cuts 
(Sec.~\ref{sec:chrg_evtsel}) eliminate most of the charged mode 
non-$\pi\pi$ background;
only a 0.09\% contribution from
the semileptonic \Kmuthree\ and \Kethree\ processes 
remains to be subtracted in the vacuum beam.
Both semileptonic background distributions are simulated,
and then normalized to data using events reconstructed
outside the invariant mass and \ptsq\ signal region.

The only substantial non-$\pi\pi$ background in the neutral 
mode is from \KLzzz\ with undetected photons 
or photons that have merged at the CsI calorimeter.
This 0.11\% background is simulated and then normalized to data 
using sidebands in the \ppn\ invariant mass distribution in 
the vacuum beam.
The decay $\KLzgg$ has a branching fraction of
$1.7\times 10^{-6}$ \cite{pdg00};
it contributes $2\times 10^{-5}$ background in the vacuum beam
and is ignored.
We also ignore 
$\Xi^{0}\to\Lambda\pi^0$ with $\Lambda\to n\pi^0$,
which contributes less than $10^{-5}$ background.

The charged and neutral decay modes both have 
misidentification background associated with 
hadronic production in the lead plate of the 
last regenerator module (Fig.~\ref{fig:regdiagram}b). 
In both the charged and neutral data samples,
this ``regenerator-hadron'' production is easily isolated
in the reconstructed $\ZK$-vertex distribution at the 
regenerator edge.
This background is almost entirely rejected by 
the \ptsq\ cut in the charged decay mode;
the remaining $10^{-5}$ background is ignored.
In the neutral decay mode, the regenerator-$\ppn$ background  is
$8\times 10^{-5}$  ($2\times 10^{-5}$) in the
regenerator (vacuum) beam, 
and is included in the background simulation.

  \subsubsection{Collimator Scattering Background} 
  \label{sec:bkg_coscat}

\begin{figure}[ht]
  \centering
  \epsfig{file=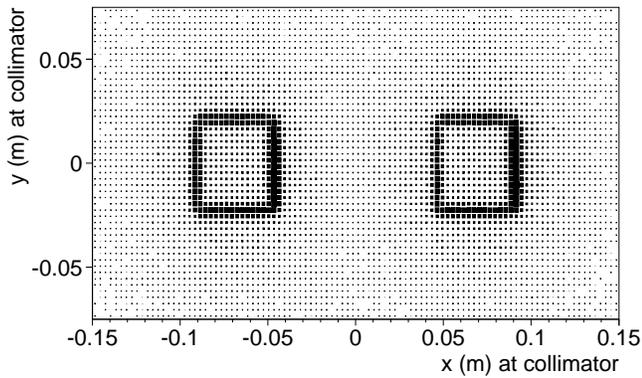,width=\linewidth}
  \caption{
       $y$ vs. $x$ distribution of the kaon trajectory projected
       back to the $z$ position of the defining collimator in
       the vacuum beam. The sample shown here is after all
       $\Kpm$ analysis cuts, except that the \ptsq\ cut is changed
       to $\ptsq > 1000~\uptsq$. 
       The square bands correspond to the
       edges of the defining collimators.
          }
  \label{fig:cosct}
\end{figure}

Scattering in the defining collimators is studied using 
\Kpm\ and $\KLpmz$ decays in  the vacuum beam. 
Figure~\ref{fig:cosct} shows the $y$ vs. $x$ 
distribution of the kaon trajectory
projected back to the $z$ position of the defining collimator for 
high $\ptsq$ vacuum beam events that satisfy all other
\Kpm\ requirements.
The square bands show kaons that scattered from the
defining collimator edges before decaying.
The events in  Fig.~\ref{fig:cosct} that lie outside the 
collimator scattering bands are mainly from semileptonic decays.
To determine the number of collimator scatters accurately,
the roughly 10\% semileptonic component is subtracted.
In the charged decay mode, the background from collimator scattering
is only 0.01\%, and is small mainly because of the \ptsq\ cut;
in the neutral decay mode, this background is about 0.1\% and
therefore requires an accurate simulation.

The MC simulation propagates each kaon to the defining collimator, 
and checks if the kaon strikes the collimator at either the
upstream end or anywhere along the 3~meter long inner surface.
Kaons that hit the collimator are traced through the steel
and allowed to scatter back into the beam.
A kaon that scatters in the collimator
is parameterized to be either pure $K_S$ or pure $K_L$,
with the relative amount adjusted to match the $z$-vertex
distribution for the collimator scattering sample shown in 
Fig.~\ref{fig:cosct}.

The MC treatment of collimator scattering is the same in both the 
vacuum and regenerator beams. 
About 1/3 of the  collimator-scattered kaons hit the 
Mask Anti (MA, Fig.~\ref{fig:RCMA}), 
and can then punch through and exit the MA as 
either a $K_L$ or $K_S$. 
Based on measurements from data, the MC includes 
a kaon punch-through probability of 60\%,
and a  $K_S$ to $K_L$ ratio of 
about 50 for kaons that exit the MA.

  \subsubsection{Regenerator Scattering Background} 
  \label{sec:bkg_regscat}

In the charged mode analysis, the regenerator scattering 
background is 0.074\% in the regenerator beam, 
and is not present in the vacuum beam. 
In the neutral mode analysis, the corresponding background levels
are 1.13\% in the regenerator beam
and 0.25\% in the vacuum beam.

\begin{figure}[ht]
  \centering
  \epsfig{file=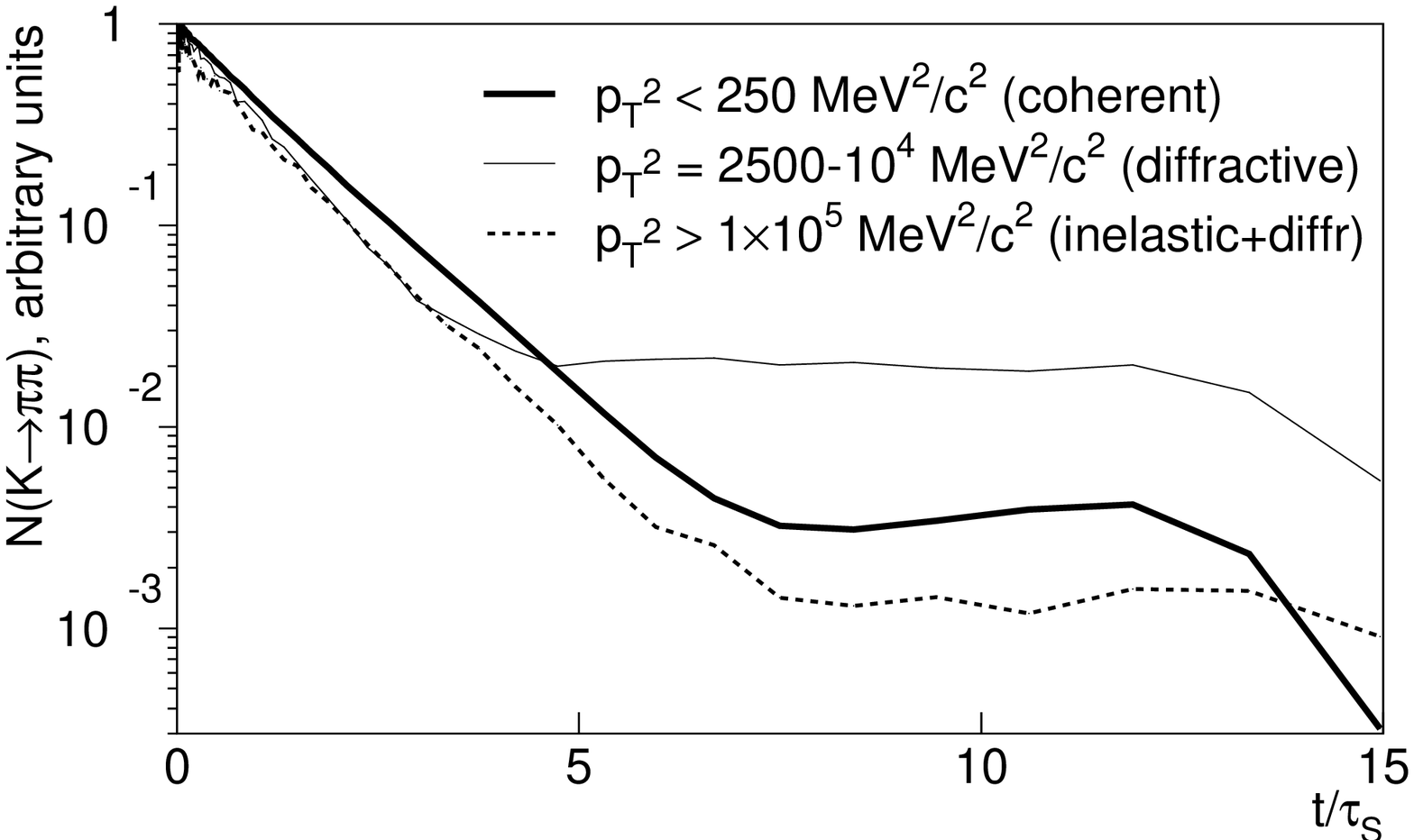,width=\linewidth}
  \caption{
     Acceptance-corrected
     distribution of the number of regenerator beam \Kpm\ decays
     vs. the number of $K_S$ lifetimes ($t/\tauS$),
     for kaon momenta between 40 and $50~\upk$.
     The proper time $t=0$ corresponds to
     the downstream edge of the regenerator.
     The \ptsq\ ranges are indicated on the plot for each curve.
     The thick curve is for coherent \Kpm\ decays.
     The thin solid curve corresponds to small \ptsq\ 
     and is mostly from diffractive scattering.
     The dashed curve is for large \ptsq\ events, which are
     due to both inelastic and diffractive scattering.
     The fluctuations are due to statistics.
     All \Kpm\ cuts except \ptsq\  have been applied.
     Backgrounds from collimator scattering and semileptonic decays 
     have been subtracted.
          }
  \label{fig:tdk_rgsct}
\end{figure}

Regenerator scattering is more complicated 
than collimator scattering, particularly in the time dependence
of \Kpp\ decays resulting from the coherent $K_L$-$K_S$ mixture.
Figure~\ref{fig:tdk_rgsct} shows the observed proper
decay time distributions for kaons that scatter 
with small $\ptsq$ ($2500{\rm -}10^4~\uptsq$) and
large $\ptsq$ ($ > 10^5~\uptsq$),
and for the unscattered \Kpm\ signal
($\ptsq < 250~\uptsq$).
Note that the proper time distribution depends strongly on the 
\ptsq\ value.  
Large-\ptsq\ scattering contributes
mainly to crossover background,
while small-\ptsq\ scattering contributes mostly to 
in-beam background. 
A detailed description of the regenerator scattering background 
is needed for the neutral decay mode because the $\Kzz$ 
sample includes events with $\ptsq$ values up to about
$3\times 10^4~\uptsq$ in the regenerator beam, and up to 
$5\times 10^5~\uptsq$ in the vacuum beam.

A \Kpp\ decay from a kaon that has scattered in the regenerator,
referred to as a ``regenerator scattering decay,''
is described in the MC using a function that is
fit to acceptance-corrected \Kpm\ data after 
subtracting collimator scattering and semileptonic decays.
The acceptance correction allows us to fit the ``true''
regenerator scattering decay distribution, 
so that the scattering simulation
can be used to predict background for both 
charged and neutral mode decays.
As described in Appendix~\ref{app:fitfun},
the fit function depends on
proper time, $\ptsq$, and kaon momentum.
To remove non-\ppc\ background from charged 
tracks produced in the regenerator lead (Fig.~\ref{fig:regdiagram}b),
the fit excludes decays within
0.2 $K_S$ lifetimes of the regenerator edge.
To avoid the $\ptsq$ tail from coherent events,
only decays with $\ptsq >2500~\uptsq$ are used in the fit.
The fit momentum region is the same as in the signal analysis:
40 to $160~\upk$.

As discussed in Section~\ref{sec:regenerator}, there are
two processes that contribute to regenerator scattering.
The first process is diffractive scattering, 
which is identical in the charged and neutral mode analyses 
because no energy is deposited in the regenerator or 
photon vetos.
The second process is inelastic scattering, 
which is slighlty different 
in the two modes because of different photon veto 
requirements.
To address this charged-neutral difference,
the fit function is based 
on a phenomenological model that has separate terms for 
diffractive and inelastic scattering.
The \ptsq\ distribution is significantly steeper for 
diffractive scattering than for inelastic scattering,
and this difference allows for the two scattering
processes to be distinguished.

Using the fit function to simulate the $\Kzz$ scattering background,
and normalizing to data events 
in the range $300 < \ring\ < 800~{\rm cm}^2$,
we find that the neutral mode veto requirements suppress
inelastic scattering by an additional 16\% compared 
to the charged mode veto requirements.
This 16\% charged-neutral difference in the inelastic component
corresponds to a 3\% charged-neutral difference in the total
regenerator scattering background.

  \subsubsection{Summary of Backgrounds} 
  \label{sec:bkg_summ}

\begin{figure}[ht]
  \centering
  \epsfig{file=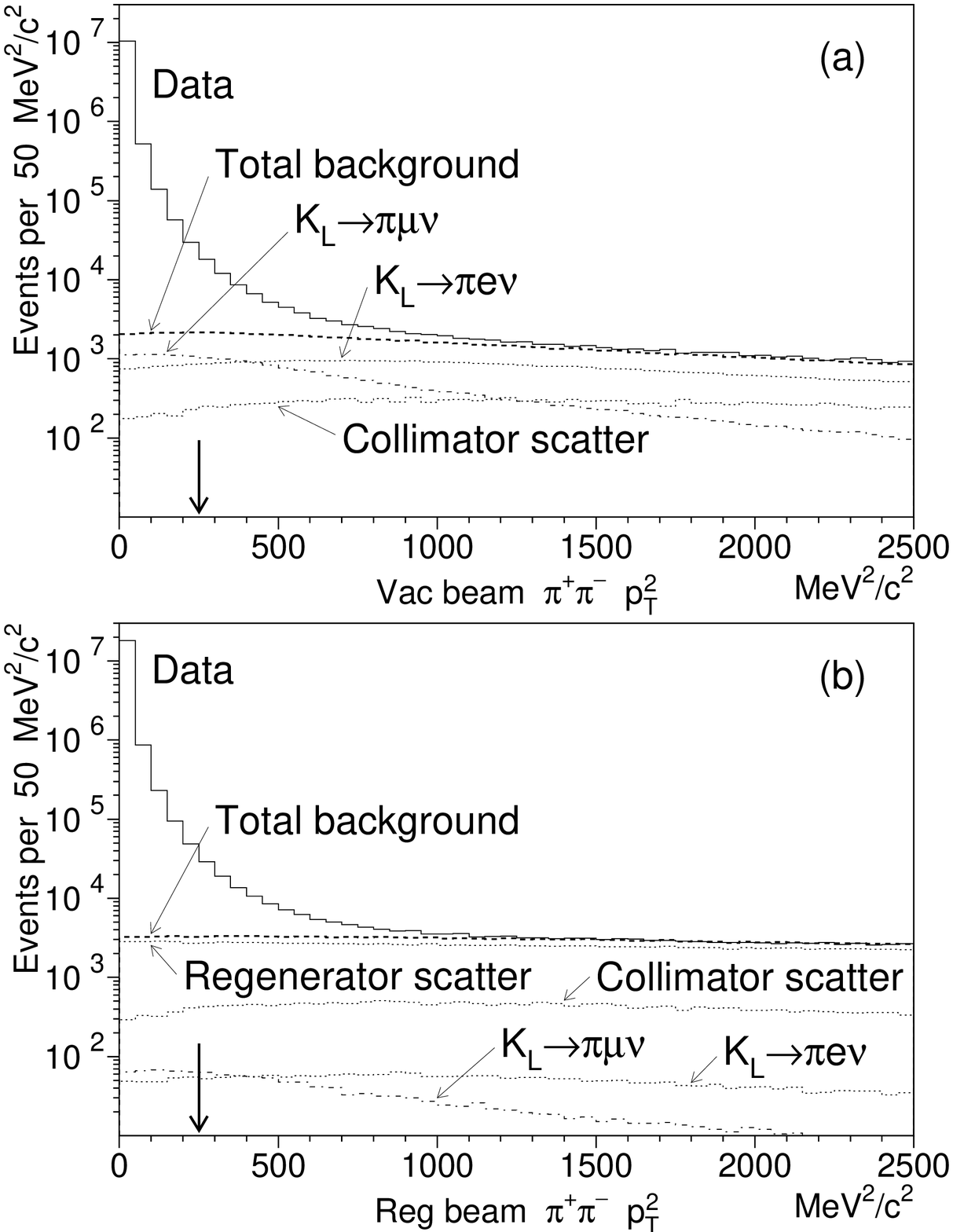,width=\linewidth}
  \caption{
     Data \ptsq\ distribution after all other \Kpm\ selection cuts for
     (a) the vacuum beam and (b) the regenerator beam.     
     The Monte Carlo predictions for the background components
     are overlaid.
     Events with $\ptsq < 250~\uptsq$ (vertical arrow)
     are included in the final \Kpm\ sample. 
          }
  \label{fig:ptsq_shapes}
\end{figure}

\begin{figure}[ht]
  \epsfig{file=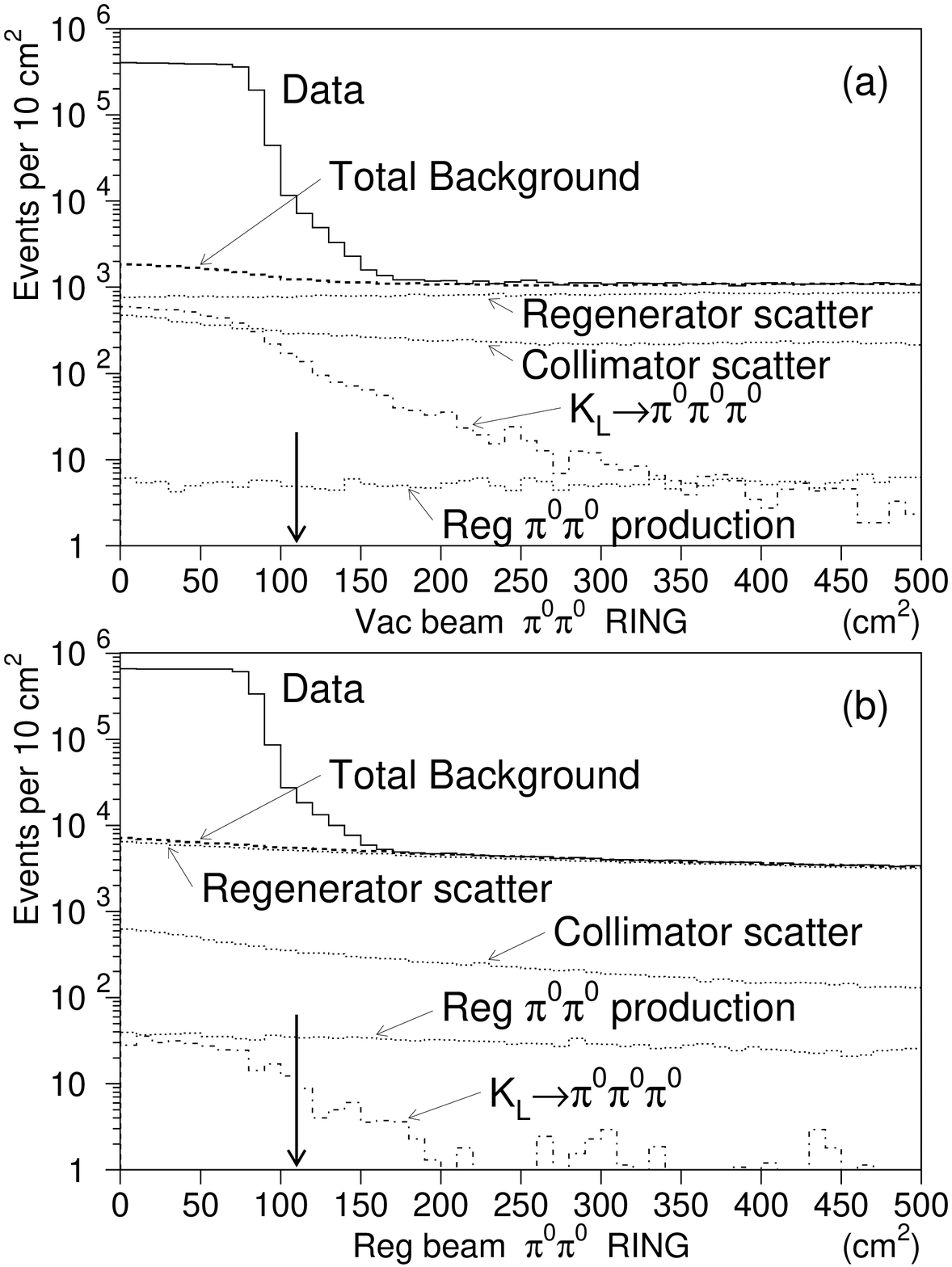,width=\linewidth}
  \caption{
     Data  $\ring$  distribution after all other 
     \Kzz\ selection cuts for
     (a) the vacuum beam and (b) the regenerator beam.
     The Monte Carlo predictions for the background components
     are overlaid. 
     Events with $\ring < 110~{\rm cm}^2$ (vertical arrow)
     are included in the final \Kzz\ sample.
          }
  \label{fig:ringvaceps}
\end{figure}

\begin{figure}[ht]
  \epsfig{file=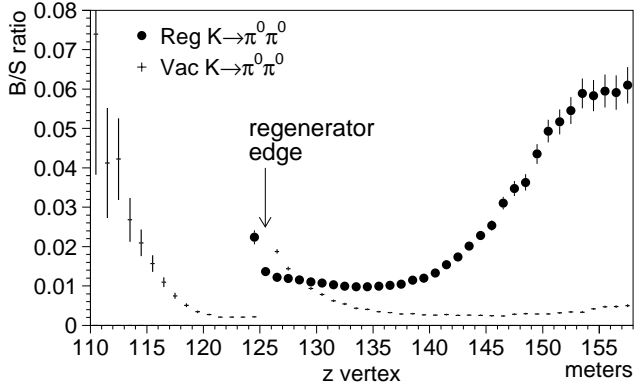,width=\linewidth}
  \caption{
     \Kzz\ background-to-signal (B/S) ratio vs. $z$-vertex in 1~meter bins.
     The regenerator (vacuum) beam is represented
     by solid dots (crosses). The vertical arrow shows
     the location of the downstream edge of the regenerator.
          }
  \label{fig:zbkgneut}
\end{figure}

Figure \ref{fig:ptsq_shapes} shows the vacuum and regenerator beam
\ptsq\ distributions after all other charged mode analysis requirements.
MC simulations of the background processes are also shown.
The regenerator beam background is mostly from regenerator scattering,
and the vacuum beam background is mostly from semileptonic decays.
Figure \ref{fig:ringvaceps} shows the neutral mode $\ring$ 
distribution in both beams, along with MC background simulations.
Figure~\ref{fig:zbkgneut} shows the neutral mode background-to-signal 
ratio (B/S)  as a function of the \Kzz\ decay vertex. 
In the regenerator beam, the main background is from
regenerator scattering.
In the vacuum beam,
the largest sources of background are 
collimator scattering between 110~m and 125~m,
crossover regenerator scattering between 125~m and 140~m,
and \KLzzz\ decays for $z>140$~m.

Table \ref{tab:bkgd} summarizes the background levels for both
decay modes. The charged mode background
level is $\sim 10^{-3}$ in both beams; 
the neutral background level is 1.2\%
in the regenerator beam and 0.5\% in the vacuum beam. 
The background subtraction results in corrections 
to $\reepoe$ of 
$-12.5\eu$ for the neutral decay mode and 
$-0.2\eu$ for the charged decay mode.

\begin{table}
  \centering
  \caption{
      \label{tab:bkgd}
       The background-to-signal ratio (B/S) for each
       subtracted background component,
       for each decay mode, and for each beam.
       The $\reepoe$ systematic uncertainty associated with
       each background component is listed in the column labeled
       ``$\sigma_{syst}$''.
       }
  \medskip
\begin{ruledtabular}
\begin{tabular}{lccc}
    Background         & Vac         & Reg      & $\sigma_{syst}$ \\
    process            & B/S (\%)    & B/S (\%) & $(\eu)$          \\         
  \hline 
    \multicolumn{4}{c}{ $\Kpm$ Analysis:} \\
    \Kethree\                          & \VacKelBkg\ & \RegKelBkg\ & 0.12 \\ 
    \Kmuthree\                         & \VacKmuBkg\ & \RegKmuBkg\ & 0.12 \\ 
    Collimator scattering              & \VacColBkg\ & \RegColBkg\ & 0.01 \\
    Regenerator scattering             & \VacSctBkg\ & \RegSctBkg\ & 0.10 \\ 
 \hline 
    Total charged                      & \VacTotBkg\ & \RegTotBkg\  & \BkgdChrgSyst \\
 \hline 
 \hline 
    \multicolumn{4}{c}{ $\Kzz$ Analysis:} \\
     $\KLzzz$                           & 0.107       & 0.003       & 0.07 \\ 
     Reg-$\ppn$ production             & 0.002       & 0.008       & 0.05 \\
     Collimator scattering             & 0.123       & 0.094       & 0.10 \\
     Regenerator scattering            & 0.252       & 1.130       & \BkgdNeutRegSyst \\ 
  \hline 
     Total neutral                 &  0.484      & 1.235           & \BkgdNeutSyst \\ 
\end{tabular}
\end{ruledtabular}
\end{table}

  \subsubsection{Systematic Uncertainties from Background} 
  \label{sec:bkg_syst}

The uncertainties in $\reepoe$ resulting from the background subtraction
are shown in the last column of Table~\ref{tab:bkgd}.
In the \Kpm\ analysis, the background contributes 
an uncertainty of $0.20\eu$ on $\reepoe$;
this uncertainty is based on changes in $\reepoe$ 
when background rejection cuts are varied
for the \ppc\ invariant mass, $E/p$,
and the minimum pion momentum.

In the \Kzz\ analysis, the background contributes 
an uncertainty of $1.07\eu$ on $\reepoe$,
and is mostly from the 5\% uncertainty on the
background level for in-beam regenerator scattering.
This background subtraction depends largely on modeling the 
acceptance for high $\ptsq$ $\Kpm$ events.

To check our understanding of the detector acceptance at high $\ptsq$,
we use $\ppc$ pairs from $\KLpmz$ decays.
Comparing the data and MC \ppc\ \ptsq\ distributions,
we limit the data-MC difference to be less than 
0.5\% per $10000~\uptsq$.
To convert this limit on the ``$\ptsq$ slope'' 
into a potential bias on $\reepoe$,
we weight the \ptsq\ distribution 
in the neutral mode background simulation by this slope;
the resulting $0.4\eu$ change in $\reepoe$ is included as
a systematic uncertainty.

Imperfections in the phenomenological parameterization and fitting 
of the $\Kpm$ scattering distribution are estimated by comparing
charged mode data to the MC simulation. The maximum
data-MC difference in the scattering distribution is $3\%$ 
of the background level, which
corresponds to a $0.50\eu$ uncertainty for $\reepoe$.

There is also an uncertainty in how the scattering distribution,
measured with \Kpm\ decays,
is used to simulate background for \Kzz\ decays.
As mentioned in Section~\ref{sec:bkg_regscat},
the observed charged-neutral difference of 3\% in the regenerator
scattering level is accounted for by a 16\% reduction
in the inelastic scattering component in the neutral mode
background simulation (Appendix~\ref{app:fitfun}).
If we ignore differences between diffractive and inelastic 
scattering,
and simply reduce the total scattering level by 3\% in the
neutral mode simulation, $\reepoe$ changes by $+0.3\eu$.
We assign $0.3\eu$ as the systematic uncertainty on $\reepoe$
to account for the uncertainty in the inelastic-to-diffractive ratio
for \Kzz\ decays,
and to account for possible charged-neutral differences in the
\ptsq\ distribution.

We have also checked the effect of variations in the 
analysis requirements on the background subtraction.
The most significant effect is from increasing the 
regenerator-veto threshold from 8~MeV to 24~MeV 
in both the \Kpm\ kaon scattering analysis
and the \Kzz\ signal analysis.
This change doubles the inelastic regenerator
scattering background
and shifts \reepoe\ by $(+0.7\pm 0.3)\eu$,
leading to an additional systematic uncertainty of $0.8\eu$.

The other neutral mode background sources (Table~\ref{tab:bkgd}) 
have a much smaller effect on the measurement than 
regenerator scattering.
The $\reepoe$  uncertainty from all
neutral mode backgrounds is $\BkgdNeutSyst\eu$.

  \subsection{Analysis Summary}
  \label{sec:ana summary}

\begin{figure} 
  \epsfig{file=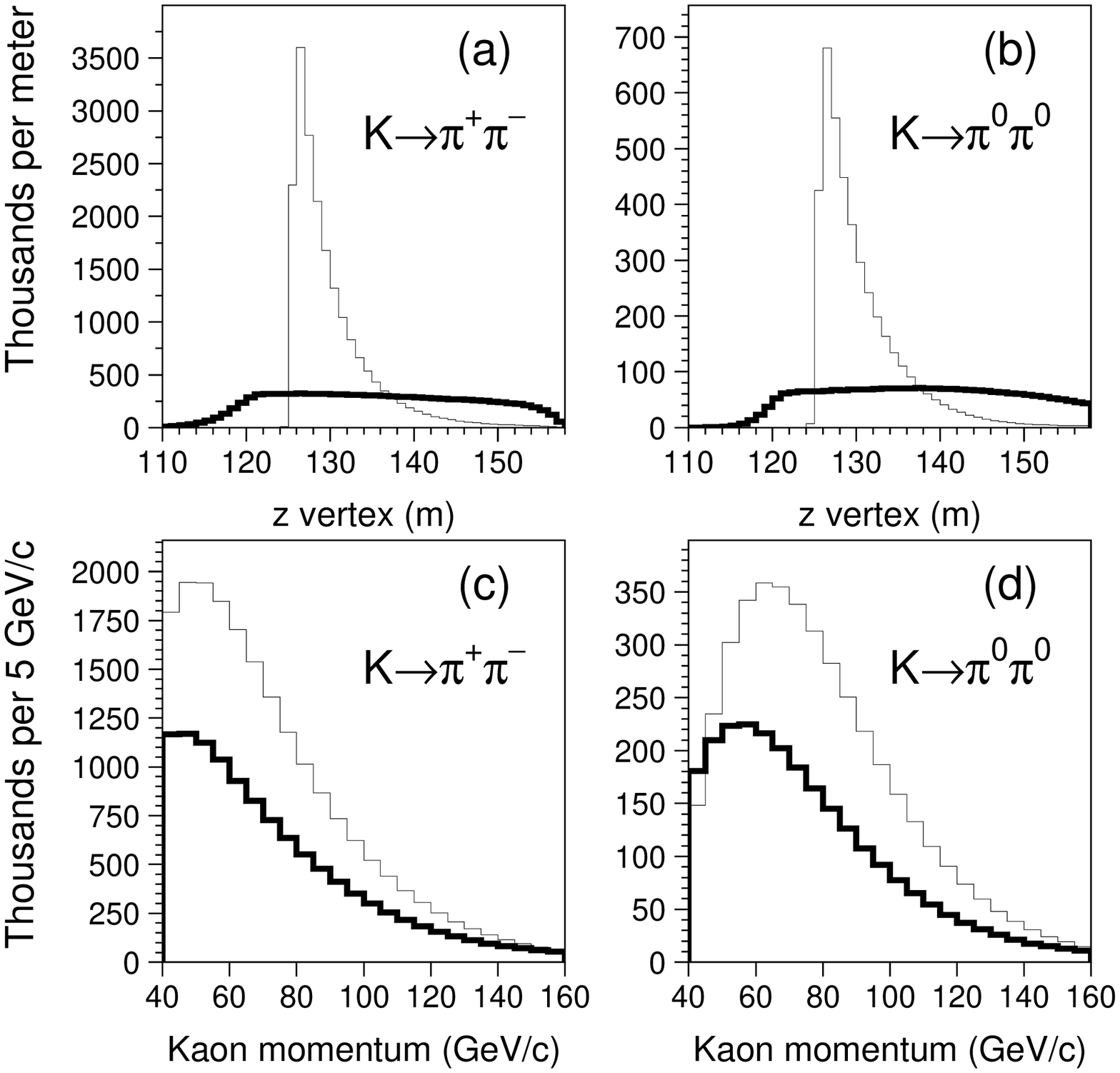,width=\linewidth}
  \caption{
      (a) $z$-vertex distribution for reconstructed \Kpm\ decays
      for the vacuum beam (thick) and regenerator beam (thin histogram).
      (b) $z$-vertex distributions for reconstructed \Kzz\ decays.
      (c) Kaon momentum distributions for reconstructed \Kpm\ decays.
      (d) Kaon momentum distributions for reconstructed \Kzz\ decays.
      All $\Kpp$ analysis cuts have been applied
      and background has been subtracted.
            }
  \label{fig:pzdata}
\end{figure}

The numbers of events after all event selection requirements 
and background subtraction are given in Table~\ref{ta:yield}.  
The measurement of $\reepoe$ is statistically limited by 
the 3.3 million $\KL\to\ppn$ decays.
Figure~\ref{fig:pzdata} shows the $\ZK$-vertex and kaon momentum
distributions for the four event samples.

\begin{table}[hbt]
\centering
\caption{
     \label{ta:yield}
     \Kpp\ event totals after all analysis requirements
     and background subtraction.
         }
\begin{ruledtabular}
\begin{tabular}{l|c|c } 
          & Vacuum Beam     & Regenerator Beam   \\  \hline
  $\Kpm$ & $11,126,243$     &  $19,290,609$  \\
  $\Kzz$ & ~$3,347,729$     &  ~$5,555,789$  \\
\end{tabular}
\end{ruledtabular}
\end{table}

 \section{Extracting Physics Parameters}
 \label{sec:extract}

To measure physics parameters with the event samples described in the
previous section, we correct for detector acceptance and perform a fit
to the data. 
The acceptance correction and the associated systematic 
error are described in 
Sections~\ref{sec:acceptance}-\ref{sec:acc_syst}. 
The fitting program used to
extract physics results from the event samples 
is described in Section~\ref{sec:fitting}. 
Systematic uncertainties associated with fitting
are discussed in Section~\ref{sec:fit_syst}.

  \subsection{The Acceptance Correction} 
  \label{sec:acceptance}

A Monte Carlo simulation is used to determine the acceptance,
which is the fraction of $\Kpp$ decays that satisfy the 
reconstruction and event selection criteria. 
The very different $K_L$ and $K_S$ lifetimes cause a difference 
between the average acceptance for decays in the two beams.
Correcting for the acceptance difference in the momentum and
$z$-vertex range used in this analysis,
the measured value of $\reepoe$ shifts by $\sim 85\eu$~.
About 85\% of this correction is the result of detector geometry, 
which is 
known precisely from optical survey and measurements with data.
The remaining part of the acceptance correction
depends on detailed detector response and resolution in the
simulation. 
Including accidental activity in the simulation
results in corrections to $\reepoe$ that are about
$-0.9\eu$ and $-0.5\eu$ 
for the charged and neutral decay modes, respectively.
As will be discussed in Section~\ref{sec:acc_syst}, 
comparisons of data and MC $z$-vertex distributions
allow us to estimate the systematic uncertainty associated 
with the acceptance correction.

\begin{figure}
  \epsfig{file=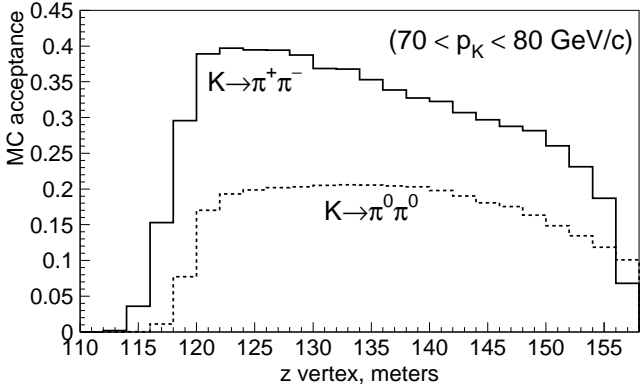, width=\linewidth}
  \caption{
     MC acceptance vs. $\ZK$-vertex (2~meter bins) for \Kpp\ decays with 
     momentum range 70-$80~\upk$.  The solid (dotted)
     histogram refers to the charged (neutral) decay mode
     as indicated on the plot.
         }
  \label{fig:mczacc}
\end{figure}

For a given range of kaon $\PK$ and $\ZK$, 
the acceptance is defined as 
\begin{equation}
  A_{p,z}=N_{p,z}^{rec}/N_{p,z}^{gen}, 
  \label{eq:acceptance}
\end{equation}
where $N^{rec}_{p,z}$ ($N^{gen}_{p,z}$) 
is the number of reconstructed (generated) Monte Carlo
events in the specified $\PK$, $\ZK$ range.
The generated $p$ and $z$ ranges are slightly larger than
the ranges used in the analysis to account for the effects
of resolution.
Fig.~\ref{fig:mczacc} shows the acceptance as a function of
$\ZK$  for $70 < \PK < 80~\upk$. 
The \Kpp\  MC samples used to calculate the detector acceptance 
correspond to 4.7 times the \Kpm\ data sample
and 10.4 times the \Kzz\  data sample.
The resulting statistical uncertainties on \reepoe\ from the acceptance
correction are
$\KtevMCStatChrg \eu$ and $\KtevMCStatNeut \eu$ for the charged 
and neutral decay modes, respectively.

As will be described in Section~\ref{sec:fitting}, 
we extract \reepoe\ using twelve  $10~\upk$
bins in kaon momentum and a single, integrated $\ZK$ bin. 
This $p$ binning reduces  sensitivity to the
momentum dependence of the detector acceptance
and to our understanding of the kaon momentum spectrum.
The use of $p$ bins also allows us to
account for the momentum dependence of the regeneration amplitude.

  \subsection{Systematic Uncertainty From The Acceptance Correction} 
  \label{sec:acc_syst}

We evaluate the quality of the 
simulation by comparing the data and 
Monte Carlo $\ZK$-vertex distributions in 
the vacuum beam, where the generated $\ZK$ distribution depends only on the 
well known $K_L$ lifetime. Imperfections in the understanding
of detector size and efficiency would change the number of
reconstructed events in a non-uniform way along the decay region,
and would result in a data-MC difference in the $\ZK$-vertex distribution.

The procedure for converting the data-MC vertex comparison 
into a systematic uncertainty on $\reepoe$ is as follows.
Since \reepoe\ is measured in $10~\upk$ kaon momentum bins, 
we weight the number of MC events in each energy bin so that
the data and MC kaon momentum distributions agree.  
We then compare the data and the weighted MC $\ZK$ distributions,
and fit a line to the data/MC ratio as a function of $\ZK$.
The slope of this line, $s$, is called an acceptance ``$z$-slope.''
To a good approximation, a $z$-slope affects the measured 
value of \reepoe\ as $s\Delta z / 6$, 
where $\Delta z$ is the difference of the mean
$z$ values for the vacuum and regenerator beam vertex distributions, 
and the factor 6 arises from converting a bias on the 
vacuum-to-regenerator ratio to a bias on $\reepoe$. 
$\Delta z = 5.6$~m and $\Delta z = 7.2$~m in the charged and
neutral $\pi\pi$ modes, respectively. 
Equation~\ref{eq:error} is used to convert the
bias on \reepoe\ to a systematic  uncertainty.
The uncertainty in $\tau_L$~\cite{pdg00}, which affects the
MC $z$-vertex distribution, 
contributes a negligible uncertainty of $0.034\eu~{\rm m}^{-1}$
to the $z$-slope.

\begin{figure}
  \epsfig{file=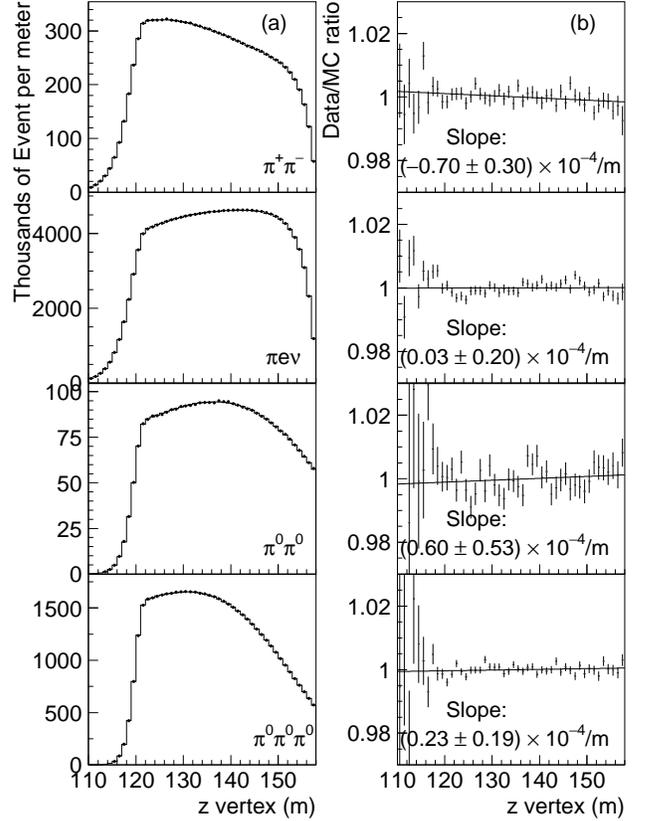, width=\linewidth}
  \caption{
     Comparison of the vacuum beam $z$ distributions for data (dots) 
     and MC (histogram). The data-to-MC ratios on the right
     are fit to a line, and the $z$-slopes (see text) are shown.
     The neutral distributions are for the combined 1996+1997 samples;
     the charged distributions are for 1997 only.
         }
  \label{fig:vtxz_slopes}
\end{figure}

Figure~\ref{fig:vtxz_slopes} shows the data-MC $z$-vertex 
comparisons for the charged and neutral $\pi\pi$ decay modes,
and for the high statistics $\KLpienu$ and \KLzzz\ modes.
The $z$-slope in \KLpm\ is \ChrgZSlope,
and leads to a systematic uncertainty of
$\ChrgZSlopeSyst \eu$ in $\reepoe$.
This charged $z$-slope has a significance of $2.3\sigma$, 
and is mostly from the first 20\% of the data sample 
(i.e., data collected at the start of the 1997 run).
The very small \Kethree\ $z$-slope is shown as a crosscheck,
but is not used to set the systematic error because of the 
different particle types in the final state.
To assign a systematic uncertainty for the neutral decay mode 
acceptance, we use $\KLzzz$ decays.
This decay mode has the same particle type in the 
final state as $\Kzz$ decays, 
and the \KLzzz\ reconstruction is more sensitive
to the effects of nearby clusters, 
energy leakage at the calorimeter edges, 
and low photon energies.
Using a sample of 50 million reconstructed $\KLzzz$ decays
in both data and MC, the $z$-slope is 
\ZSlopeKzzz\
leading to a neutral mode acceptance uncertainty of 
$\NeutZSlopeSyst \eu$ on $\reepoe$.
The \Kzz\ $z$-slope  of  \ZSlopeKzz\
is consistent with the $z$-slope in \KLzzz\ decays.
The uncertainty in
the $z$-dependence of the acceptance for each decay mode 
is included in the summary of systematic uncertainties
shown in Table~\ref{tb:syst_reepoe}.

 \subsection{Fitting Decay Distributions}
 \label{sec:fitting}

For pure $K_L$ and $K_S$ beams, the event yields (Table~\ref{ta:yield})
and acceptance for each mode (Fig.~\ref{fig:mczacc}) 
would be sufficient to determine $\reepoe$ from the 
acceptance-corrected double ratio.
The regenerator, however, produces a coherent $K_S$-$K_L$ mixture, 
so that a simple double ratio underestimates
$\reepoe$ by a ${\rm few}\eu$. 
A proper treatment of the regenerator, as well as \tgtKS\
in both beams, is included in a fitting program.
In addition to extracting $\reepoe$, this fitting program is also 
used to make measurements of the kaon parameters 
$\delm$, $\tauS$, $\phipm$, and $\delphi$;
the fitting procedure described below applies to both $\reepoe$
and these kaon parameters.
Each fit has different conditions related to the $\ZK$-binning
and CPT assumptions, 
which are summarized in Table~\ref{tb:fits}.

The fitting procedure is to minimize the $\chi^2$ between 
background-subtracted data yields and a prediction function,
and uses the MINUIT~\cite{minuit} program.
The prediction function (${\cal P}$) for each beam 
and decay mode is
\begin{equation}
   {\cal P}_{p,z} = \npredict \times A_{p,z}~,
   \label{eq:predict fun}
\end{equation}
where $\npredict$ is the calculated number of $\Kpp$
decays in the specified $p,z$ range,
and $A_{p,z}$ is the detector acceptance determined by a Monte Carlo
simulation  (Eq.~\ref{eq:acceptance}).
Note that $\npredict$ includes full propagation of the kaon state 
from the target up to the decay point,
as in the MC simulation (Sec.~\ref{sec:mc_kaonprop}).

For all fits, the prediction function is computed in 
$1~\upk$ $p$-bins and 2~meter $z$-bins.
To evaluate the $\chi^2$ in the $\reepoe$ fit,
the event yields and prediction function
are integrated in $10~\upk$-wide $p$-bins,
and each $p$-bin is integrated over the the 
full $z$-range from 110~m to 158~m.
For the  other kaon parameter fits, 
the event yields and prediction function are 
integrated over $10~\upk \times 2~{\rm m}$  $p$-$z$ bins.

To simplify the discussion that follows, the \tgtKS\
component is ignored.
For a pure $K_L$ beam, the number of $\Kpp$ decays is 
\begin{equation}
    \npredict  \propto
     {\cal F}(p)  
          \left| \eta    \right|^2 e^{-t/\tauL},
  \label{eq:dnvac_dt}
\end{equation}
where $t={m_K}(z-z_{\rm reg})/p$ is the measured proper time 
relative to decays at the regenerator edge,
$\eta = \eta_{+-}~(\eta_{00})$ for charged (neutral) decays,
${\cal F}(p)$ is the kaon flux,
and $\tauL$ is the $K_L$ lifetime.
The vacuum beam decay distribution is determined by $\tauL$;
the total event yield is proportional to $|\eta|^2$
and the kaon flux.

\begin{figure}
  \centering
  \epsfig{file=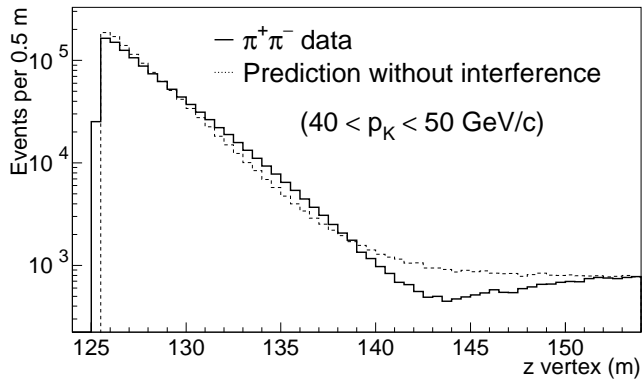,width=\linewidth}
  \caption{
       $\ZK$ decay distribution of $\Kpm$ decays in the regenerator beam,
       for the restricted momentum range 40-$50~\upk$.
       The MC prediction (dashed) is without the interference term
       that is proportional to ``$2\magrho|\eta|$'' in 
       Eq.~\ref{eq:dnreg_dt}.
         }
  \label{fig:decay_dnstream_reg}
\end{figure}

For a pure $K_L$ beam incident on the \ktev\ regenerator,
the number of decays downstream of the regenerator is
\begin{eqnarray}
  \npredict & \propto & 
     {\cal F}_{R}(p)  \treg(p)
    \left[
          \left| \rho(p) \right|^2 e^{-t/\tauS} + 
          \left| \eta    \right|^2 e^{-t/\tauL} +   
          \right.     \nonumber \\ 
     &  &  \left.
             2 | \rho | | \eta |
             \cos( \delm t + \phi_{\rho} - \phi_{\eta} )
                      e^{ -t/\tau_{avg} }
    \right] ~,
  \label{eq:dnreg_dt}
\end{eqnarray}
where $\phi_{\eta} = \arg(\eta)$,
$\magrho$ and $\phi_{\rho}$ are the magnitude and phase of the
coherent regeneration amplitude
\footnote{
Approximate expressions for the regeneration amplitude
resulting from fits to \ktev\ data are
$\magrho  \simeq   0.03 ({\PK}/70)^{-0.54}$ 
and
$\phi_{\rho} \simeq  [-34 + 12( {\PK}/140 -1 )^2]\degs$.
},
$1/\tau_{avg} \equiv (1/\tauS + 1/\tauL)/2$,
${\cal F}_{R}(p)$ is the kaon flux upstream of the regenerator,
and $\treg(p)$ is the kaon flux transmission through the regenerator.
The prediction function accounts for decays inside the
regenerator by using the effective regenerator edge
(Fig.~\ref{fig:regdiagram}b) as the start of the decay region.
All three terms in Eq.~\ref{eq:dnreg_dt} are important,
as illustrated in Fig.~\ref{fig:decay_dnstream_reg},
which shows interference effects in the regenerator-beam 
$\ZK$-vertex distribution.

\begin{table}
\caption{
       \label{tb:fits}
     Fit conditions used to analyze $\Kpp$ data.
     ``$z$-binning'' refers to using 2~m $z$-bins in
     the regenerator beam. 
     ``Assume CPT''  means that Eq.~\ref{eq:sw} is a fit constraint.
     Free parameters common to all fits, but not shown in the table,
     include the regeneration parameters $\fminus$ and $\alpha$,
     and the kaon flux in each $10~\upk$ momentum bin,
     ${\cal F}(p_1)-{\cal F}(p_{12})$.
  }
\begin{ruledtabular}
\begin{tabular}{ l || l | l | l }
             &  \multicolumn{3}{c}
                {\underline{Fit Conditions} }    \\
             &  \multicolumn{3}{c}{}                   \\
Fit          &               & Assume   &  Free          \\
Type         &  $z$ binning   &   CPT    &  Parameters    \\
\hline\hline   
$\reepoe$    &  No  &  Yes &   $\reepoe$           \\      \hline
\delm,\tauS  &  Yes &  Yes &  \delm,~\tauS         \\      \hline
\phipm\      &  Yes &  No  &  \delm,~\tauS,~\phipm  \\      \hline
\delphi\     &  Yes &  No  &  \delm,~\tauS,~\phipm   \\
             &      &      &  \reepoe,~\imepoe       \\
\end{tabular}
\end{ruledtabular}
\end{table}

Next, we discuss how the various factors in 
Eqs.~\ref{eq:dnvac_dt}-\ref{eq:dnreg_dt} are treated
in the fits.
The average vacuum-to-regenerator kaon flux ratio (${\cal F}/{\cal F}_R$) 
and the average regenerator transmission ($\treg$) 
cancel in the $\reepoe$ fit as explained in 
Appendix~\ref{app:exp details}.
To account for the momentum dependence of the regeneration
amplitude in the fits (see below),
we need to know the momentum-dependence of 
${\cal F}/({\cal F}_R\treg)$;
it is measured from the vacuum-to-regenerator ratio of $\KLpmz$ decays,
and is found to vary linearly by 
$(+7.0 \pm 0.7)$\% between 40 and $160~\upk$.
This variation in ${\cal F}/({\cal F}_R\treg)$ is mostly
from the momentum dependence of the regenerator transmission,
and to a lesser extent from the movable absorber transmission.

The $K_L$ lifetime is taken from \cite{pdg00}.
The values of $\delm$ and $\tauS$ are fixed to 
our measurements (Sec.~\ref{sec:taus_delm}) for the $\reepoe$ fit,
and are floated in the fits for the other kaon parameters.
In the fits that assume CPT symmetry,
$\phipm$ and $\phizz$ are set equal to the superweak phase:
\begin{equation}
  \phi_{\eta} = \phisw = \tan^{-1}(2\delm/\delg).  
  \label{eq:sw}
\end{equation}

The final component of the prediction function is the
regeneration amplitude. 
We use a model that relates $\rho$ to the difference between 
the forward kaon-nucleon 
scattering amplitudes for $\Kz$ and $\Kzbar$:
\begin{equation}
  \rho \propto f_{-} \equiv \hbar \frac{ f(0) - \bar{f}(0) }{p},
\end{equation}
where $f(0)$ and $\bar{f}(0)$ are the forward scattering
amplitudes for \Kz\ and \Kzbar, respectively, 
and $p$ is the  kaon momentum.
Additional factors that contribute to $\rho$ are
described in \cite{prd:731}.

For an isoscalar target and high kaon momentum,
$f_-$ can be approximated by a single Regge trajectory
corresponding to the $\omega$ meson. In that case,
Regge theory \cite{pr:gilman} predicts that the
magnitude of $f_-$ should vary with kaon momentum as a  power law:
\begin{equation}
  \left| f_{-}(p)\right| = \fminus 
     \left(
         \frac{ p }{ 70~\upk} 
     \right)^{\alpha}.
  \label{eq:rho_pk}
\end{equation}
The complex phase of $f_{-}$ can be determined from its momentum
dependence through an integral dispersion relation, with the
requirement that the forward scattering amplitudes be analytic
functions. 
This ``analyticity'' requirement  yields a constant phase
for a power-law momentum dependence:
\begin{equation}
  \phi_{f_{-}}  = - \frac{\pi}{2} ( 2 + \alpha). 
     \label{eq:analyticity} 
\end{equation}

In practice, the kaon-nucleon interactions in carbon are screened
due to rescattering processes. 
The effects of screening modify the momentum dependence of 
$|f_{-}(p)|$ as well as its phase. 
Screening corrections are evaluated using Glauber theory 
formalism~\cite{glauber55,glauber66} 
for diffractive scattering,
and using various models~\cite{regmodels} for inelastic scattering.
The screening corrections in the prediction function
used in the $\reepoe$ fit result in a $10$\% correction to $\alpha$,
and a $0.36\eu$ shift in $\reepoe$.

   \subsection{Systematic Uncertainties from Fitting}
   \label{sec:fit_syst}

Uncertainties from the fitting procedure
are summarized in Table~\ref{tb:syst_epefit} and discussed below.
These uncertainties  are mainly related to regenerator properties, 
and contribute $\FitSyst\eu$ to the $\reepoe$ uncertainty.

The uncertainty on the momentum dependence of the 
regenerator transmission 
corresponds to a $\AttSyst \eu$ uncertainty on $\reepoe$. 
The sensitivity to \tgtKS\ is checked by floating
the $\Kz/\Kzbar$ flux ratio (Sec.~\ref{sec:mc_kaonprop})
in the fit;
this changes the \tgtKS\ component by $(2.5\pm 1.6)$\% of itself, 
and leads to a systematic uncertainty of $0.12\eu$ on $\reepoe$.

The dependence of $\reepoe$ on \delm\ and \tauS\ is
\begin{eqnarray} 
  \Delta \reepoe &= & \left( +\ReepoeDelmSyst \eu \right) \times
                      \frac{ \delm - \KtevDelm}{\KtevDelmTerr}
                      \nonumber \\ 
          & & + \left( -\ReepoeTausSyst \eu \right) \times \frac{
                      \tauS - \KtevTaus}{\KtevTausTerr},  
     \label{eq:reepoe_dmts}
\end{eqnarray}
where \delm\ and \tauS\ are in units of 
$10^{6}~\hbar {\rm s}^{-1}$ and 
$10^{-12}~{\rm s}$, 
respectively. 
Each numerator in Eq.~\ref{eq:reepoe_dmts} is the difference
between the true value and the \ktev\ measurement 
(Sec.~\ref{sec:taus_delm});
the denominators are the total \ktev\ uncertainties
on $\delm$ and $\tauS$.
Since the \ktev\ measurements of \delm\ and \tauS\ are anti-correlated, 
the  systematic uncertainty on \reepoe\ due to 
variations in these parameters is $\ReepoeDelmTausSyst \eu$.

There are also uncertainties in $\reepoe$ associated with
the analyticity relation and screening correction used to 
predict the regeneration phase $\phi_{\rho}$.
It has been argued that the analyticity assumption is good to 
$0.35^{\circ}$ in the E773 experiment\cite{analphi},
which included kaon momenta down to $30~\upk$.
A smaller deviation from analyticity is expected with the $40~\upk$
minimum momentum cut used in this analysis; this leads to a
$0.25^{\circ}$ uncertainty in $\phi_{\rho}$ and a $0.07\eu$
uncertainty in $\reepoe$.
Using different screening models in the fit leads to a 
$\reepoe$ uncertainty of $0.15\eu$.

The fitting program uses the same $K_S/K_L$ flux ratio for
the charged and neutral decay modes.  
Since the 1996 $K\to\pi^+\pi^-$ sample is excluded,
we consider the possibility of
a change in the kaon flux ratio between 1996 and 1997. 
The kaon flux ratio depends only on the physical properties
of the movable absorber and regenerator.
The density of these two elements could change between the two
years because of a possible few degree temperature difference,
leading to a systematic uncertainty of $0.05\eu$ on $\reepoe$.

\begin{table}
  \centering
  \caption{
     \label{tb:syst_epefit}
      Systematic uncertainties in $\reepoe$ from fitting.
         }
  \medskip
\begin{ruledtabular}
\begin{tabular}{lc} 
                             &  $\reepoe$ Uncertainty  \\
  Source of Uncertainty      &     ($\eu$)          \\  \hline
  Regenerator transmission        & $0.19$  \\
  Target-\KS\                     & $0.12$  \\
  $\delm$ and $\tauS$             & $\ReepoeDelmTausSyst$  \\
  Regenerator screening                 & $0.15$  \\
  $\phi_{\rho}$ (analyticity)           & $0.07$  \\
  1996 vs. 1997 $K_S/K_L$ flux ratio    & $0.05$  \\
  $\tauL$~\cite{pdg00}                  & $0.02$  \\
 \hline 
  Total                           & $\FitSyst$  \\
\end{tabular} 
\end{ruledtabular}
\end{table}

\bigskip

Although fitting uncertainties in the \reepoe\ measurement
are a small part of the total uncertainty,
they are more significant in the measurements of
$\delm$, $\tauS$, and $\phipm$ 
(Table~\ref{tb:syst_kparfit}).
Uncertainties in the regenerator transmission and
in the analyticity assumption contribute the largest uncertainties
in the measurements of $\delm$ and $\tauS$.
The uncertainty in the regenerator screening model contributes
an uncertainty of $0.75\degs$ in the $\phipm$ measurement.
Fitting uncertainties have a negligible effect on the measurement
of \delphi\  because of cancellations between the charged and 
neutral decay modes.

\begin{table}
  \centering
  \caption{
    \label{tb:syst_kparfit}
     Fitting uncertainties in
     \delm, \tauS, \phipm, and \delphi.
     ``$\sigma_{syst}$'' refers to  systematic uncertainty.
         }
\begin{ruledtabular}
\begin{tabular}{lcccc} 
           & \multicolumn{4}{c}{$\bf \sigma_{syst}$ for:} \\
Source of    & $\delm$        & $\tauS$        & ~~$\phipm$~~& ~~$\delphi$ ~~\\
Uncertainty  & $(\times 10^6\hbar/s)$ 
                              & $(\tausunits)$ & $(\degs)$ & $(\degs)$ \\
\hline 
Regen. transmission         & $10.0$  & $0.020$ & $0.07$ & $0.01$    \\
Target-\KS\                 &  $1.4$  & $0.017$ & $0.13$ & $0.01$    \\
Regen.  screening           &  $3.0$  & $0.020$ & $0.75$ & $0.03$    \\
$\phi_{\rho}$ (analyticity) &  $8.1$  & $0.030$ & $0.25$ & $0.00$    \\
$\tauL$                     &  $0.0$  & $0.001$ & $0.00$ & $0.00$    \\
 \hline 
Total                       & $13.3$ & $0.045$ & $0.80$ & $0.03$ \\
\end{tabular}
\end{ruledtabular}
\end{table}

 \section{Measurement of $R\lowercase{e}(\epe)$ }
 \label{sec:reepoe_measure}

The \ktev\ measurement of $\reepoe$ uses
the background-subtracted \Kpm\ and \Kzz\ samples in the
vacuum and regenerator beams,
the prediction for the \Kpp\ acceptances using the Monte Carlo,
and the fitting program.
Section~\ref{sec:reepoe_results} presents the 
$\reepoe$ result and a summary of the systematic uncertainties.
Section~\ref{sec:crosschecks}
presents several crosschecks,
including a ``reweighting'' technique which does not
use a Monte Carlo acceptance correction.

  \subsection{The $\reepoe$ Result}
  \label{sec:reepoe_results}

There are 48 measured quantities that enter into the $\reepoe$ fit:
the observed numbers of \Kpm\ and \Kzz\ decays
in the vacuum and regenerator beams,
each in twelve $10~\upk$ wide $p$ bins.
Within each momentum bin we use the $\ZK$-integrated yield
from 110~m to 158~m.
There are 27 fit parameters including 24 kaon fluxes,
two regeneration parameters, and $\reepoe$. Therefore, 
the number of degrees of freedom in the fit is $48-27 = 21$.
CPT symmetry is assumed (Eq.~\ref{eq:sw}),
and the values of \delm\ and \tauS\ are from
our measurements described in Sec.~\ref{sec:taus_delm}.

For the combined 1996 and 1997 datasets, we obtain
\begin{eqnarray}
\begin{array}{ccc}
  \reepoe & = & \left( \KtevXReepoe \pm \KtevStat \right) \eu
              \\
   \fminus & = & \KtevAmp70andErr
              \\
   \alpha &  = & \KtevPwrSlpandErr
          \label{eq:reepoe fit results}   \\
   \chi^2/dof & = & 27.6/21 ~,
\end{array}
\end{eqnarray}
where the errors reflect the statistical uncertainties.
Including the systematic uncertainty,
\begin{eqnarray}
  \reepoe & = & \left[ \KtevXReepoe \pm \KtevStat~({\rm stat}) \pm
  \TotSystMC~({\rm syst}) \right] \eu \nonumber \\
          & = & \left( \KtevReepoe \pm \KtevTErr \right)  \eu ~,
  \label{eq:reepoe_result}
\end{eqnarray}
where the contributions to the systematic uncertainty are summarized in
Table~\ref{tb:syst_reepoe}.  
The systematic uncertainties from the charged and neutral decay modes 
contribute $1.26\eu$ and $2.00\eu$, respectively.
The largest uncertainties are from the
CsI energy reconstruction ($1.47\eu$),
neutral mode background subtraction ($1.07\eu$),
$z$-dependence of the acceptance in the charged decay mode 
($\ChrgZSlopeSyst\eu$),
and the charged mode Level~3 filter ($\L3CHRGSYST\eu$).

\begin{table}
\caption{
     \label{tb:syst_reepoe}
      Summary of systematic uncertainties in $\reepoe$.
      Uncertainties from the charged and neutral decay modes are
      presented in more detail in
      Tables~\ref{tb:chrg_syst} and \ref{tb:neut_syst}.
         }
\begin{ruledtabular}
\begin{tabular}{ldd} 
		 & \multicolumn{2}{c}{$\reepoe$ Uncertainty ($\eu)$ } \\
                 & \multicolumn{2}{c}{ from: }         \\
   Source of uncertainty & \multicolumn{1}{c}{$\Kpm$} 
                         & \multicolumn{1}{c}{$\Kzz$}  \\
\hline 
  Trigger                      &  $0.58$   & $0.18$  \\
  CsI energy, position recon   &   -       & $1.47$  \\  
  Track reconstruction         &  $0.32$   &  -   \\
  Selection efficiency         &  $0.47$   & $0.37$  \\
  Apertures                    &  $0.30$   & $0.48$  \\
  Background                   &  $0.20$   & $1.07$  \\
  $\ZK$-dependence of acceptance &  $\ChrgZSlopeSyst$   
                                 &  $\NeutZSlopeSyst$  \\
  MC statistics                &  $\KtevMCStatChrg$   & $\KtevMCStatNeut$  \\
  Fitting                 & \multicolumn{2}{d}{$\FitSyst$} \\  \hline
   {TOTAL}                & \multicolumn{2}{d}{$\TotSystMC$} \\ 
\end{tabular}
\end{ruledtabular}
\end{table}

 \subsection{$\reepoe$ Crosschecks}
 \label{sec:crosschecks}

  \subsubsection{Consistency Among Data Subsets}
  \label{sec:subset_checks}

\begin{figure}[ht]
  \centering
  \epsfig{file=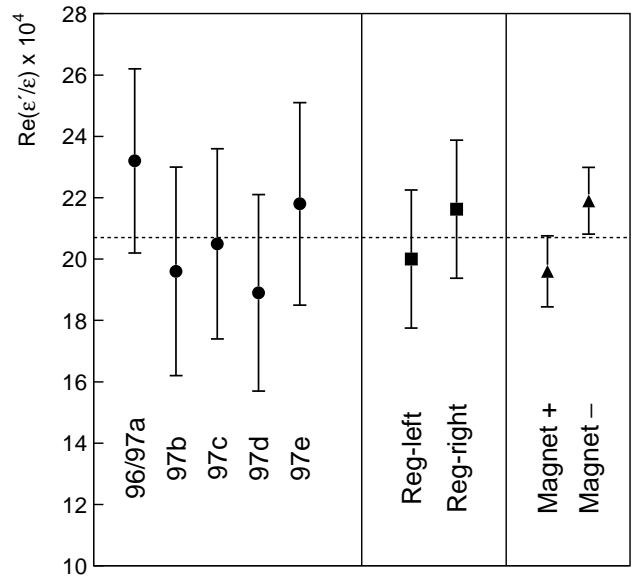,width=\linewidth}
  \caption{
        \reepoe\ consistency checks. 
        The first five data points
        correspond roughly to one-month time periods.
        Reg-Left and Reg-Right correspond to the two regenerator positions.
        Magnet$+$ and Magnet$-$ correspond to the two magnet polarities, 
        with the \Kzz\ sample common to both results.
        The uncertainties shown within each category are independent
        statistical errors.
             }
  \label{fig:crosscheck}
\end{figure}

We have performed several crosschecks of our result
by dividing the \Kpp\ samples into  subsets
and checking the consistency of $\reepoe$ and other parameters
among the different subsets.
Figure~\ref{fig:crosscheck} shows the $\reepoe$ result
in roughly month-long time periods, 
in each regenerator position,
and for the two magnet polarities. 
These comparisons all show good agreement.
The first data point labeled 96/97a corresponds to the 
current analysis applied to the sample used in our 
previous publication \cite{prl:pss}; 
this reanalysis is discussed in Appendix~\ref{sec:prl99}.   

\begin{figure}
  \centering
  \epsfig{file=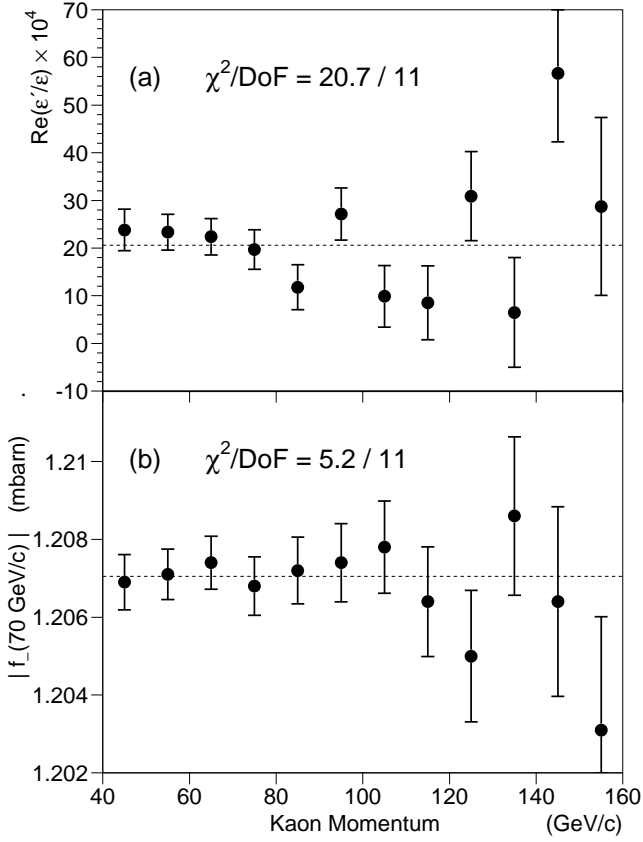,width=\linewidth}
  \caption{
       Results of twelve independent fits in $10~\upk$ wide
       $p$ bins for:
       (a) $\reepoe$ and (b) $\fminus$.
       Within each kaon momentum bin, $\reepoe$ and  $\fminus$
       are extracted from the same fit.
       Only statistical errors are shown.
       Each dashed line shows the average of all momentum bins.
          }
  \label{fig:reepoe_pk}
\end{figure}

To check the dependence on kaon momentum,
12 separate fits are done in $10~\upk$ momentum bins.
The free parameters in each fit are $\reepoe$
and $\fminus$, which is proportional to the regeneration 
amplitude at $70~\upk$.
The momentum dependence of the regeneration amplitude 
within each $10~\upk$ bin is described by the
power-law, $\rho \sim  p^{\alpha}$,
where $\alpha$ is fixed to the value found in the nominal fit 
(Eq.~\ref{eq:reepoe fit results}).
The $\chi^2$ per degree of freedom is 20.7/11 for \reepoe\ vs. $\PK$
(Fig.~\ref{fig:reepoe_pk}a)
and 5.2/11 for $\fminus$ vs. $\PK$ (Fig.~\ref{fig:reepoe_pk}b);
the combined $\chi^2/{\rm dof}$ is 25.9/22~.
The scatter of $\reepoe$ in the higher momentum
bins is not present in the kaon parameter measurements 
(Fig.~\ref{fig:kparvspk}),
and a linear fit to $\reepoe$ vs. $p$ has 
no significant slope ($<0.8\sigma$).
The 140-$150~\upk$ bin, 
which accounts for 6.3 of the $\chi^2$ in Fig.~\ref{fig:reepoe_pk}a,
also contributes 6.7 to the $\chi^2$ in the
nominal fit (Eq.~\ref{eq:reepoe fit results}),
with  nearly equal contributions from all four
\Kpp\ samples.

Another crosscheck is that $\alpha$, which describes
the regeneration power law (Eq.~\ref{eq:rho_pk}),
should be the same for both the charged and neutral decays.
A separate fit in each decay mode results in
\begin{eqnarray}
\begin{array}{ccc}
      \alpha_{+-} & = & -0.5421 \pm 0.0009~({\rm stat}) \\
      \alpha_{00} & = & -0.5445 \pm 0.0017~({\rm stat})
\end{array}
\end{eqnarray}
which agree to within $1.2\sigma$.

The \tgtKS\ correction is checked in a separate
$\reepoe$ fit that uses only those events with kaon momenta
below $100~\upk$ and a $\ZK$-vertex farther than 124~m
from the BeO target. 
This sample has a negligible $K_S$ component, and is therefore 
described by essentially pure $K_L$ beams entering the decay region.
Using this sub-sample, the change in $\reepoe$ from the 
nominal result (Eq.~\ref{eq:reepoe_result}) 
is $(+0.85\pm 0.89)\eu$, where the error
reflects the uncorrelated statistical uncertainty.

  \subsubsection{$\reepoe$ From Reweighting Technique}
  \label{sec:rewgt}

As a final crosscheck of our ``standard'' analysis, 
we also measure $\reepoe$
using a reweighting technique that does not depend on a Monte Carlo
acceptance correction.
This technique is similar to that used by the NA48 experiment 
\cite{na48:reepoe}.
The ``local acceptance'' of \Kpp\ decays in each $p$-$z$ bin 
($1~\upk \times 2~{\rm m}$)
is nearly identical in both beams, 
with the only difference arising from the effects of
accidental activity.
In this method, a weight is applied to
vacuum beam events such that the
regenerator and weighted vacuum beam events have the same 
statistical sampling of decay vertex and kaon energy.  
With the same local acceptance in the two beams as a function
of $p$ and $z$, 
an ideal weight function eliminates differences
in the reconstruction efficiencies and resolutions for \Kpp\
in the two beams.
The weight factor, which is applied event-by-event,
is the {\it a priori} ratio of the regenerator 
beam and vacuum beam decay rates, 
\begin{equation}
   W(p,z)=\frac{ {d\Gamma_{reg}/dt}(p,z)} { {d\Gamma_{vac}/dt}(p,z)}~.
   \label{eq:wgt factor}
\end{equation}
The functions $d\Gamma_{vac}/dt$ and $d\Gamma_{reg}/dt$
are similar to those given by Equations~\ref{eq:dnvac_dt} 
and~\ref{eq:dnreg_dt}, respectively,
with the modification that they are constrained to vanish upstream 
of the regenerator.

\begin{figure}
\epsfig{file=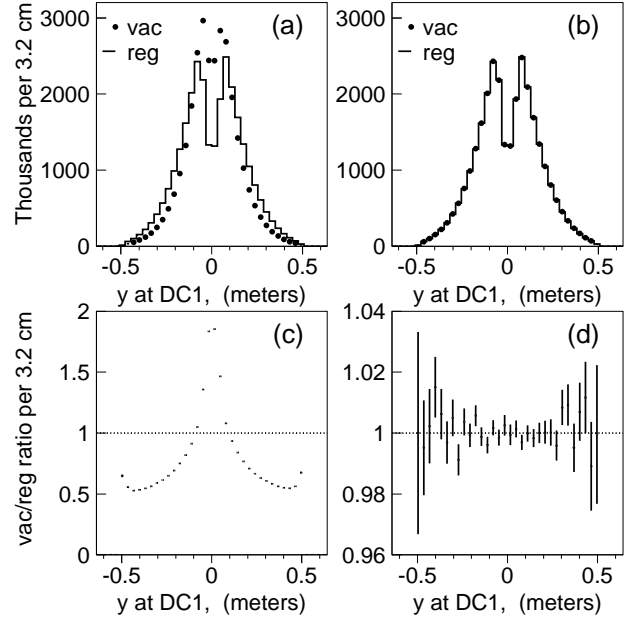,width=\linewidth}  
\caption{
     (a) $\Kpm$ track illumination in the DC1 $y$-view
     for vacuum (dots) and regenerator (histogram) beams.
     (b) The same track illumination after applying the 
     weight factor to the vacuum beam.
     The vacuum-to-regenerator beam ratios are shown
     for each case in (c) and (d); note the different
     vertical scales for the ratios.
           }      
\label{fig_rwt_illum}
\end{figure}

The $z$ distribution in each beam, without the weight factor, 
is shown in Fig.~\ref{fig:pzdata}a-b,
and the pion track $y$-illumination at the first drift chamber
is shown in Fig.~\ref{fig_rwt_illum}a. 
The differences between the vacuum and regenerator beam
distributions are due to the average acceptance difference 
coming from the different $K_L$ and $K_S$ lifetimes.
The effect of $W(p,z)$ is that the vacuum beam distributions
match those in the regenerator beam  (Fig.~\ref{fig_rwt_illum}b).
The main drawback to this reweighting method
is that the statistical uncertainty is increased by a factor of 1.7
compared to the standard analysis,
because of the loss of vacuum beam events 
through the reweighting function.
In addition, the reweighting technique is more sensitive to
the neutral energy reconstruction because
the weight factor (Eq.~\ref{eq:wgt factor}) depends
on kaon energy.

The event reconstruction and selection are very similar to that of the
standard analysis.  
In the $\Kzz$ mode, the event selection cuts are identical.  
The most significant difference in the
reweighting analysis is the energy scale correction,
where the absolute energy scale is corrected
as opposed to the relative data-MC scale.  
The energy scale correction is derived
from the difference between the $\gamma\gamma$ and
$\pi^+\pi^-$ $z$-vertex reconstructed in $\KLpmz$.

For the $\Kpm$ selection, the reweighting analysis differs
from the standard analysis by adding a center-of-energy 
``$\ring$'' cut (Eq.~\ref{eq:ring_def}).
This cut is the same for neutral and charged events, 
and eliminates the need to correct for the effect 
of kaon scattering in the movable absorber upstream of the 
regenerator (Fig.~\ref{fig:beamline}).

$\reepoe$ is extracted in a fit which compares the
background-subtracted yields to a prediction function.
The prediction function is the same as 
Eqs.~\ref{eq:dnvac_dt}-\ref{eq:dnreg_dt},
except that in the vacuum beam,
both the data and the prediction function 
include the weight factor.
As in the standard fit, there are 48 measured inputs,
which correspond to the numbers of \Kpp\ events in each $10~\upk$ wide
kaon momentum bin, for both decay modes and both regenerator
positions.
The free parameters in the fit include 24 kaon fluxes,
the regeneration amplitude and power-law slope (Eq.~\ref{eq:rho_pk}),
and $\reepoe$.
In the reweighting fit, the two regeneration parameters also appear 
in the vacuum beam fit functional via the reweighting function, 
where they are not varied. We
find that the value of $\reepoe$ is quite insensitive to the 
regeneration parameters used in the reweighting function.
As in the standard \reepoe\ fit, 
$\delm$ and $\tauS$ are fixed to the values from
the measurements in Sec.~\ref{sec:taus_delm}.
There are a total of 27 free parameters and 21 degrees of
freedom in the reweighting method fit.

The systematic errors for the reweighting analysis
are shown in Table~\ref{tab:rewt_syst}. 
In the neutral decay mode, the largest systematic uncertainty 
results from the sensitivity to the minimum
photon cluster energy. 
The other large source of systematic uncertainty
results from accidental activity.
The level of accidental activity in the detector is slightly
different for the two beams;
this affects the local acceptance differently in the vacuum
and regenerator beams and is not 
accounted for by the weight factor.
In the standard analysis, accidental effects are accounted for
in the Monte Carlo acceptance correction.

To determine the uncorrelated statistical uncertainty between the
standard and reweighting analyses, a large number of Monte Carlo 
samples are generated and fit with both methods;
the uncorrelated uncertainty results mainly from the effective
loss in statistics in the reweighting method. 
The uncorrelated systematic error is mainly from the 
uncertainty in the minimum cluster energy requirement
in the reweighting analysis; 
there are also contributions from the acceptance correction 
in the standard analysis and accidental effects in the
reweighting analysis.  

The reweighting and standard analyses were both applied to the part
of the 1997 data sample that was not used in the previous publication,
corresponding to roughly 3/4 of the total sample.
There is good agreement between the two analyses:
the difference between the reweighting and standard 
$\reepoe$ results is 
\begin{eqnarray}
  \Delta[\reepoe] 
     & = & [+1.5 \pm 2.1~({\rm stat}) \pm 3.3~({\rm syst})] \eu
       \nonumber    \\
     & = & (+1.5 \pm 3.9) \eu ~,
\end{eqnarray}
where the errors reflect the uncorrelated uncertainty between the
two methods. 

\begin{table}[bth]
\caption{
       \label{tab:rewt_syst} 
      Systematic Uncertainties in \reepoe\
      from the reweighting analysis.
      }
\begin{ruledtabular}
\begin{tabular}{ l|l| c} 
        & Source of           & \reepoe\ Uncertainty  \\
Sample  & Uncertainty         & ($\eu$)               \\ \hline
Neutral & Backgrounds         & 1.31 \\
        & Reconstruction      & 2.93 \\
        & Trigger             & 0.41 \\
        & Accidental Bias     & 1.46 \\ \hline
Charged & Backgrounds         & 0.16 \\
        & Reconstruction      & 0.87 \\
        & Trigger             & 1.09 \\
        & Accidental Bias     & 0.81 \\
        & Regenerator location& 0.20 \\ \hline
Common  & Ring Number cut     & 0.70 \\
        & Reweighting Parameters & 0.30\\ \hline
Total   &                     & 3.98 \\ 
\end{tabular}
\end{ruledtabular}
\end{table}

 \section{Measurements of Kaon Parameters}
 \label{sec:kaonpar}

The regenerator beam decay distribution
allows measurements of the kaon parameters
$\tauS$ and $\delm$,
and CPT tests based on measurements of \phipm\ and $\delphi$.
The main difference compared to the $\reepoe$ fit is that
we fit the shape of decay distribution
instead of the integrated yield.
All of the fits discussed below use $2$~meter wide $\ZK$-bins in 
the regenerator beam from 124~m to 158~m.
In the vacuum beam, one $\ZK$-bin from 110~m to 158~m
is used to determine the
kaon flux in each $10~\upk$ momentum bin.

A $\ZK$-binned fit increases the sensitivity 
to migrations in $\ZK$. 
To allow for such migrations near the regenerator edge,
we include an extra  ``$\ZK$-shift'' parameter which is
the shift in the effective regenerator edge
relative to the nominal value calculated in
Section~\ref{sec:regenerator}.
In all fits, the charged and neutral data are consistent 
with no $\ZK$-shift at the regenerator edge.

Systematic errors for the kaon parameter measurements 
are evaluated in a manner similar to the $\reepoe$ analysis,
and are summarized in Table~\ref{tb:syst_kpar}. 
The sensitivity to  $\ZK$ migration is most pronounced
in the $\delphi$ uncertainty related to CsI energy reconstruction.
The complicated $z$ dependence of the
regenerator scattering background
(Fig.~\ref{fig:zbkgneut})
also contributes significant uncertainties to results
obtained from $z$-binned fits.

\begin{table}
  \centering
  \caption{
    \label{tb:syst_kpar}
     Systematic uncertainties in \delm, \tauS, \phipm, and \delphi.
     ``$\sigma_{syst}$'' refers to  systematic uncertainty.
      The uncertainty associated with fitting is common to both
      decay modes.
      The ``combined total'' uncertainty is explained in the text.
         }
\begin{ruledtabular}
\begin{tabular}{lcccc} 
           & \multicolumn{4}{c}{$\bf \sigma_{syst}$ for:} \\
Source of    & $\delm$        & $\tauS$        & ~~$\phipm$~~& ~~$\delphi$~~ \\
Uncertainty  & $(\times 10^6\hbar/s)$ & $(\tausunits)$
                                               & $(\degs)$ &  $(\degs)$~ \\
\hline 
        \multicolumn{5}{c}{\bf $\Kpm$ Analysis:} \\
Trigger                   & $0.2$  & $0.004$ & $0.10$ & $0.02$  \\
Track reconstruction      & $0.6$  & $0.032$ & $0.02$ & $0.02$  \\
Selection efficiency      & $3.2$  & $0.011$ & $0.35$ & $0.06$  \\
Apertures                 & $2.8$  & $0.038$ & $0.76$ & $0.09$  \\
Background                & $0.8$  & $0.002$ & $0.01$ & $0.01$  \\
Acceptance                & $1.2$  & $0.026$ & $0.14$ & $0.06$  \\
MC statistics             & $2.6$  & $0.012$ & $0.28$ & $0.05$  \\
Fitting                   & $\KtevDelmFITerr$ & 
                            $\KtevTausFITerr$ & $0.80$ & $0.03$  \\
\hline
\Kpm\ Total               & $14.3$ & $0.074$ & $1.20$ & $0.14$ \\
\hline\hline 
        \multicolumn{5}{c}{\bf $\Kzz$ Analysis:} \\
Trigger                    & $0.4$  & $0.013$ &  $-$   & $0.03$ \\
CsI reconstruction         & $8.1$  & $0.094$ &  $-$   & $0.37$ \\
Selection efficiency       & $5.0$  & $0.035$ &  $-$   & $0.06$ \\
Apertures                  & $2.2$  & $0.040$ &  $-$   & $0.14$ \\
Background                 & $7.0$  & $0.030$ &  $-$   & $0.14$ \\
Acceptance                 & $2.0$  & $0.030$ &  $-$   & $0.05$ \\ 
MC statistics              & $3.3$  & $0.016$ &  $-$   & $0.06$ \\
Fitting                    & $\KtevDelmFITerr$ & 
                             $\KtevTausFITerr$ &  $-$   & $0.03$ \\
\hline
\Kzz\ Total                & $18.3$ & $0.126$ &  $-$    & $0.43$ \\
\hline \hline 
                           &         &          &        &           \\
Combined Total             &  $14.2$ &  $0.069$ &   $-$  &  $\DelPhiTOTerr$ \\
\end{tabular}
\end{ruledtabular}
\end{table}

  \subsection{Measurement of $\delm$ and $\tau_S$}
  \label{sec:taus_delm}

To measure $\delm$ and $\tauS$, we fit the charged and neutral
modes separately and then combine results according to the statistical
and uncorrelated systematic errors.
We assume CPT symmetry (Eq.~\ref{eq:sw}) by dynamically setting
the value of $\phi_{\eta}$ equal to the superweak phase 
using the floated values of $\Delta m$ and $\tauS$. 
The fit values of $\tau_S$ and $\delm$ are independent
of the value of $\reepoe$. 

For each charged and neutral mode fit, 
there are 216 measured input quantities.
The number of vacuum beam decays in each $10~\upk$ momentum bin
gives 12 inputs; the number of regenerator beam decays in
each $2~{\rm m} \times 10~\upk$ bin adds $17\times 12=204$ inputs.
The floated parameters include the kaon flux in each of 12 momentum bins,
the magnitude and phase of the regeneration amplitude,
a $\ZK$-shift parameter, $\delm$, and $\tauS$;
these 17 floated parameters lead to 199 degrees of freedom.
The results of separate fits to the charged and neutral mode 
data are shown in Table~\ref{tab:delm_taus}. 
The difference between the charged and neutral mode results,
after accounting for the common systematic uncertainty
described below,
is $1.6\sigma$ for $\delm$ and $0.1\sigma$ for $\tauS$.

\begin{table}
  \centering
  \caption{
      \label{tab:delm_taus}
       \delm\ and \tauS\ results for the regenerator beam 
       charged and neutral data samples.
       The first uncertainty is statistical; the second is systematic.
          }
\begin{ruledtabular}
\begin{tabular}{ l|c|c|c}
 decay   &  \delm\                  & \tauS\          &                  \\
 mode    & $(\delmunits)$           & $(\tausunits)$  &  $\chi^2/$dof \\ 
 \hline  
   \ppc\ & $5266.7\pm ~5.9\pm 14.3$ & $89.650\pm 0.028\pm 0.074$ & 228/199 \\ 
   \ppn\ & $5237.3\pm 10.6\pm 18.3$ & $89.637\pm 0.050\pm 0.126$ & 195/199 \\
\end{tabular}
\end{ruledtabular}
\end{table}

Systematic errors arising from data analysis are larger 
in the neutral decay mode than in the charged mode, 
primarily because of larger background and uncertainties
in the CsI energy reconstruction.
The systematic uncertainties due to regeneration properties 
(screening, attenuation, and analyticity)
are more significant than in the $\reepoe$ analysis
because there is no cancellation between 
charged and neutral mode data. 
These uncertainties in the regeneration properties
are common to the charged and neutral mode fits,
and are applied to the final result after averaging.
The common systematic uncertainty is 
$\KtevDelmFITerr \delmunits$ on $\delm$, 
and $\KtevTausFITerr \tausunits$ on \tauS.

We combine the charged and neutral mode results weighted by
the statistical uncertainty and the independent parts of the 
systematic uncertainty. 
The results are 
\begin{eqnarray}
  \delm & = & ( \KtevDelm  \pm \KtevDelmTerr ) \delmunits~,     \\ 
  \tauS & = & ( \KtevTaus  \pm \KtevTausTerr ) \tausunits ~,
\end{eqnarray}
which correspond to a superweak phase of
\begin{equation}
   \phisw = (\PhiSW \pm \PhiSWErr )\degs. 
\end{equation}

  \subsection{Measurement of $\phipm$ and $\phipm-\phisw$}

The fit for \phipm\ is similar to the $\delm$-$\tauS$ fit. 
The main difference is that we remove the 
CPT assumption (Eq.~\ref{eq:sw})
and float $\phi_{+-}$ in addition to $\delm$ and $\tauS$. 
The fit is performed to \Kpm\ data only.
Compared to the $\delm$-$\tauS$ fit,
we have the same number of measured inputs (216) and
one additional free parameter ($\phipm$), 
for a total of $216-18=198$ degrees of freedom.

There is a large correlation among  
$\phipm$, $\delm$, and $\tauS$,
which is illustrated in Fig.~\ref{fig:phipmcorr}
(Appendix~\ref{app:kparcor}).
The  $\delm$-$\tauS$ correlation
is much stronger than in a fit using the CPT assumption,
and therefore results in a larger
statistical uncertainty on $\delm$ and $\tauS$.
The $\phipm$ statistical uncertainty in our fit is 2.4 times larger 
than in a fit with a fixed value of $\tauS$.

The $0.76\degs$ systematic uncertainty from apertures
(Table~\ref{tb:syst_kpar}) 
is mainly from the cell separation cut at the drift chambers
(Sec.~\ref{sec:chrg_syst}).
Different screening models result in a $0.75\degs$
uncertainty on $\phipm$ (Table~\ref{tb:syst_kparfit}).
The value of $\phipm$ depends on the regeneration phase $\phi_{\rho}$; 
a $0.25\degs$ uncertainty from the analyticity assumption
leads to a $0.25\degs$ error on $\phipm$.
The total systematic uncertainty on \phipm\ is 
$\PhipmAllSyst\degs$.

The results of the fit are:
\begin{equation}
 \begin{array}{lcl}
  \phipm & = & \left[ \PhipmAll \pm \PhipmAllErr~\mbox{(stat)} 
                                \pm \PhipmAllSyst~\mbox{(syst)}
               \right]\degs \\
         & = & (\PhipmAll \pm \PhipmAllTerr )\degs \\
  \delm  & =&  \left[\PhiAllDM \pm \PhiAllDMErr~\mbox{(stat)}\right] 
                       \delmunits \\ 
  \tauS & =&   \left[\PhiAllTs \pm \PhiAllTsErr~\mbox{(stat)}\right] 
                        \tausunits   \\  
  \chi^2/\nu & = & \PhiAllChi / \PhiAllDOF ~.
  \label{eq:phipm_fit}
 \end{array}
\end{equation}

Next, we fit the deviation from the superweak phase,
$\phipm-\phisw$, which is a direct test of CPT symmetry.
Compared to the $\phipm$ fit shown above, 
the fit for $\phipm - \phisw$ 
results in slightly reduced statistical
and systematic uncertainties
because the value of $\phisw$ is computed dynamically 
using the floated values of $\delm$ and $\tauS$ 
(Eq.~\ref{eq:sw}),
and is less sensitive to the correlations.
The result of this fit is
\begin{eqnarray}
\begin{array}{lcl}
   \phipm-\phisw & = & 
        \left[  \dPhiSW \pm \dPhiSWSTATerr~\mbox{(stat)} 
                     \pm \dPhiSWSYSTerr ~\mbox{(syst)} \right]\degs  \\
        & = & ( \dPhiSW\pm \dPhiSWTOTerr)\degs~,
\end{array}
\end{eqnarray}
and the $\chi^2$ is the same as for the $\phipm$ fit
(Eq.~\ref{eq:phipm_fit}).

  \subsection{Measurement of $\delphi$}

The measurement of  $\delphi$ 
is performed in a simultaneous fit to neutral and charged mode data. 
The number of measured inputs is 432, which is simply twice the number 
used in the $\delm$-$\tauS$ fits, since both charged and neutral 
modes are used in the same fit.
The floated parameters include the charged and neutral 
kaon fluxes in each of 12 momentum bins ($12+12=24$), 
the regeneration amplitude and phase,
one $\ZK$-shift term in charged and one in neutral,
the real and imaginary parts of $\epe$,
$\delm$, $\tauS$, and $\phipm$;
these 33 floated parameters lead to 399 degrees of freedom.
Note that the fit uses $\imepoe$ instead of \delphi\ as 
a free parameter (Eq.~\ref{eq:delphimpe}).

The fit for $\delphi$ benefits from the cancellation of uncertainties 
in the regenerator properties.
Also, there is little correlation with the other kaon 
parameters such as $\delm$, $\tauS$, and the phase of $\epsilon$.
Consequently, systematic uncertainties due to 
regenerator properties are small.  
The largest systematic uncertainty of  $0.37\degs$ is from the 
CsI cluster reconstruction in the neutral mode analysis.

There is  a  correlation between the real and imaginary
parts of $\epsilon'/\epsilon$, 
with a  correlation coefficient of $-0.565$.
As a result, the statistical uncertainty for $\reepoe$ is increased 
by about $20\%$ compared to the standard fit 
that sets $\imepoe=0$.

The results of the fit without CPT assumptions are
  \begin{eqnarray}
    \imepoe  & = & 
   \left[ \IMEPOE \pm \IMEPOEStatErr~\mbox{(stat)} 
                  \pm \IMEPOESystErr~\mbox{(syst)} \right] \eu
    \nonumber \\
            & = & \IMEPOEpmErr
    \nonumber \\
        \reepoe & = & [+22.5\pm1.9~(\mbox{stat})]\eu
    \nonumber \\
       \chi^{2}/\nu  &  = &     425 / 398 ~.
    \nonumber  
  \end{eqnarray}
In terms of $\delphi$, the result is
\begin{equation}
\begin{array}{lcl}
    \delphi & = & 
    \left[ \DelPhi \pm \DelPhiSTATerr~\mbox{(stat)} 
                   \pm \DelPhiSYSTerr~\mbox{(syst)} 
    \right]\degs \\
            & = & \left( \DelPhi \pm \DelPhiTOTerr \right)\degs. \\
\end{array}
\end{equation}

  \subsection{Kaon Parameter Crosschecks}
  \label{sec:kaonpar_checks}

\begin{figure}
  \epsfig{file=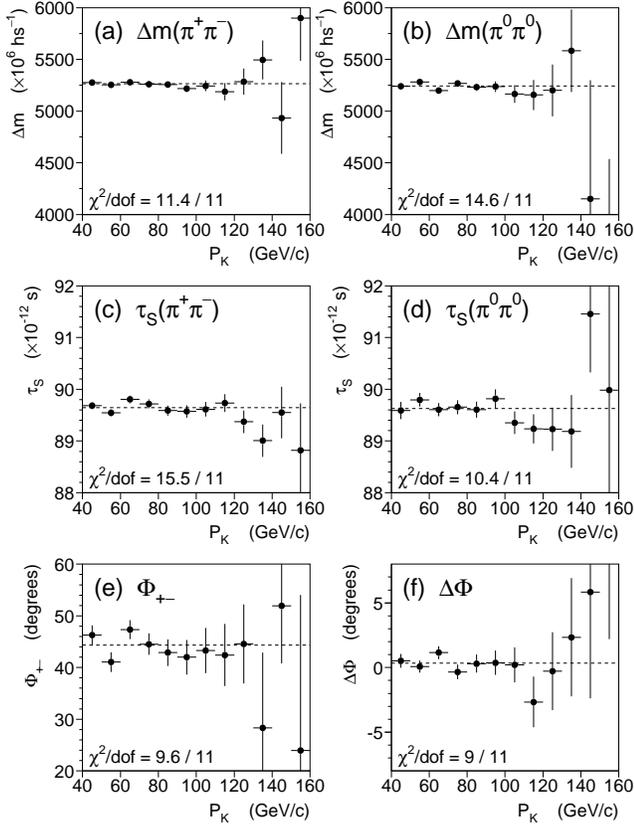,width=\linewidth}
  \caption{
       Kaon parameters as a function of kaon momentum and
       decay mode are shown
       for: 
       (a) $\delm$  from \ppc, (b) $\delm$ from \ppn,
       (c) $\tauS$  from \ppc, (d) $\tauS$ from \ppn,
       (e) $\phipm$, and
       (f) $\delphi$.
       Each dashed horizontal line is the average of all momentum bins,
       and the $\chi^2/$dof noted on each plot is for the consistency
       relative to the dashed line.
              }      
  \label{fig:kparvspk}
\end{figure}

The \Kpp\ samples are divided into various subsets,
among which we check the consistency of
\delm, \tauS, \phipm, and $\delphi$.
For all four measurements, we find good agreement
between five month-long time periods, 
the two regenerator positions,
and the two magnet polarities.
The consistency of \delm, \tauS, \phipm, and \delphi\
as a function of kaon momentum is shown in 
Fig.~\ref{fig:kparvspk}.
There is good agreement among the 12 momentum bins in both the
charged and neutral decay modes.
Allowing each parameter to have a slope as a function
of kaon momentum, 
the significance of each slope is between
$0.5\sigma$ and $1.5 \sigma$,
consistent with no momentum dependence.

\begin{figure}
  \epsfig{file=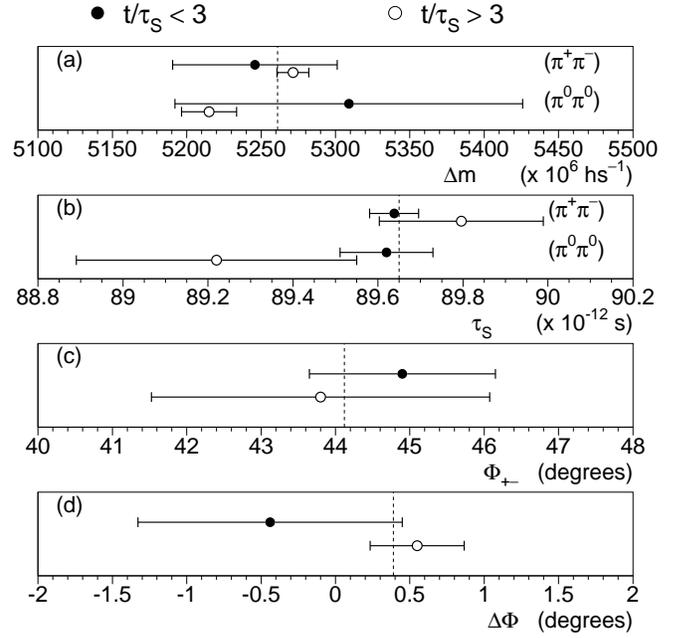,width=\linewidth}
  \caption{
       Dots (open circles) show the kaon parameter measurements 
       using regenerator beam data with proper time less (greater) 
       than three $K_S$ lifetimes relative to the regenerator edge.
       For (a) $\delm$ and (b) $\tauS$, 
       the charged and neutral mode results are shown separately.
       For each pair of measurements (dot and open circle), 
       the error bars reflect
       statistical uncertainties and are independent.
       Each vertical dashed line shows the nominal result
       using all of the data.
              } 
  \label{fig:kparvstaus}
\end{figure}

To check the dependence on proper decay time,
the regenerator beam \Kpp\  samples are divided into subsets
with proper time less than and greater than 3 $K_S$ lifetimes 
relative to the regenerator edge.
This test is also sensitive to the significant background variations
as a function of decay vertex (Fig.~\ref{fig:zbkgneut}).
The entire vacuum beam samples are used to determine the
kaon flux in each $10~\upk$ momentum bin, and the 
statistical uncertainty from the vacuum beam data
is subtracted for these comparisons.
For the sample with decays near the regenerator edge,
85\% of the \Kpp\ decay rate is from the $K_S$ term that 
is proportional to $\magrho^2$ in Eq.~\ref{eq:dnreg_dt};
for the other sample, 42\% of the \Kpp\ decay rate
is from the $K_S$ term.
Fig.~\ref{fig:kparvstaus} shows consistent results
between the two proper time ranges for the 
kaon parameter measurements.
Note that the measurement of $\phipm$, 
which has strong correlations with \delm\ and \tauS, 
is more sensitive to early decay times;
the measurement of \delphi, which is very weakly correlated with 
\delm\ and \tauS, is more sensitive to later decay times.

  \section{Conclusions}
  \label{sec:conclude}

In this paper, we report an improved measurement of direct
CP violation in the decay of the neutral kaon:
\begin{equation}
 \begin{array}{lcl}
    \reepoe & = & 
   \left[ \KtevReepoe \pm \KtevStat~({\rm stat}) \pm
  \TotSystMC~({\rm syst}) \right] \eu  \\
          & = & \left( \KtevReepoe \pm \KtevTErr \right) \eu ~.
 \end{array}
\end{equation}
This result, which supersedes reference \cite{prl:pss},
is consistent with the measurement from the
NA48 collaboration \cite{na48:reepoe,na48:blois02}
$\reepoe = (15.3 \pm 2.6)\eu$.
The average of our result and measurements from 
\cite{prl:731,pl:na31,na48:reepoe}
gives $\reepoe  = (17.2 \pm 1.8)\eu$ with a $13\%$ confidence level.

In addition, we report new measurements of 
the $K_L$-$K_S$ mass difference and the $K_S$ lifetime:
\begin{equation}
\begin{array}{lcl}
  \delm  & = & ( \KtevDelm  \pm \KtevDelmTerr ) \delmunits ~, \\
  \tauS  & = & ( \KtevTaus  \pm \KtevTausTerr ) \tausunits ~,
   \label{eq:delmtauS}
\end{array}
\end{equation}
where CPT symmetry is assumed.
Although these results are consistent with individual previous 
measurements used in the PDG averages~\cite{pdg00}, 
our results each differ from the 
PDG averages by more than two standard deviations.
The \ktev\  $\tauS$ measurement is consistent with a recent
NA48 measurement \cite{na48:taus},
and both results are much more precise than the
PDG average.

Finally, we measure phase differences
\begin{equation}
\begin{array}{lcl}
                 & \phipm - \phisw  & =  
                     (\dPhiSW \pm \dPhiSWTOTerr)\degs \\
  \delphi \equiv & \phizz - \phipm     & =  
                     (\DelPhi \pm \DelPhiTOTerr )\degs, 
\end{array}
\end{equation}
which are consistent with the CPT-symmetry prediction of zero.
These phase differences are extracted from fits in which
\delm\ and \tauS\ are free parameters to avoid
the CPT assumptions used to extract the nominal values
in Eq.~\ref{eq:delmtauS}.
The $\delphi$ result can be expressed in terms of 
$\imepoe$:
\begin{equation}
  \imepoe =  \IMEPOEpmErr ~.
\end{equation}

  \section{Acknowledgments}

We gratefully acknowledge the support and effort of the Fermilab
staff and the technical staffs of the participating institutions for
their vital contributions.  This work was supported in part by the U.S. 
Department of Energy, The National Science Foundation and The Ministry of
Education and Science of Japan. 
In addition, A.R.B., E.B. and S.V.S. 
acknowledge support from the NYI program of the NSF; A.R.B. and E.B. from 
the Alfred P. Sloan Foundation; E.B. from the OJI program of the DOE; 
K.H., T.N. and M.S. from the Japan Society for the Promotion of
Science; and R.F.B. from the Funda\c{c}\~{a}o de Amparo \`{a} 
Pesquisa do Estado de S\~{a}o Paulo.   P.S.S. 
acknowledges support from the Grainger Foundation.


                            \appendix


 \section{Principles of the \ktev\ $\reepoe$ Measurement}
 \label{app:exp details}

\def\tlts{\frac{\tau_L}{\tau_S}}

A simplified treatment of the \ktev\  measurement technique 
is presented here to illustrate some of the important cancellations 
that reduce systematic uncertainties in the $\reepoe$ measurement.
The measurement is based on the number of reconstructed
\Kpm\ and \Kzz\ decays in the vacuum and regenerator beams.
To illustrate how these four quantities are related to $\reepoe$,
it is convenient to ignore the \KSKL\ interference in both
beams.
In this simplified case,  the four measured quantities are  
related to experimental parameters as follows:
\begin{eqnarray}
     N(vac~\ppc) & \simeq & 
     \tlts B_S^{+-} {\cal F}_V^{+-} {\cal A}_V^{+-} \vert \etapm\vert^2 
     \frac{L}{\gamma c\tauL}
             \label{eq:dumfun1} \\
     N(reg~\ppc) & \simeq & 
     ~~~B_S^{+-}{\cal F}_R^{+-} {\cal A}_R^{+-} {\magrho}^2  \treg
             \label{eq:dumfun2} \\
     N(vac~\ppn) & \simeq & 
     \tlts B_S^{00}{\cal F}_V^{00} {\cal A}_V^{00}  \vert \etazz\vert^2 
     \frac{L}{\gamma c\tauL}
             \label{eq:dumfun3} \\
     N(reg~\ppn) & \simeq & 
     ~~~B_S^{00}{\cal F}_R^{00} {\cal A}_R^{00}  {\magrho}^2  \treg ~,
             \label{eq:dumfun4}
\end{eqnarray}
where the $N$'s are the observed number of \Kpp\ decays in each beam,
$B_S^{+-(00)}$ is the branching fraction of $K_S\to\ppc(\ppn)$,
${\cal F}_V^{+-(00)}$ are the vacuum beam kaon fluxes,
${\cal F}_R^{+-(00)}$ are the kaon fluxes
just upstream of the regenerator,
the  ${\cal A}$'s are the acceptances determined by 
a Monte Carlo simulation,
$L\sim 40~{\rm m}$ is the length of the useful decay region,
$\treg \sim 0.18$ is the kaon flux transmission through the regenerator,
$\gamma = E_K/M_K \sim 140$ is the average kaon boost,
and $\rho\sim 0.03$ is the regeneration amplitude for forward scatters.
The factor $L/\gamma c\tauS$ is not present
because essentially all of the $K_S$ decay within the decay region.
The charged mode vacuum-to-regenerator single ratio is
\begin{equation}
   r_{+-} = 
       \left\vert \frac{\etapm}{\rho} \right\vert^2 
       \tlts                                     \cdot
       \frac{ {\cal F}_V^{+-}}{ {\cal F}_R^{+-}} \cdot
       \frac{ {\cal A}_V^{+-}}{ {\cal A}_R^{+-}} \cdot
       \frac{L}{\gamma c\tauL}                   \cdot
       \frac{1}{\treg}
    \label{eq:single ratio}
\end{equation}
and similarly the neutral mode single ratio is obtained
with $+-$ replaced by $00$. 
To get a typical value of the single ratio, use 
$\vert\eta/\rho\vert \sim 0.07$,           
$\tauL/\tauS = 580$,                       
${\cal F}_V/{\cal F}_R \simeq 2.32$        
due to the movable absorber (Fig.~\ref{fig:detector}),
${\cal A}_V/{\cal A}_R \sim 0.8$           
and $L/\gamma c\tauL \sim 0.02$;           
this gives $r\sim 0.6$, which shows that \ktev\ is
designed to collect roughly the same statistics in the vacuum
and regenerator beams. 
The statistical precision on \reepoe\ is limited by the
number of vacuum beam \Kzz\ decays; 
with ${\cal F}_V^{00} \simeq 2$~MHz and 
${\cal A}_V^{00} \sim 0.05$,
the rate of $\KLzz$ is $\sim 2$~Hz.  
The 5\% acceptance used here is defined relative to all
kaons and includes a livetime factor of 0.7;
it is therefore smaller than the acceptance shown in
Figure~\ref{fig:mczacc} that is defined within a specific
momentum and $\ZK$-vertex range.

%
%
%

The desired quantity, $\vert\etapm/\etazz\vert^2$, 
is proportional to the experimentally measured 
double ratio, $r_{+-}/r_{00}$.  
The factors 
$\tauL/\tauS$, $\treg$,  $\magrho$, and  $L/\gamma c\tauL$
cancel in the double ratio.
For the kaon flux cancellation,
we use the constraint that the vacuum-to-regenerator 
kaon flux ratio is the same for
both the charged and neutral decay modes,
\begin{equation}
     R_F \equiv {\cal F}_{V+-}/{\cal F}_{R+-} = 
                {\cal F}_{V00}/{\cal F}_{R00} \simeq 2.32.
   \label{eq:vacreg}
\end{equation}
Note that Eq.~\ref{eq:vacreg} requires equal live-time for
the vacuum and regenerator beams;
the charged and neutral mode live-times,
which are  more difficult to control experimentally,
do not have to be the same.
If the beam collimation system results in 
small kaon-flux differences between the two
neutral beams, 
switching the regenerator position between the two beams 
ensures that Eq.~\ref{eq:vacreg} is satisfied.
The only quantities in the double ratio that do not cancel
are the acceptances,
for which we use a detailed Monte Carlo simulation.

The quantities $r_{+-}$ and $r_{00}$ can be measured
at different times provided that the following are 
the same for both measurements:
(i) the kaon transmission in both the 
movable absorber and regenerator,
(ii) the regeneration amplitude $\rho$ 
(iii) the distance between the primary BeO target
and regenerator.

When Equations~\ref{eq:dumfun1}-\ref{eq:dumfun4}
are modified to account for \KSKL\ interference,
there is no algebraic solution for $\reepoe$;
results are extracted from a fit described in 
Section~\ref{sec:fitting}.

 \section{Function for Kaon Scattering in the Regenerator}
 \label{app:fitfun}

This appendix describes the function used to model
the \Kpp\ decay distribution for kaons that scatter
in the regenerator.
The function depends on decay time, $\ptsq$, and kaon momentum,
and is also used to
separate diffractive and inelastic contributions.
The functional form is
\begin{equation}
   N_{\rm regscat}  \propto  \sum_{j=1}^{6}
          A_j e^{\alpha_j p_t^2}
         \vert \hat{\rho}_j e^{\Lambda_S t} + \eta e^{\Lambda_L t} \vert^2
          ~,
   \label{eq:fitfun}
\end{equation}
where $A_j$, $\alpha_j$, $\vert\hat{\rho}_j\vert$ and 
$\phi_{\hat{\rho}_j}$ are the 24 fit parameters,
$\Lambda_{S,L} = i m_{S,L} - \frac{1}{2}\Gamma_{S,L}$,
and $t$ is the proper time of the decay.
Note that each $\hat{\rho}_j$ is an independent 
regeneration amplitude for scattering.
The terms in Eq.~\ref{eq:fitfun} can be roughly associated with
the known properties of kaon scattering in carbon, hydrogen and lead.
In addition to these 24 parameters, there are two 
extra parameters that describe the momentum dependence of
the phase ($\phi_{\hat{\rho}_j}$) and $\ptsq$ slope ($\alpha_j$)
associated with diffractive scattering from the lead
at the downstream edge of the regenerator.
Of the 26 parameters in Eq.~\ref{eq:fitfun}, 
8 are fixed based on previously measured
properties of kaon scattering.
An additional 12 parameters are used to float the 
momentum dependence of $N_{\rm regscat}$ in $10~\upk$ bins;
the momentum dependence for scattered kaons varies by only a few 
percent compared to that of unscattered kaons.
The total number of free parameters in the regenerator scattering
function is $18+12 = 30$.

\begin{figure}
  \epsfig{file=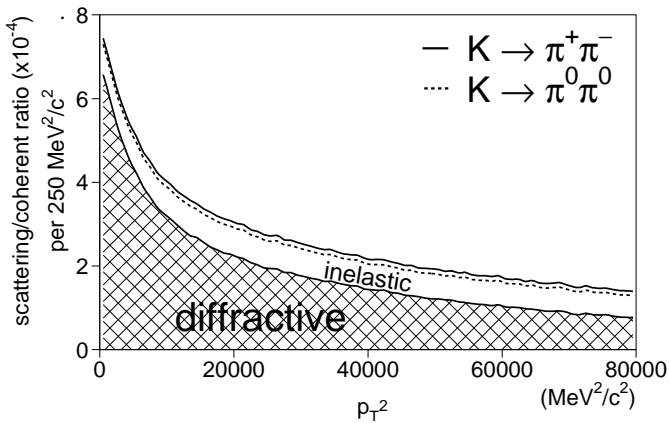,width=\linewidth}
  \caption{ 
       Regenerator scattering \ptsq\ distribution (solid curve)
       from a fit to acceptance-corrected \Kpm\ data.
       The normalization 
       corresponds to the measured 0.074\% scattering-to-coherent 
       background level in the charged decay mode
       (Table~\ref{tab:bkgd}).
       The hatched (clear) region shows the diffractive 
       (inelastic) component.
       The total (diffractive + inelastic) scattering-to-coherent        
       ratio is shown for \Kpm\ (solid) and \Kzz\ (dashed).
       The 3\% charged-neutral difference reflects the       
       additional 16\% suppression of inelastic scattering
       in the neutral mode analysis.
         }
  \label{fig:regscat}
\end{figure}

The $\alpha_j$ parameters, 
which describe the exponential $\ptsq$ dependence, 
are used to distinguish between inelastic and diffractive scattering. 
The term with the broadest \ptsq\ distribution has 
$\alpha^{-1} = 2.4\times 10^5~\uptsq$, and is identified
with inelastic scattering.
The other terms have much steeper $\ptsq$ distributions,
with $5000 < \alpha^{-1} < 70000 ~\uptsq$,
and are associated with diffractive scattering.
Figure~\ref{fig:regscat} shows the diffractive and inelastic
contributions to regenerator scattering.

After determining the shapes of the diffractive and inelastic
scattering distributions, the next step is the absolute
normalization of the background relative to the 
coherent \Kpp\ signal.
In the charged analysis, we define $\nscat{+-}$ 
to be the number of reconstructed \Kpm\ events 
with $\ptsq > 2500~\uptsq$,
and $\ncoh{+-}$ to be the number of coherent events 
with $\ptsq < 250~\uptsq$ 
(Fig.~\ref{fig:ptsq_shapes}).
The scattering level is adjusted in the simulation 
so that the reconstructed $\nscat{+-}/\ncoh{+-}$ ratio
is the same in the data and MC.
Note that $\nscat{+-}$ in data is obtained after subtracting
collimator scatters and semileptonic decays.

In the neutral mode analysis, we define similar quantities
in the regenerator beam:
$\nscat{00}$ is the number of reconstructed \Kzz\
events with  $300 < \ring < 800~{\rm cm}^2$ 
(after subtracting the other background components)
and $\ncoh{00}$ is the number with
$\ring < 110~{\rm cm}^2$ (Fig.~\ref{fig:ringvaceps}).
Using the scattering-to-coherent ratio determined
with acceptance-corrected \Kpm\ decays, 
the simulation over-predicts the  
$\nscat{00}/\ncoh{00}$ ratio in data by 3\%;
this difference results from the additional veto requirements 
in the neutral mode analysis, and is illustrated
in Fig.~\ref{fig:regscat}.
To match the $\nscat{00}/\ncoh{00}$ ratio in data,
the neutral mode scattering simulation is adjusted by a
16\% reduction in the inelastic scattering level;
note that the diffractive scattering level is not affected
by the veto cuts, and is therefore the same in the charged
and neutral analyses.

If we ignore the difference between diffractive and inelastic
scattering,
the simulated $\nscat{00}/\ncoh{00}$ ratio can be adjusted
by a 3\% reduction in the total scattering level.
Compared with the nominal 16\% reduction in the inelastic level,
a global 3\% adjustment 
gives equally good data-MC agreement in the $\ring$
distribution from 300 to 800~cm$^2$
because the diffractive and inelastic distributions are very 
similar for $\ptsq > 30000~\uptsq$.
In the coherent signal region, however, 
the diffractive and inelastic scattering distributions 
are different, 
leading to a 0.01\% difference in the background prediction between
these two ways of adjusting of the $\nscat{00}/\ncoh{00}$ ratio.
This difference is used in evaluating the systematic uncertainty
on $\reepoe$ in Section~\ref{sec:bkg_syst}.

 \section{Discussion of the Previous \ktev\ Measurement of $\reepoe$}
 \label{sec:prl99}

The $\reepoe$ result reported in Section~\ref{sec:reepoe_results}
includes a full reanalysis of our
previously published data sample \cite{prl:pss}.  
The reanalysis of that data sample gives
\begin{eqnarray}
  \reepoe  & = & 
   [ \KtevReepoeA  \pm  3.0~({\rm stat}) 
                   \pm   2.9~({\rm syst}) ] \eu \nonumber \\
          & = &
   [ \KtevReepoeA  \pm  \KtevTErrA~ ]\eu   \\
         &   & (\mbox{Reference~\cite{prl:pss}~sample}) ~,   \nonumber
\end{eqnarray}
with $\chi^2/dof = 18.7/21$.
The $\reepoe$ shift is $-4.8 \eu$ relative to 
the previous result in \cite{prl:pss}, 
and is larger than the previous systematic uncertainty of $2.8 \eu$. 
The shifts in $\reepoe$ resulting from changes in the
analysis are summarized in 
Table~\ref{tab:prl99 shift}.
The change in the regenerator scattering background 
for \Kzz\ is mainly from correcting
an error in the kaon scattering function.
The other shifts are due to improvements and are consistent
with the systematic uncertainties assigned in \cite{prl:pss}.
All of the changes in Table~\ref{tab:prl99 shift} 
are from independent sources which are described below.

\begin{table}[th]
\caption{
   \label{tab:prl99 shift}
    Sources of the $\reepoe$ shift between reference \cite{prl:pss} 
    and re-analysis of the same data sample. 
    The systematic $\sigma$'s ($\sigma_{syst}$) are from \cite{prl:pss}.
       }
\begin{ruledtabular}
\begin{tabular}{ l| c | c } 
Source of        & $\reepoe$ shift  & Number              \\ 
change           & $(\eu)$          & of $\sigma_{syst}$  \\ \hline
Regenerator scattering & $-1.7$   & 2.1  \\
Collimator scattering  & $-0.2$   & 0.6  \\
Screening                & $-0.3$   & 1.5  \\
Regenerator transmission & $-0.3$   & 1.5  \\
Regenerator edge         & $-0.4$   & 1.6  \\
Neutral analysis       & $+0.1$   & 0.1  \\
Mask Anti MC           & $+0.3$   & 1.3  \\
Absorber scattering MC & $-0.6$   & 0.6  \\
\ktev\ \delm\ and \tauS\      
                       & $-0.5$   & 
   3.1\footnote{Reference \cite{prl:pss} used PDG98 values and errors
               for \delm\ and \tauS.}      
                        \\
MC fluctuation         & $-1.1$   & 1.0 \\
\hline
Total                  & $-4.8$   & 1.7  \\
\end{tabular}
\end{ruledtabular}
\end{table}

\begin{itemize}
  \item The regenerator scattering function
        includes two extra scattering terms (Eq.~\ref{eq:fitfun})
        and two extra parameters to simulate subtle 
        momentum-dependent features.
  \item  Collimator scattering (Sec.~\ref{sec:bkg_coscat}) is measured 
         after subtracting semileptonic decays, 
         and the simulation includes two extra parameters to better 
         describe the $K_L$ $\ptsq$-dependence.
  \item  In the $\reepoe$ fit (Sec.~\ref{sec:fitting}), 
         a nuclear screening correction is used.
  \item  The energy dependence of the regenerator transmission 
         is measured with four times more $\KLpmz$ data.
  \item  The calculation of the effective regenerator edge in the 
         neutral decay mode (Fig.~\ref{fig:regdiagram}b) 
         includes kaon decays upstream of the last regenerator-lead piece
         instead of considering only decays in the last piece of lead.
         The effective regenerator edge in charged mode is based on a 
         better technique to measure the veto threshold.
  \item  Additional neutral mode veto cuts on hits in the trigger 
         hodoscope  and drift chambers reduce sensitivity to the
         transverse energy cut on CsI clusters.
  \item  The Mask Anti simulation allows kaons and photons 
         to punch through the lead-scintillator outside the
         beam-hole regions (Fig.~\ref{fig:RCMA}).
  \item  The absorber scattering simulation is improved to match beam 
         shapes in a special run without the regenerator, 
         but with the movable absorber still in place.
  \item  We use our measurements of \delm\ and \tauS\ that each 
         differ by more than $2\sigma$ from the 
         PDG averages \cite{pdg00}.
\end{itemize}
The changes listed above account for $-3.7\eu$ of the shift in
$\reepoe$. 
We attribute the remaining shift of $-1.1\eu$
to a $1 \sigma$ fluctuation between the two independent MC 
samples used for the acceptance correction.

The largest systematic uncertainty in \cite{prl:pss} 
was based on the $z$-dependence of the acceptance for \KLpm\ decays.
Unfortunately, this data-MC discrepancy still remains in the
reanalysis.

 \section{Correlations Among Kaon Parameter Measurements}
 \label{app:kparcor}

The determination of correlations among different kaon parameters, 
including correlations arising from systematic uncertainties, 
is based on the following $\chi^2$:
\begin{equation}
\begin{array}{lcl}
 \label{eq:syst}
 \chi^2(X,\alpha_i) & = & (X-\overline{X}) C^{-1} (X-\overline{X}) 
                           + \sum_i \alpha_i^2 \\
     & &  \\
  \hfill \overline{X} & = & \overline{X_0} + \sum_i \beta_i \alpha_i ~.
\end{array}
\end{equation}
$X$ is the vector of the measured physics quantities from the fit
({\it e.g.}, \delm\ and \tauS ),
and $C$ is the covariance matrix including 
statistical uncertainties for both data and MC.
The $\alpha_i$ are fit parameters representing the 
number of ``$\sigma$'' associated with each
source of systematic uncertainty that causes correlations
among the physics quantities 
($\alpha_i = 1$ corresponds to $1\sigma$ change).
$\overline{X_0}$ is the vector of physics quantities
obtained in the nominal fit with all $\alpha_i = 0$,
and $\beta_i$ are the changes in the central value of $X$ 
corresponding to a $1\sigma$ 
change in the systematic source $i$.

The total uncertainty in a physics quantity
obtained from the $\chi^2$ in Eq.~$\ref{eq:syst}$ 
is equivalent to adding 
statistical and systematic uncertainties
in quadrature. 
The minimization of this $\chi^2$ accounts for
correlations among the  physics quantities
for each source of systematic uncertainty.
We ignore correlations among the different
sources of systematic uncertainty; 
these correlations are reduced by the 
grouping of systematic uncertainties shown in
Table~\ref{tb:syst_kpar}.

\begin{figure}[ht]
 \epsfig{file=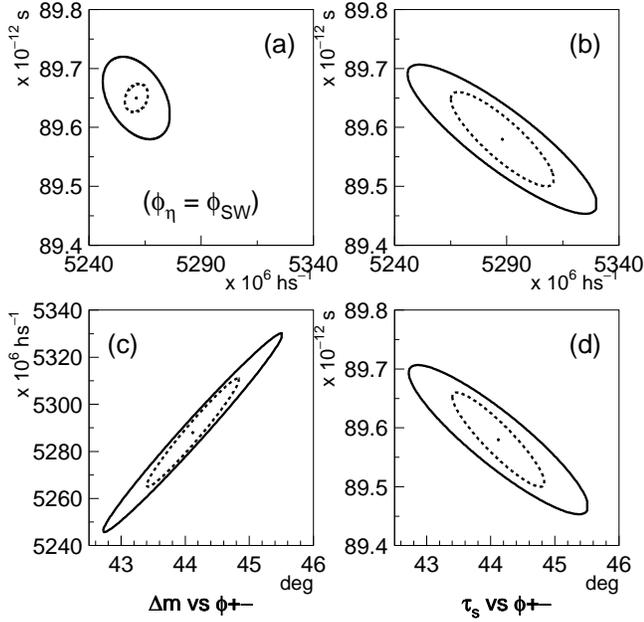, width=\linewidth}
\caption{
   \label{fig:phipmcorr} 
    $1\sigma$ contours of statistical uncertainty (dashed) and 
    total uncertainty (solid) for 
    (a) $\delm$-$\tauS$ fit (combined charged+neutral),
    and for (b)-(d) correlations from the $\phipm$ fit.
        }
\end{figure}

\begin{figure}[ht]
 \epsfig{file=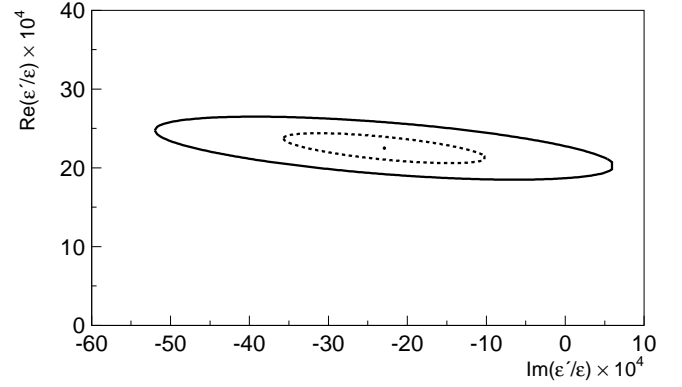, width=\linewidth}
\caption{
    \label{fig:imrep} 
    $1\sigma$ contours of statistical uncertainty (dashed) and 
    total uncertainty (solid) for $\reepoe$ vs. $\imepoe$
    from the fit with no CPT assumptions related to 
    $\phipm$ and $\phizz$.
       }
\end{figure}

The total uncertainties and correlation coefficients for 
all of the fits are given in Table~\ref{tb:syst_kparcorr}.
Note that the statistical and systematic uncertainties
are larger in the fits in which CPT symmetry is not assumed.
Figure~\ref{fig:phipmcorr}  shows $1\sigma$ 
contours of statistical and total uncertainties for the
measurements of \delm, \tauS, and \phipm.
For the $\delm$-$\tauS$ fit (Fig.~\ref{fig:phipmcorr}a),
systematic uncertainties have a significant effect on
the $\delm$-$\tauS$ correlation.
For the $\phipm$ fit (Fig.~\ref{fig:phipmcorr}b-d),
systematic uncertainties have a very small effect
on the correlations.
Figure~\ref{fig:imrep} shows the correlation between
the real and imaginary parts of $\epe$ from the fit
without CPT assumptions;
systematic uncertainties have a small effect on the correlation.


\begin{table}[ht]
\caption{
   \label{tb:syst_kparcorr}
   Total uncertainties and correlation coefficients for 
   the three fits. The fit parameters and CPT assumptions
   are listed above each set of values.
        }
\begin{ruledtabular}
\begin{tabular}{lccc}
\multicolumn{3}{l}{I. $\delm$-$\tauS$ Fit with CPT Assumption:} \\
        & \delm\  &  \tauS\  &   \\
\hline
Total Error  & $\KtevDelmTerr\delmunits$  
             & $\KtevTausTerr\tausunits $  
             &                              \\ \hline 
Correlation coefficients: &           &       &         \\
~~~ \delm                 &   1.      &       &         \\
~~~ \tauS                 & $-0.396$  &   1.  &         \\
\hline\hline
        & & & \\
\multicolumn{3}{l}{II. $\delm$-$\tauS$-$\phipm$ Fit without CPT Assumption:} \\
        & \delm\  &  \tauS\ & \phipm\ \\
\hline
Total Error  & $42\delmunits$  & ~~$0.13\tausunits$ & $1.40\degs $  \\
\hline
Correlation coefficients: &           &           &      \\
~~~ \delm                 &   1.      &           &         \\
~~~ \tauS                 & $-0.874$  &   1.      &         \\
~~~ \phipm\               & $+0.987$  & $-0.898$  &   1.    \\
\hline\hline
        & & & \\
\multicolumn{3}{l}{III. $\epe$ Fit without CPT Assumption:} \\
        & \reepoe\  &  \imepoe\  &  \\
\hline
Total Error  & $ 4.0\eu$   & $\IMEPOETotErr\eu$  &  \\
\hline
Correlation coefficients:  &           &         &   \\
~~~  \reepoe\              &   1.      &         &   \\
~~~  \imepoe\              & $-0.647$  &   1.    &   \\
\end{tabular}
\end{ruledtabular}
\end{table}

\newpage

%
%


\end{document}